\documentclass[natbib,smallextended]{svjour3}       
\usepackage{graphicx}
\usepackage{booktabs}  
\usepackage{xspace}
\usepackage{lscape}

\usepackage[colorlinks=true,citecolor=blue,urlcolor=blue,breaklinks]{hyperref}

\graphicspath{{figs/},{upload1/},{../}}

\journalname{Living Reviews in Solar Physics}

\newcommand{\dv}[2]{\frac{{\mathrm d}#1}{{\mathrm d}#2}}
\newcommand{\pdv}[2]{\frac{\partial #1}{\partial #2}}
\newcommand{\vc}[1]{\mathbf{#1}}
\newcommand{\sgn}{{\mathrm{sgn}\,}}
\newcommand{\Rmax}{R_{\mathrm{max}}}
\newcommand{\Rmin}{R_{\mathrm{min}}}
\newcommand{\tmax}{t_{\mathrm{max}}}
\newcommand{\tmin}{t_{\mathrm{min}}}

\newcommand{\R}{$R$\xspace}
\newcommand{\RZ}{$R_Z$\xspace}
\newcommand{\RB}{$R_B$\xspace}

\newcommand{\sqkms}{{km$^2/$s}\xspace}
\newcommand{\Rsun}{R_\odot}

\def\lesssim{\mathrel{\mathchoice {\vcenter{\offinterlineskip\halign{\hfil
 $\displaystyle##$\hfil\cr<\cr\sim\cr}}}
 {\vcenter{\offinterlineskip\halign{\hfil$\textstyle##$\hfil\cr
 <\cr\sim\cr}}}
 {\vcenter{\offinterlineskip\halign{\hfil$\scriptstyle##$\hfil\cr
 <\cr\sim\cr}}}
 {\vcenter{\offinterlineskip\halign{\hfil$\scriptscriptstyle##$\hfil\cr
 <\cr\sim\cr}}}}}

\newcommand{\ppd}[2]{\frac{\partial^2 #1}{\partial {#2}^2}}
\newcommand{\varv}{v}

\newenvironment{descr}{\begin{description}}{\end{description}}
\newenvironment{nquote}{\bigskip\strut\hfill\begin{minipage}{0.8\textwidth}}{\end{minipage}\bigskip}

\newcommand{\newtext}{\bf }

\newcommand{\newsection}[1]{\section{#1 {\bf [NEW!]} }}
\newcommand{\newsubsection}[1]{\subsection{#1 {\bf [NEW!]} }}
\newcommand{\newsubsubsection}[1]{\subsubsection{#1 {\bf [NEW!]} }}
\newcommand{\oldtext}{\sl }

\renewcommand{\newtext}{}
\renewcommand{\newsubsection}[1]{\subsection{#1}}
\renewcommand{\newsubsubsection}[1]{\subsubsection{#1}}
\renewcommand{\oldtext}{}

\newcommand{\revone}{}

\begin{document}

%
%
%
%
%
%
%

\title{Solar cycle prediction%
}

\author{Krist\'of Petrovay}

\institute{K. Petrovay \at 
E{\"o}tv{\"o}s Lor\'and University, Department of Astronomy \\
Budapest, Hungary\\
\email{K.Petrovay@astro.elte.hu}
}

\date{Received: date / Accepted: date}

\maketitle

\begin{abstract} 
A review of solar cycle prediction methods and their performance is given,
including {\newtext early forecasts for Cycle~25.} The review focuses on those
aspects of the solar cycle prediction problem that have a bearing on dynamo
theory. The scope of the review is further restricted to the issue of predicting
the amplitude (and optionally the epoch) of an upcoming solar maximum no later
than right after the start of the given cycle. 

Prediction methods form three main groups. \emph{Precursor methods} rely on the
value of some measure of solar activity or magnetism at a specified time to
predict the amplitude of the following solar maximum. {\newtext The choice of a
good precursor often implies considerable physical insight: indeed, it has become
increasingly clear that the transition from purely empirical precursors to 
\emph{model-based methods} is continuous. Model-based approaches can be further
divided into two groups: predictions based on surface flux transport models and
on consistent dynamo models.

The implicit assumption of precursor methods }
is that each numbered solar cycle is a consistent unit in itself, while solar
activity seems to consist of a series of much less tightly intercorrelated
individual cycles. \emph{Extrapolation methods,} in contrast, are based on the
premise that the physical process giving rise to the sunspot number record is
statistically  homogeneous, i.e., the mathematical regularities underlying its
variations are the same at any point of time, and therefore it lends itself to
analysis and forecasting by time series methods. 

In their overall performance during the course of the last few solar cycles,
precursor methods have clearly been superior to extrapolation methods. One
method that has yielded predictions consistently in the right range during the
past few solar cycles is the polar field precursor. Nevertheless, some
extrapolation methods may still be worth further study.  {\newtext Model based
forecasts are quickly coming into their own, and, despite not having a long
proven record, their predictions are received with increasing confidence by the
community.}

\end{abstract}

\setcounter{tocdepth}{3}
\tableofcontents



\section{Introduction}
\label{section:introduction}

Solar cycle prediction is an extremely extensive topic, covering a very wide
variety of proposed prediction methods and prediction attempts on many different
timescales, ranging from short term (month--year) forecasts of the runoff of the
ongoing solar cycle to predictions of long term changes in solar activity on 
centennial or even millennial scales. As early as 1963, Vitinsky published a
whole monograph on the subject, later updated and extended
(\citealt{Vitinsky:book1}, \citeyear{Vitinsky:book2}). More recent overviews of
the field or aspects of it include \cite{Hathaway:prediction.review},
\cite{Kane:pred23rev}, \cite{Pesnell}, and {\newtext the first edition of this review
\citep{Edition1}.} 
In order to narrow down the scope of the review, we constrain our field
of interest in two important respects.

Firstly, instead of attempting to give a general review of all prediction
methods suggested or citing all the papers with forecasts, here we will focus on
those aspects of the solar cycle prediction problem that have a bearing on
dynamo theory. We will thus discuss in more detail empirical methods that,
independently of their success rate, have the potential of shedding some light
on the physical mechanism underlying the solar cycle, as well as the prediction
attempts based on solar dynamo models.

Secondly, we will here only be concerned with the issue of \textit{predicting the
amplitude} (and optionally the epoch) \textit{of an upcoming solar maximum no later
than right after the start of the given cycle}. This emphasis is also motivated
by the present surge of interest in precisely this topic, prompted by the
unusually long and deep recent solar minimum and by sharply conflicting
forecasts for the maximum of the incipient solar Cycle~24.

As we will see, significant doubts arise both from the theoretical and
observational side as to what extent such a prediction is possible at all
(especially before the time of the minimum has become known). Nevertheless, no
matter how shaky their theoretical and empirical backgrounds may be, forecasts
\textit{must} be attempted. Making verifiable or falsifiable predictions is
obviously the core of the scientific method in general; but there is also a more
imperative urge in the case of solar cycle prediction. Being the prime
determinant of space weather {\revone  and space climate}, solar activity
clearly has enormous technical, scientific, and financial impact on activities
ranging from space exploration to civil aviation and everyday communication.
Political and economic decision makers expect the solar community to provide
them with forecasts on which feasibility and profitability calculations can be
based. Acknowledging this need, {\newtext around the time of solar minimum} the
Space Weather Prediction Center of the US National Weather Service does present
annually or semiannually updated ``official'' predictions of the upcoming
sunspot maximum, emitted by a Solar Cycle Prediction Panel of experts. The
unusual lack of consensus {\revone in} the early meetings of this panel during the
{\newtext previous} minimum (\citealt{SWPC2009upd}), as well as the concurrent,
more frequently updated but wildly varying predictions of a NASA MSFC team
(\citealt{MSFC2009}) {\newtext made evident the deficiencies of prediction
techniques available at the time. In view of this, Cycle~24 provided  us with
some crucial new insights into the physical mechanisms underlying cyclic solar
activity. In preparation to convene the Solar Cycle~25 Prediction Panel a new
call for predictions was issued in January 2019 (\citealt{SWPC2019call}). }

While a number of indicators of solar activity exist, by far the most commonly
employed is still the smoothed relative sunspot number \R; the ``Holy Grail'' of
sunspot cycle prediction attempts is to get \R right for the next maximum. We,
therefore, start by briefly introducing the sunspot number and inspecting its
known record. Then, in Sects.~\ref{sect:precursor},
\ref{sec:Extrapolation-Methods}, and \ref{sec:Model-Based-Predictions} we
discuss the most widely employed methods of cycle predictions.
Section~\ref{sect:summary} presents a summary evaluation of the past
performance of different forecasting methods, {\newtext while 
Sect.~\ref{sect:cyc25} finally collects some early forecasts for Cycle~25
derived by various approaches.} 

\newsubsubsection{What's new in this edition?}

For readers familiar with the 1st edition of this review, who would prefer to go
through the ``new stuff'' only, here I briefly list the new or thoroughly
rewritten sections.

The revision of the sunspot number series that took place in 2015 is one topic
that had to be discussed in detail. The new Sect.~\ref{sec:SSNrevision} is
devoted to this subject but other parts of Sects.~\ref{sect:wolfno} and
\ref{sec:indicators} have also been subjected to a major revision.

For reasons explained in Sect.~\ref{sect:classif}, the overall structure of
the review has been given a major overhaul: the section on extrapolation methods
has now been placed \emph{after} the section on model-based approaches. 

Researches into the origins of the sudden change in the behaviour of our Sun 
from Cycle~23 to 24 led to important, although still poorly understood
realizations, now discussed in a dedicated subsection (Sect.~\ref{sect:portents}). And the stellar rise in popularity of the polar precursor
led to such an amount of exciting new research that Sect.~\ref{sect:polar},
discussing this method, had to be completely rewritten and expanded;
furthermore, it gave rise to another section on ``The quest for a precursor of
the polar precursor'' (Sect.~\ref{sec:quest}), containing mostly new or
thoroughly updated material.

The section on model-based predictions now includes a presentation of the
approach based on surface flux transport models (Sect.~\ref{sec:SFT}). In the field of
dynamo-based cycle prediction the major novelty in this decade was the
development of nonaxisymmetric models capable to account for the emergence of
individual active regions: a new subsection is now devoted to this topic (Sect.
\ref{sec:model_nonaxi}).

Finally, the updated summary evaluation and the overview of early predictions for
Cycle~25 obviously cover mostly new results (Sect.~\ref{sect:summary}
and \ref{sect:cyc25}).

\subsection{The sunspot number (SSN)} 
\label{sect:wolfno}

Despite its somewhat arbitrary construction, the series of relative sunspot 
numbers constitutes the longest homogeneous global indicator of solar activity
determined by direct solar observations and {\newtext by methods that were,
until recently, perceived to be carefully controlled}. For this reason, their
use is still predominant in studies of solar activity variation. 

{\newtext 
As aptly noted by \cite{Clette+:SSNrevision}, until recently the sunspot number
series was ``assumed to be carved in stone,  i.e., it was considered largely as a
homogeneous, well-understood and thus immutable data set. This feeling was
probably reinforced by the stately process through which it was produced by a
single expert center at the Z\"urich Observatory during 131 years.''

This perception has now been shattered by the major revision of the official
sunspot number series that took place in 2015, and opened the way to further
periodic revisions.

In what follows, the original series, the revision, and transformed versions of
the series will be discussed in turn.

}

\subsubsection{Version 1.0}

As defined originally by \cite{Wolf:relno.def}, the relative sunspot
number is
\begin{equation}
R_Z=k(10\,g+f)\,,  
\end{equation}
where $g$ is the number of sunspot groups (including solitary spots), $f$ is the
total number of all spots visible on the solar disc, while $k$ is a correction
factor depending on a variety of circumstances, such as instrument parameters,
observatory location, and details of the counting method. Wolf, who decided to
count each spot only once and did not count the smallest spots, used $k=1$.
{\newtext He also introduced a hierarchical system for the determination of \RZ
where data from a list of auxiliary observers were used for days
when the primary observer (Z\"urich) could not provide a value.} 

The counting system employed {\newtext in Z\"urich} was changed by Wolf's
successors to count even the smallest spots, attributing a higher weight (i.e.,
$f>1$) to spots with a penumbra, depending on their size and umbral structure.
{\newtext (The exact details and timing of these changes are incompletely
documented and controversial, see discussion in the next subsection.)} As the
changes in the counting and the regular use of a larger telescope naturally
resulted in higher values, the Z\"urich correction factor was set to $k=0.6$ for
subsequent determinations of \RZ to ensure continuity with Wolf's work.
(\citealt{Waldmeier:book}; see also {\newtext \citealt{Izenman},}
\citealt{Kopecky_:relativeno}, \citealt{Hoyt_Schatten:GSN}, {\newtext
\citealt{Friedli:Wolf}} for further discussions on the determination of \RZ).

In addition to introducing the relative sunspot number, 
\cite{Wolf:cycle.length.activity} also used earlier observational records
available to him to reconstruct its monthly mean values since 1749. In this way,
he reconstructed 11-year sunspot cycles back to that date: {\newtext hence, the
now universally used numbering of solar cycles starts with the first complete
cycle in the monthly \RZ series}. In a later work he also determined annual mean
values for each calendar year going back to 1700. {\newtext This reconstruction
and calibration work took place in several steps, so the \RZ record was very
much a project in the making until the end of the 19th century (see
\citealt{Clette+:SSNrevision}). It was only from the early 20th century that the
series came to be regarded as ``carved in stone''.}

In 1981, the observatory responsible for the official determination of the
sunspot number changed from Z\"urich to the Royal Observatory of Belgium in
Brussels. The website of the SIDC\footnote{\url{http://sidc.oma.be}}
(originally  Sunspot Index Data Center, recently renamed Solar Influences Data
Analysis Center) is now the most authoritative source of archive sunspot number
data. {\newtext The department of SIDC formally responsible for the sunspot
number series is WDC-SILSO (World Data Centre for Sunspot Index and Long-term
Solar Observations).} It has become customary to denote the original Z\"urich
series with \RZ (``the Z\"urich sunspot number''), and its continuation by the
SIDC from 1981 to 2015 with $R_i$ (International Sunspot Number, ISN). The new,
revised series is conventionally denoted by $S_N$.

It must be kept in mind that {\newtext since the middle of the 20th century,}
the sunspot number is also regularly determined by other institutions: {\newtext
the most widely used} such variants are informally known as the American sunspot
number (collected by AAVSO and available from the National Geophysical Data
Center\footnote{\url{http://www.ngdc.noaa.gov/ngdc.html}})  and the Kislovodsk
Sunspot Number (available from the web page of {\newtext the Kislovodsk Mountain
Astronomical Station of} Pulkovo
Observatory\footnote{\url{http://en.solarstation.ru/}}). 



Given that \RZ is subject to large fluctuations on a time scale of days to
months, it has become customary to use annual mean values for the study of
longer term activity changes. To get rid of the arbitrariness of calendar years,
the standard practice\footnote{Alternative proposals have been put
forward by \cite{Munozjara+:polarfacproxy} and \cite{Podladchikova+:smoothing};
who pointed out that Eq.~(\ref{eq:Rdef}) does a poor job of filtering out high-frequency variations, and
use better or even optimized weight factors for the $R_{\mathrm{m},i}$'s 
instead.}
is to use 13-month boxcar averages of the monthly averaged sunspot numbers,
wherein the first and last months are given half the weight of other months:
\begin{equation}
  R=\frac1{24}\left(R_{\mathrm{m},-6}+2\sum_{i=-5}^{i=5} R_{\mathrm{m},i} +R_{\mathrm{m},6}\right)\,,
  \label{eq:Rdef}
\end{equation}
%
$R_{\mathrm{m},i}$ being the mean monthly value of the daily sunspot number
values for \textit{i}th calendar month counted from the present month. It is
this running mean \R that we will simply call ``the sunspot number'' throughout
this review and what forms the basis of most discussions of solar cycle
variations and their predictions. 

In what follows, $\Rmax^{(n)}$ and $\Rmin^{(n)}$ will refer to the maximum and
minimum value of \R in cycle $n$ (the minimum being the one that starts the
cycle). Similarly, $\tmax^{(n)}$ and $\tmin^{(n)}$ will denote the epochs when
\R takes these extrema.

\newsubsubsection{Revision}
\label{sec:SSNrevision}

The process that led to the 2015 revision was started by Leif Svalgaard
(Stanford) who pointed out a number of inhomogeneities in the series, rooted in
changes in the base data and processing techniques. Starting from 2011, at
Svalgaard's initiative, a series of four workshops on the sunspot number were held
by the community involved. The 2015 revision is the result of
this process. The motivation for and
the detailed process of the revision was described by \cite{Clette+:SSNrevision}
and \cite{Clette+:SSNv2} and discussed in a number of papers in a topical issue
of \textit{Solar Physics} (vol.~291, issue 9--10; \citealt{Clette+:topical}). 

The revision included dropping the $k=0.6$ scaling factor traditionally
applied to the Z\"urich data, so all values increased by a factor of
$\textstyle\frac
53$. In addition to this trivial rescaling and some other minor changes, 
three major corrections were implemented.

\textit{(a) The Locarno drift after 1981.} The determination of the International
Sunspot Number $R_i$ by the SIDC did not follow Wolf's hierarchical system,
taking into account observations from all network stations and only dropping
outliers. Nevertheless, in order to ensure continuity, the Locarno solar
observatory (Z\"urich's successor) still had a special role as a pilot station,
all other observers being calibrated to Locarno's scale. A slow time-varying
drift in the Locarno data came to light during the revision process and has been
corrected in the new series. This change is apparently uncontroversial and was
made with the full consensus of all actors involved
(\citealt{Clette+:Locarnodrift}).

\textit{(b) The ``Waldmeier jump'' from 1947.} Plotting the original Z\"urich sunspot numbers
against other sunspot-related indices such as sunspot areas or group numbers, or
even against non-weighted sunspot numbers determined by non-Z\"urich observers,
a jump was discovered which was suggested to originate in the introduction of a
new weighting method (in use in Locarno until the present day) by the new
Z\"urich director, Max Waldmeier and his largely new staff. Under current solar
conditions the weighting results in a 15--20\% inflation of the sunspot
numbers (\citealt{Svalgaard+:weighting}). This assumed inflation of the series
has now been corrected.

\textit{(c) The Schwabe--Wolf transition (1849--64)} The upward correction of
14\% applied in this period relies primarily on a comparison of the original
sunspot number series with group sunspot numbers (the result being apparently
insensitive to which group number series is used). The presumed cause  of the
discrepancy is that in this period the sunspot number was determined by Wolf
using small portable telescopes, while Schwabe also continued his observations.
For days not covered by his own observations Wolf used Schwabe's data without
marking these out. It was only in 1861 that, upon cross-correlating their data
Wolf determined a correction factor $k=1.25$ for Schwabe, which he also applied
retrospectively to the pre-1849 observations of Schwabe (and, by inference, of
earlier observers calibrated to Schwabe). However, the correction factor was
apparently not considered for his own observations (mixed with Schwabe's) in the
period before the early 1860s.

The revised series, introduced from 1 July 2015, is now considered version 2.0
of the sunspot number series. Further corrections, with proper version tracking,
are expected as early data may contain other inconsistencies, and the
corrections applied in v2.0 were somewhat crude. In particular, recomputation of
the whole series from observational data, wherever available, is planned. The
process has now been placed under the {\ae}gis of the IAU, with a dedicated
Working Group ``Coordination of Synoptic Observations of the Sun'' focusing on
the validation and accreditation process of new SSN versions.

A further possible bias in the series that remains to be corrected may concern
the counting of sunspot groups. While in earlier parts of the series physical
closeness of spots was considered a sufficient criterion, since the mid-20th
century evolutionary information is also taken into account, sometimes resulting
in the division into several groups of what would have been considered as a
single group by early observers. \cite{Svalgaard+:weighting} estimate that this
effect may have inflated Waldmeier's sunspot numbers by 4--5\% relative to
earlier counts, while the effect on the late 18th century sunspot reconstruction
of the SSN by Wolf based on the drawings of Staudacher may be even larger,
reaching 25\% (\citealt{Svalgaard:Staudach.recount}). On the other hand,
\cite{Izenman} notes that Waldmeier's authoritative 1965 edition of the $R_Z$
series does contain slight corrections also to the data published previously, in
1925 by Wolfer. (One might also consider this as a surreptitious earlier minor
``revision'' of the SSN.) This shows that Waldmeier himself was very much
concerned with the long-term homogeneity of the data, already taking some
measures to homogenize the data processing.

Amidst all the revision fervour, some caveats may still be in order. On the one
hand, for the current generation of solar physicists it is abundantly clear that
our Sun is capable of rather sudden unexpected changes and that a varying ratio
between different activity indices  (even just those related to sunspots)  can
be a real, physical feature  (\citealt{Georgieva+:antiV2}; see also
Sect.~\ref{sect:portents} below). Suggestions for revisions based purely on
variations or jumps in the ratio of the sunspot number to other indices are
therefore to be treated with {\revone strong caution  as it may potentially bias
understanding of the physical processes involved}. Second, having superior
instruments does not entitle us to think we are smarter than our predecessors
were, or that we know their data better than they themselves did. Information
lost and unknown considerations may well explain practices that, in retrospect,
seem incorrect to us. Even in the case of the already implemented corrections
\textit{(b)} and \textit{(c)} some nagging questions do remain and need further
exploration (\citealt{Clette+:SSNv2}).

{\oldtext 

The significant disagreements between determinations of the SSN by various
observatories, observers and methods are even more pronounced in the case of
historical data, especially prior to the mid-19th century. In particular, the
controversial suggestion that a whole solar cycle may have been missed in the
official sunspot number series at the end of the 18th century is taken by some
as glaring evidence for the unreliability of early observations. Note, however,
that independently of whether the claim for a missing cycle is well founded or
not, there is clear evidence that this controversy is mostly due to the very
atypical behaviour of the Sun itself in the given period of time, rather than to
the low quality and coverage of contemporary observations. These issues will be
discussed further in Sect.~\ref{sect:evenodd}.

}

\subsubsection{Alternating series and nonlinear transforms} 

Instead of the ``raw'' sunspot number series $R(t)$ many researchers prefer to
base their studies on some transformed index $R'$. The motivation behind this is
twofold.

(a) The strongly peaked and asymmetrical sunspot cycle profiles strongly deviate
from a sinusoidal profile; also the statistical distribution of sunspot numbers
is strongly at odds with a Gaussian distribution. This can constitute a problem
as many common methods of data analysis rely on the assumption of an
approximately normal distribution of errors or nearly sinusoidal profiles of
spectral components. So transformations of \R (and, optionally, \textit{t}) that
reduce these deviations can obviously be helpful during the analysis. In this
vein, e.g., Max Waldmeier often based his studies of the solar cycle on the use
of logarithmic sunspot numbers $R'=\log R$; many other researchers use
$R'=R^\alpha$ with $0.5\leq\alpha<1$, the most common value being $\alpha=0.5$.

(b) As the sunspot number is a rather arbitrary construct, there may
be an underlying more physical parameter related to it in some
nonlinear fashion, such as the toroidal magnetic field strength $B$,
or the magnetic energy, proportional to $B^2$. It should be emphasized
that, contrary to some claims, our current understanding of the solar
dynamo does \textit{not} make it possible to guess what the underlying
parameter is, with any reasonable degree of certainty. In particular,
the often used assumption that it is the magnetic energy, lacks any
sound foundation. If anything, on the basis of our current best
understanding of flux emergence we might expect that the amount of
toroidal flux emerging from the tachocline should be $\int
|B-B_0|\,dA$ where $B_0$ is some minimal threshold field strength for
Parker instability and the surface integral goes across a latitudinal
cross section of the tachocline (cf.\ \citealt{Ruzmaikin:biasing}; 
{\newtext 
indeed, a generalized linear $R$--$B$ link involving a threshold field strength
has now also been used in the dynamo models of
\citealt{Pipin+Sokoloff:Waldmeierdyn} and \citealt{Pipin:GOdynamo}).
} 
As, however, the lifetime of any given sunspot group is finite and proportional
to its size (\citealt{Petrovay_vDG:decay1}; \citealt{Henwood_}), instantaneous
values of \R or the total sunspot area should also depend on details of the
probability distribution function of $B$ in the tachocline. This just serves to
illustrate the difficulty of identifying a single physical governing parameter
behind \R.

One transformation that may still be well motivated from the physical point of
view is to attribute an alternating sign to even and odd Schwabe cycles: this
results in the  the \textit{alternating sunspot number series} $R_\pm$. The idea is
based on Hale's well known polarity rules, implying that the period of the solar
cycle is actually 22 years rather than 11 years, the polarity of magnetic fields
changing sign from one 11-year Schwabe cycle to the next. In this
representation, first suggested by \cite{Bracewell}, usually odd cycles are
attributed a negative sign. This leads to slight jumps at the minima of the
Schwabe cycle, as a consequence of the fact that for a 1\,--\,2 year period around
the minimum, spots belonging to both cycles are present, so the value of \R
never reaches zero; in certain applications, further twists are introduced into
the transformation to avoid this phenomenon. 

After first introducing the alternating series, in a later work
\cite{Bracewell:trafo} demonstrated that introducing an underlying ``physical''
variable \RB such that
\begin{equation}
  R_\pm = 100\left( R_B/83\right)^{3/2}
  \label{eq:Bracewell}
\end{equation}
(i.e., $\alpha=2/3$ in the power law mentioned in item (a) above)
significantly simplifies the cycle profile. Indeed, upon introducing a
``rectified'' phase variable\footnote{The more precise condition defining $\phi$
is that  $\phi=\pm\pi/2$ at each maximum and $\phi$ is quadratically related to
the time since the last minimum.} $\phi$ in each cycle to compensate for the
asymmetry of the cycle profile, \RB is a nearly sinusoidal function of $\phi$.
The empirically found 3/2 law is interpreted as the relation between the
time-integrated area of a typical sunspot group vs. its peak area (or peak \RZ
value), i.e., the steeper than linear growth of \R with the underlying physical
parameter \RB would be due to the larger sunspot groups being observed
longer, and therefore giving a disproportionately larger contribution to the
annual mean sunspot numbers. If this interpretation is correct, as  suggested by
Bracewell's analysis, then \RB should be considered proportional to the total
toroidal magnetic flux emerging into the photosphere in a given interval. (But
the possibility must be kept in mind that the same toroidal flux bundle may
emerge repeatedly or at different heliographic longitudes, giving rise to
several active regions.)

\subsection{Other indicators of solar activity}
\label{sec:indicators}

\newsubsubsection{Group sunspot number (GSN)}

{\oldtext 

Reconstructions of \R prior to the early 19th century are increasingly
uncertain. In order to tackle problems related to sporadic and often unreliable
observations, \cite{Hoyt_Schatten:GSN} introduced the \textit{Group Sunspot
Number} (GSN) as an alternative indicator of solar activity. In contrast to the
SSN, the GSN only relies on counts of sunspot groups as a more robust indicator,
disregarding the number of spots in each group. Furthermore, while \RZ was
determined for any given day from a single observer's measurements (a hierarchy 
of auxiliary observers was defined for the case if data from the primary observer
were unavailable), the GSN uses a weighted average of all observations available
for a given day. 
}
A further advantage is that, in addition to the published series, all the raw
data upon which it is based are made public.

The GSN series published by \cite{Hoyt_Schatten:GSN} remained unchanged until
the 2010s when it was taken under revision in concert with the SSN revision
discussed above. As in the case of the GSN there is no generally accepted
responsible organization, i.e., no ``official'' series, revisions were undertaken
by several teams, leading to conflicting results.\footnote{Note that
\cite{Hoyt_Schatten:GSN} included a coefficient of 12.08 in the definition of
their GSN in order to bring it the same scale as the SSN; this
coefficient has been omitted in the more recent reconstructions, sometimes
leading to some confusion regarding the scale applied.}

The common denominator of all efforts to reconstruct the GSN is (or should be)
the common set of observations upon which the construction of the series is
based. This observational data set has been greatly extended in the past two
decades thanks to the discovery and/or publication of many previously
inaccessible historical sources. An update of the database providing a good
basis for subsequent efforts to construct a GSN series was compiled by
\cite{Vaquero+:database}. This archive is now available from the SILSO site and
from the Historical Archive of Sunspot Observations
(HASO)\footnote{\url{http://haso.unex.es/}}. (In principle, a regular upgrade is
planned, with version numbers, v1.0 referring to Hoyt \& Schatten, but the
project is currently stuck at version 1.12 dated May 2016.)

The original method of \cite{Hoyt_Schatten:GSN} was subject to a random drift of
the mean group number over long timescales. While consistent use of the
Greenwich Photoheliograph Results, available from 1874, helped to avoid such a
drift in the 20th century, a drift already appears from the late 19th century
back, owing to the still evolving techniques of photography used at Greenwich.
As a result, group numbers before this period {\revone may be} systematically
lower than what the SSN would suggest. This was the main issue motivating a
revision of the GSN series. 

New GSN series were compiled by \cite{Svalgaard+:newGSN} and by
\cite{Usoskin+:newGSN} using two alternative methods: the backbone method and a
method based on active day fraction (ADF) statistics. The backbone method
resulted in significantly elevated GSN values before about 1900, while the ADF
method resulted in a series closer to the original Hoyt \& Schatten values. Both
of these methods have been subject to criticisms (\citealt{Cliver:GSNdebate}; 
\citealt{Willamo+:newGSN}, \citeyear{Willamo+:ADFproblems}). Finally,
\cite{Chatzistergos+:newGSN} came up with a variety of the backbone method with
an improved methodology for the fitting of successive backbones, resulting in an
intermediate series. At the time of writing, this ``ultimate backbone'' GSN
series, available at
CDS\footnote{\url{http://vizier.cfa.harvard.edu/viz-bin/VizieR?-source=J/A+A/602/A69}
\\ (or via the paper's ADS link).}, seems to be the most recommendable version for
further analysis.

{\oldtext 

The GSN series has been reproduced for the whole period since 1611 and it is
generally agreed that for the period 1611\,--\,1818 it is a more reliable
reconstruction of solar activity than the relative sunspot number. Yet there
have been relatively few attempts to date to use this data series for solar
cycle prediction. 

}

\begin{figure}[htbp]
\centerline{\includegraphics[width=\textwidth]{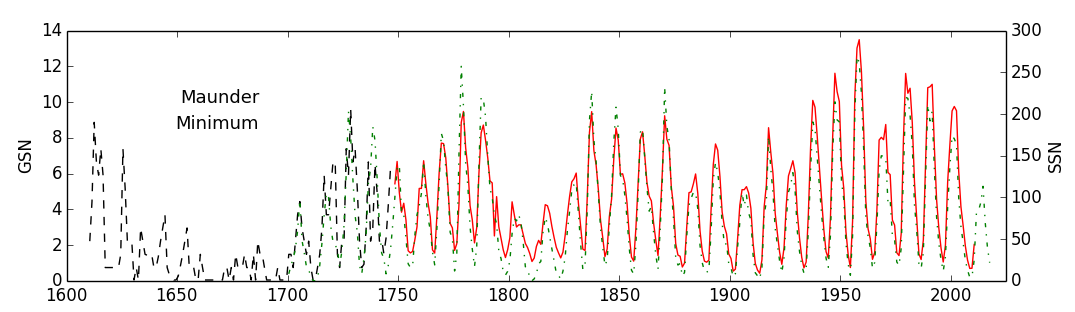}}
\caption{Annual means of the group sunspot number reconstructed by
\cite{Chatzistergos+:newGSN} (solid red curve). Values before 1749 (dashed
black) were taken from the reconstruction by \cite{Svalgaard+:newGSN},
multiplied by a fiducial factor 0.85 to align the two curves in 1750 and to 
bring the GSN and SSN (dash-dotted green) into better agreement in the early
18th
century.
}
\label{fig:GSNrecord}
\end{figure}

\newsubsubsection{Other sunspot data}


The classic source of sunspot area and position data is the Greenwich
Photoheliographic Results (GPR)
catalogue\footnote{\url{https://www.ngdc.noaa.gov/stp/solar/solardataservices.html}},
covering the period 1874--1976. The official continuation of the GPR is the
Debrecen Photoheliographic Data (DPD)
catalogue\footnote{\url{http://fenyi.solarobs.csfk.mta.hu/en/databases/Summary/}},
commissioned by the IAU and containing data from 1973 (\citealt{Baranyi+:DPD}; 
\citealt{Gyori+:DPD}). The Debrecen database also includes a revised/enriched
version of the GPR in the same format as the DPD. Another GPR extension, with
USAF/NOAA data, covering the period {\revone 1874--2019 is available from 
a website maintained by NASA MSFC 
staff.\footnote{\url{http://solarcyclescience.com/activeregions.html}}}
Sunspot data from many other observatories are also available at the NGDC site. 

Recent years have seen a surge in the digitization and processing of sunspot
drawings made before the photographic era. A major role in this work has been
played by a team in Potsdam led by Rainer Arlt (\citealt{Arlt:Staudacher};
\citealt{Arlt+:Schwabe}; \citealt{Diercke+:Sporer}).  As a result sunspot
positions (butterfly diagrams) have now been reconstructed for the period
1826--1880 from drawings by Schwabe and Sp\"orer (\citealt{Leussu+:bfly1};
\citealt{Leussu+:bfly2}); for the period 1749--1796 from drawings by Staudacher
(\citealt{Arlt:bfly.Staudacher}); for the period 1670-1711 from scattered
information (\citealt{Vaquero+:Maunderdata}; \citealt{Neuhauser:Kirch}); and for
the period 1611--1631 from drawings by Scheiner and Galileo
(\citealt{Arlt+:Scheiner}; \citealt{Vokhmyanin+:Galileo}).

{\oldtext

Instead of the sunspot number, the total area of all spots observed on the solar
disk might seem to be a less arbitrary measure of solar activity. However, these
data have been available since 1874 only, covering a much shorter period of time
than the sunspot number data. In addition, the determination of sunspot areas,
especially farther from disk center, is not as trivial as it may seem, resulting
in significant random and systematic errors in the area determinations. Area
measurements performed in two different observatories often show discrepancies
reaching $\sim$~30\% for smaller spots (cf.~the figure and discussion in
Appendix~A of \citealt{Petrovay_:decay2}).  } Despite these difficulties,
attempts at reconstructing sunpot areas have also been made
(\citealt{Carrasco+:areas}; \citealt{Senthamizh:Schwabe.areatilt}), and
\cite{Murakozy:areaindex} recently proposed a new activity index based on a
calibration of the emerged magnetic flux to sunspot areas.

\newsubsubsection{Other activity indices}

\begin{figure}[htbp]
\centerline{\includegraphics[width=\textwidth]{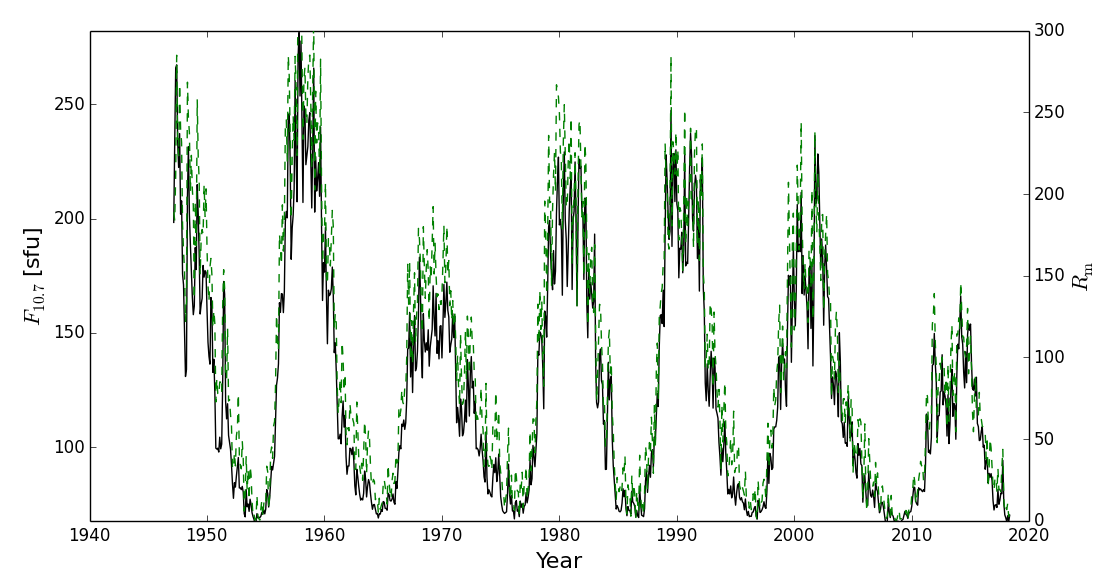}}
\caption{Monthly values of the 10.7~cm radio flux in solar flux units for the
period 1947\,--\,2017. The solar flux unit is defined as
$10^{-22}$~W/m$^2$~Hz. The dashed green curve shows the monthly mean relative
sunspot number $R_{\mathrm{m}}$ for comparison. Data
are from the NRC Canada (Ottawa/Penticton).}
\label{fig:radioflux}
\end{figure}

{\oldtext

A number of other direct indicators of solar activity have become available from
the 20th century 
}
(see \citealt{Ermolli:ISSI} for a recent review). 
{\oldtext
These include, e.g., various plage indices or the 10.7~cm solar radio flux --
the latter is considered a particularly good and simple to measure indicator of
global activity (see Fig.~\ref{fig:radioflux}). As, however, these data sets
only cover a few solar cycles, their impact on solar cycle prediction has been
minimal.
} 
A promising exception from this is the {\revone nearly} three centuries long
record of the solar EUV flux, recently reconstructed from the diurnal variation
of the  geomagnetic field by \cite{Svalgaard:EUVreconstr}.

{\oldtext
Of more importance are \textit{proxy indicators} such as geomagnetic
indices (the most widely used of which is the \textit{aa} index), the
occurrence frequency of aurorae or the abundances of cosmogenic
radionuclides such as $^{14}$C and $^{10}$Be {\revone in natural archives}. 
For solar cycle
prediction uses such data sets need to have a sufficiently high
temporal resolution to reflect individual 11-year cycles. For the
geomagnetic indices such data have been available since 1868, while an
annual $^{10}$Be series covering 600 years has been published by
\cite{Berggren_:10Be_600yrs}. Attempts have been made to reconstruct
the epochs and even amplitudes of solar maxima during the past two
millennia from oriental naked eye sunspot records and from auroral
observations (\citealt{secular.book}; \citealt{Nagovitsyn:reconstr}),
but these reconstructions are currently subject to too many
uncertainties to serve as a basis for predictions. Isotopic data with
lower temporal resolution are now available for up to 50\,000 years in
the past; while such data do not show individual Schwabe cycles, they
are still useful for the study of long term variations in cycle
amplitude.  Inferring solar activity parameters from such proxy data
is generally not straightforward. } (See review by
\citealt{Usoskin:LRSP2}.)

\subsection{The solar cycle and its variation}
\label{sect:cycle}

\begin{figure}[htbp]
\centerline{\includegraphics[width=\textwidth]{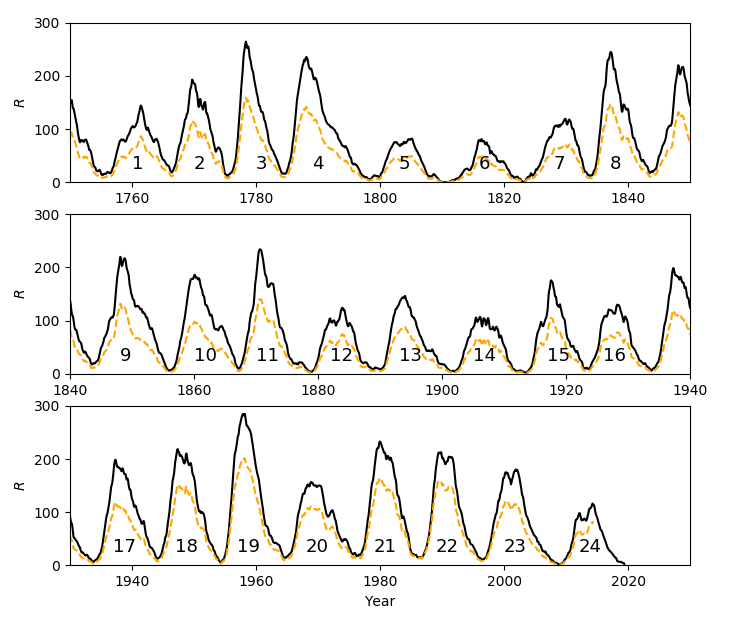}}
\caption{The variation of the monthly smoothed relative sunspot number \R during
the period 1749\,--\,2009, with the conventional numbering of solar cycles,
{\revone for SSN version 2 (black solid) and for SSN version 1 (yellow
dashed)}.}
\label{fig:SSNrecord}
\end{figure}

The series of \R values determined as described in Sect.~\ref{sect:wolfno} is
plotted in Fig.~\ref{fig:SSNrecord}. It is evident that the sunspot cycle is
rather irregular. The mean length of a cycle (defined as lasting from minimum to
minimum) is 11.0~years (median 10.9~years), with a standard deviation of
1.16~years; actual cycle lengths scatter in the range 9.0--13.6 years. 
{\newtext 
Note that cycle lengths measured between successive maxima show a wider scatter,
in the range 7.3 and 16.9 years. This is partly due to the fact that many cycles
show a double maximum, the two sub-peaks being separated by 1--2 years.
}
The mean cycle amplitude in terms of $R$  is 179 (median 183)\footnote{{\revone
Here and in the rest of this paper $R$ will normally refer to Version 2 values,
unless explicitly noted.}}, with a standard deviation of 57. It is this wide
variation that makes the prediction of the next cycle maximum such an
interesting and vexing issue.

\subsubsection{Secular activity variations}
\label{sect:secular}

Inspecting Fig.~\ref{fig:SSNrecord} one can discern an obvious long term
variation. For the study of such long term variations, the series of cycle
parameters is often smoothed on time scales significantly longer than a solar
cycle: this procedure is known as \textit{secular smoothing}. One popular method is
the so-called \textit{Gleissberg filter} or \textit{12221 filter}
(\citealt{Gleissberg:12221}). For instance, the Gleissberg filtered
amplitude of cycle $n$ is given by
\begin{equation}
\langle \Rmax\rangle_{\mathrm{G}}^{(n)} = \frac 18\left( \Rmax^{(n-2)}
+2\Rmax^{(n-1)} +2\Rmax^{(n)}  +2\Rmax^{(n+1)} +\Rmax^{(n+2)}\right) .
\end{equation}

\begin{figure}[htbp]
\centerline{\includegraphics[width=\textwidth]{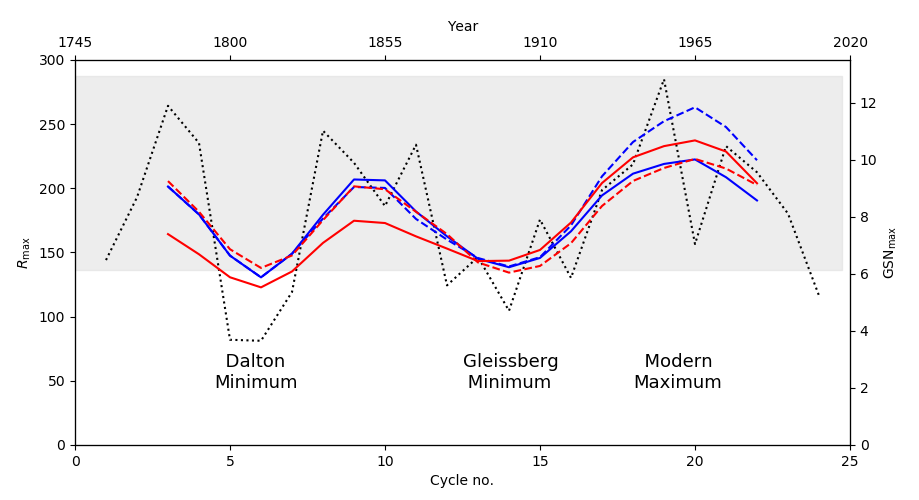}}
\caption{Amplitudes of the sunspot cycles (dotted) and their Gleissberg filtered
values (blue solid), plotted against cycle number. 
{\revone The shaded area marks a $\pm 2\sigma$ band around the mean amplitude
for cycles 17--23 comprising the Modern Maximum.}
For comparison, Gleissberg
filtered cycle amplitudes are also shown for unrevised [v1.0] SSN data (blue
dashed) and for two GSN reconstructions (red solid:
\citealt{Chatzistergos+:newGSN}; red dashed: \citealt{Svalgaard+:newGSN}). (In the case
of the two GSN series, amplitudes and dates for cycle maxima were determined
from the 121 filtered annual data.)
}
\label{fig:Gleissbgfilt}
\end{figure}

The Gleissberg filtered series of solar cycles is plotted in
Fig.~\ref{fig:Gleissbgfilt}.  
{\newtext
 The most obvious feature of the
variation is a cyclic modulation of the cycle amplitudes on a timescale of
$\sim$9--10 solar cycles. This so-called Gleissberg cycle will be discussed
further in Sect.~\ref{sect:Gleissberg}. The first minimum of this cycle
plotted in Fig.~\ref{fig:Gleissbgfilt}, known as the ``Dalton Minimum'', is
formed by the unusually weak cycles 5, 6, and 7. The second secular minimum
consists of a rather long series of moderately weak cycles 12\,--\,16,
occasionally referred to as the [last] ``Gleissberg Minimum'' -- but note that
most of these cycles are less than $1\,\sigma$ below the long-term average 
value {\revone given at the start of Section~\ref{sect:cycle}}.

Finally the last secular maximum of the cycle comprises the series of strong
cycles 17--23 in the second half of the 20th century: the ``Modern Maximum''. In
addition to this cyclic modulation there is a tendency for an overall secular
increase of solar activity in the figure: the Modern Maximum is clearly stronger
than previous maxima. However, the strength of this secular increase in the
activity level as well as the amount by which the Modern Maximum exceeds
previous maxima of the Gleissberg cycle clearly depends on the reconstruction of
the measure of activity chosen. The revision of the sunspot numbers has greatly
reduced the amount of secular increase compared to Version 1.0, in agreement
with the GSN reconstruction by \cite{Svalgaard+:newGSN}. On the other hand the
most recent GSN reconstruction (\citealt{Chatzistergos+:newGSN}) shows a marked
long-term increasing trend again. The cosmogenic record rather unequivocally
indicates that the persistently high level of solar activity characterizing the
second half of the 20th century had no precedent for thousands of years in
the history of solar activity (cf.\ Table~3 in \citealt{Usoskin:LRSP2}). The
currently hotly debated problem of the strength of the Modern Maximum has
important implications, e.g., for the understanding of the role of solar forcing
in global warming (\citealt{lean2008}; \citealt{chylek2014}; 
\citealt{Nagy+:APMO}; \citealt{Owens+:LittleIceAge}). In this context it is
important to stress that a secular increase in solar activity from the late 19th
century (beginning of terrestrial global temperature record) to the mid-20th
century is unquestionably present in all solar activity reconstructions. 
{\revone The decreasing trend displayed in the last few decades in the smoothed
cycle amplitude series is also potentially important in this respect.}

}

While the Dalton and Gleissberg minima are {\revone just} local minima in the
ever changing Gleissberg filtered SSN series, the conspicuous lack of sunspots
in the period
1640\,--\,1705, known as the \textit{Maunder Minimum} (Fig.~\ref{fig:GSNrecord})
quite obviously represents a qualitatively different state of solar activity.
Such extended periods of low activity are usually referred to as \textit{grand
minima.} 
{\newtext
Ever since the rediscovery of the Maunder Minimum in the late 19th
century {\revone (\citealt{Maunder1894}; \citealt{Eddy1976})} 
its reality and significance has time to time been brought into
question. Recent studies have shown that the 11/22-year solar cycle persisted
during the Maunder Minimum, but at a greatly suppressed level 
(\citealt{Usoskin+:Maunder.bfly};  \citealt{Vaquero+:Maunder};
\citealt{Asvestari+:Maunder}).
}
 A number of possibilities have been proposed to
explain the phenomenon of grand minima, including chaotic behaviour of the
nonlinear solar dynamo (\citealt{Weiss_:chaotic.dynamo}), stochastic
fluctuations in dynamo parameters (\citealt{Moss_:stochastic.dynamo1};
\citealt{Moss_:stochastic.dynamo2}); a bimodal dynamo with stochastically
induced alternation between two stationary states (\citealt{Petrovay:bimodal})
{\newtext
or {\revone stochastic effects like fluctuations in AR tilt 
(\citealt{Karak+Miesch:grandminima}) or ``rogue'' sunspots 
(\citealt{Py_Nagy:Jaipur}).}}

The analysis of long-term proxy data, extending over several millennia further
showed that there exist systematic long-term statistical trends and periods such
as the so called  secular and supersecular cycles (see
Sect.~\ref{sect:spectral}).

\subsubsection{Does the Sun have a long term memory?}
\label{sect:memory}


The existence of long lasting grand minima and maxima suggests that the sunspot
number record {\revone may} have a \textit{long-term memory} extending over several
consecutive cycles. Indeed, elementary combinatorical calculations show that the
occurrence of phenomena like the Dalton minimum (3 of the 4 lowest maxima
occurring in a row) in a random series of 24 recorded solar maxima has
a rather low probability (5\%). This conclusion is
{\revone supported} by the analysis of long-term proxy data, extending over 
several millennia, which showed that the occurrence of grand minima and 
{\newtext (perhaps)} grand maxima is
more common than what would follow from Gaussian statistics
(\citealt{Usoskin_Solanki}; {\revone \citealt{Wu+:9millennia}}).

It could be objected that for sustained grand minima or maxima a memory
extending only from one cycle to the next would suffice. 
This intercycle memory explanation of persistent secular activity minima and
maxima, however, would imply a good correlation between the amplitudes of
subsequent cycles, which is not the case
(cf.\ Sect.~\ref{sect:minimax} below). With the known poor
cycle-to-cycle correlation, strong deviations from the long-term mean
would be expected to be damped on time scales short compared to,
e.g., the length of the Maunder minimum. This suggests that the
persistent states of low or high activity {\revone may be} 
due to truly long term memory effects extending over several cycles.

{\newtext

In an analysis of the GSN series for the period 1799--2011 \cite{Love+Rigler}
found that the sequence of cycle maxima (and also of time-integrated activity in
each cycle), including the Modern Maximum, would not be an unlikely result of
the accumulation of multiple random-walk steps in a lognormal random walk of
cycle amplitudes where $\ln R$ performs a Gaussian random walk with mean
stepsize 0.39 (or 0.28 for the integrated activity). This analysis, however,
does not extend to the Maunder Minimum; and in any case, such a random walk
should ultimately take the values of $R$ up to arbitrarily high values in
sufficiently long time, whereas the cosmogenic record clearly shows that the
level of activity is bounded from above.

}

Further evidence for a long-term memory in solar activity comes from the
persistence analysis of activity indicators. The parameter determined in such
studies is the Hurst exponent $0<H<1$. Essentially, $H$ is the steepness of the
growth of the total range $\cal R$ of measured values plotted against the number
$n$ of data in a time series, on a logarithmic plot: ${\cal R}\propto n^H$.  For
a Markovian random process with no memory $H=0.5$. Processes with $H>0.5$ are
persistent (they tend to stay in a stronger-than-average or weaker-than-average
state longer), while those with $H<0.5$ are anti-persistent (their fluctuations
will change sign more often). 

Hurst exponents for solar activity indices have been derived using
rescaled range analysis by many authors (\citealt{Mandelbrot_:Hurst};
\citealt{Ruzmaikin:Hurst}; \citealt{Komm:DopplerHurst};
\citealt{Oliver_Ballester:Hurst}; \citealt{Kilcik_};
{\newtext
\citealt{Rypdal+}).} All studies coherently yield a value
$H=0.85\mbox{\,--\,}0.88$ for time scales exceeding a year or so, and
somewhat lower values ($H\sim 0.75$) on shorter time scales. Some
doubts regarding the significance of this result for a finite series
have been raised by \cite{Oliver_Ballester:noHurst}; however,
\cite{Qian_Rasheed} have shown using Monte Carlo experiments that for
time series of a length comparable to the sunspot record, $H$ values
exceeding 0.7 are statistically significant. 

A complementary method, essentially equivalent to rescaled range analysis is
detrended fluctuation analysis. Its application to solar data
(\citealt{Ogurtsov:Hurst}) has yielded results in accordance with the $H$
values quoted above.

The overwhelming evidence for the persistent character of solar activity and for
the intermittent appearance of secular cyclicities, however, is not much help
when it comes to cycle-to-cycle prediction. It is certainly reassuring to know
that forecasting is not a completely idle enterprise (which would be the case
for a purely Markovian process), and the long-term persistence and trends may
make our predictions statistically somewhat different from just the long-term
average.  There are, however, large decadal scale fluctuations superposed on the
long term trends, so the associated errors will still be so large as to make the
forecast of little use for individual cycles.

\subsubsection{Waldmeier effect and amplitude--frequency correlation}
\label{sect:Waldmeier}

\begin{nquote}
{\sl  ``Greater activity on the Sun goes with shorter periods, and less with
longer periods. I believe this law to be one of the most important relations
among the Solar actions yet discovered.''}\\
\strut\hfill{\citep{Wolf:cycle.length.activity}}
\end{nquote}

\begin{figure}[htbp]
\centerline{\includegraphics[width=0.5\textwidth]{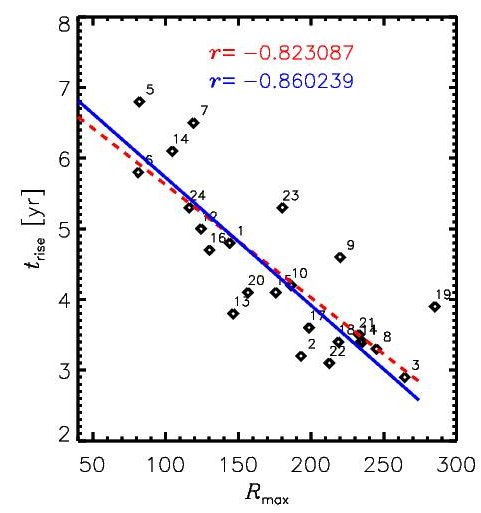}
\includegraphics[width=0.5\textwidth]{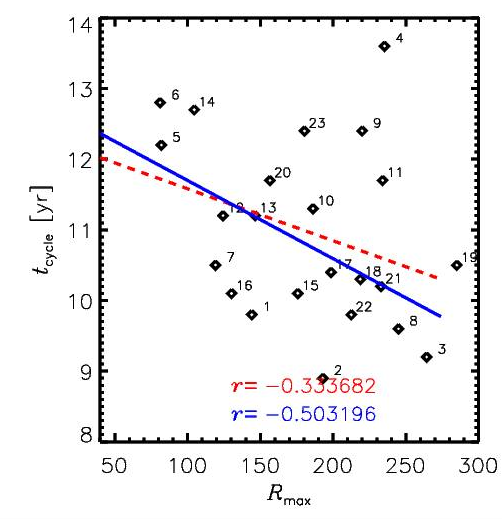}}
\caption{Monthly smoothed sunspot number \R at cycle maximum plotted against the rise
time to maximum (left) and against cycle length (right). Cycles are labeled with
their numbers. In the plots the red dashed lines are linear regressions
to all the data, while the blue solid lines are fits to all data except
outliers. Cycle~19 is considered an outlier on both plots, cycle~4 on the right
hand plot only.  The corresponding correlation coefficients are shown.}
\label{fig:waldmeier}
\end{figure}

It is apparent from Fig.~\ref{fig:SSNrecord} that the profile of sunspot cycles
is asymmetrical, the rise being steeper than the decay. Solar activity maxima
occur 3 to 4 years after the minimum, while it takes another 7\,--\,8 years to reach
the next minimum. It can also be noticed that the degree of this asymmetry
correlates with the amplitude of the cycle: to be more specific, the length of
the rise phase anticorrelates with the maximal value of \R
(Fig.~\ref{fig:waldmeier}), while the length of the decay phase shows weak or no
such correlation.\footnote{\newtext 
Note that \cite{Osipova:2Waldmeier} recently constructed two separate group
number series for small and large sunspot groups and they found that the
Waldmeier effect applies better to large spots.
}

Historically, the relation was first formulated by \cite{Waldmeier:effect} as an
inverse correlation between the rise \textit{time} and the cycle amplitude;
however, as shown by \cite{Tritakis:Waldmeier}, the total rise time is a weak
(inverse logarithmic) function of the rise rate, so this representation makes
the correlation appear less robust. (Indeed, when formulated with the rise time
it is not even present in some activity indicators, such as sunspot areas --
cf.\ \citealt{Dikpati_:Waldmaier}; 
{\newtext \citealt{Ogurtsov+Lindholm}.)} 
As pointed out by
\cite{Cameron_:Waldmeier}, the weak link between rise time and slope is due to
the fact that {\revone subsequent cycle are known to overlap by 1--2 years, so} 
in steeper rising cycles the minimum will occur earlier, thus partially compensating 
for the shortening due to a higher rise rate. The effect
is indeed more clearly seen when the rate of the rise is used instead of the
rise time (\citealt{Lantos}; \citealt{Cameron_:Waldmeier})
{\newtext
or if the rise time is redefined as the time spent from 20 to 80\% of the
maximal amplitude (\citealt{Karak+:Waldmeier}).
}

Nevertheless, when coupled with the nearly nonexistent correlation between the
decay time and the cycle amplitude, even the weaker link between the rise time
and the maximum amplitude is sufficient to forge a weak inverse correlation
between the total cycle length and the cycle amplitude
(Fig.~\ref{fig:waldmeier}). This inverse relationship was first noticed by
\cite{Wolf:cycle.length.activity}.

A stronger inverse correlation was found between the cycle amplitude and the
length of the \textit{previous} cycle by \cite{Hathaway:periodampl}. This
correlation is also readily explained as a consequence of the Waldmeier effect,
as demonstrated in a simple model by \cite{Cameron_:prediction}; 
{\newtext the same probably holds for the correlations reported by
\cite{Hazra+:decaycorr}.
} 
Note that in a
more detailed study \cite{Solanki_:cycle.length} found that the correlation
coefficient of this relationship has steadily decreased during the course of the
historical sunspot number record, while the correlation between cycle amplitude
and the length of the \textit{third} preceding cycle has steadily increased. The
physical significance (if any) of this latter result is unclear.


In what follows, the relationships found by \cite{Wolf:cycle.length.activity},
\cite{Hathaway:periodampl}, and \cite{Solanki_:cycle.length}, discussed above,
will be referred to as ``$\Rmax\mbox{\,--\,}t_{\mathrm{cycle},n}$ correlations'' with
$n = 0, -1$ or $-3$, respectively.

Modern time series analysis methods offer several ways to define an
instantaneous frequency $f$ in a quasiperiodic series. One simple approach was
discussed in the context of Bracewell's transform, Eq.~(\ref{eq:Bracewell}),
above. \cite{Mininni_:vanderpol} discussed several more sophisticated methods to
do this, concluding that G\'abor's analytic signal approach yields the best
performance. This technique was first applied to the sunspot record by
\cite{Palus_Novotna}, who found a significant long term correlation between the
smoothed instantaneous frequency and amplitude of the signal. On time scales
shorter than the cycle length, however, the frequency--amplitude correlation has
not been convincingly proven, and the fact that the correlation coefficient is
close to the one reported in the right hand panel of Fig.~\ref{fig:waldmeier}
indicates that all the fashionable gadgetry of nonlinear dynamics could
achieve was to recover the effect already known to Wolf. It is clear from this
that the ``frequency--amplitude correlation'' is but a secondary consequence of
the Waldmeier effect.


Indeed, an anticorrelation between cycle length and amplitude is characteristic
of a class of stochastically forced nonlinear oscillators and it may also be
reproduced by introducing a stochastic forcing in dynamo models
(\citealt{Stix:Waldmeier}; \citealt{Ossendrijver_:stoch.dynamo}; 
\citealt{Charbonneau_Dikpati}). In some such models the characteristic
asymmetric profile of the cycle is also well reproduced
(\citealt{Mininni_:vanderpol}; \citealt{Mininni_:vanderpol2}). The predicted
amplitude--frequency relation has the form
\begin{equation}
  \log\Rmax^{(n)} = C_1+ C_2f \, 
  \label{eq:stochRf}
\end{equation}
{\revone where $f\sim(\tmin^{(n+1)}-\tmin^{(n)})^{-1}$ is the frequency.}

Nonlinear dynamo models including some form of $\alpha$-quenching also have the
potential to reproduce the effects described by Wolf and Waldmeier without
recourse to stochastic driving. In a dynamo with a Kleeorin--Ruzmaikin type
feedback on $\alpha$, \cite{Kitiashvili_:nonlin.dynamo} are able to
qualitatively reproduce the Waldmeier effect. Assuming that the sunspot number
is related to the toroidal field strength according to the Bracewell transform,
Eq.~(\ref{eq:Bracewell}), they find a strong link between rise time and
amplitude, while the correlations with fall time and cycle length are much
weaker, just as the observations suggest. They also find that the form of the
growth time--amplitude relationship differs in the regular (multiperiodic) and
chaotic regimes. In the regular regime the plotted relationship
suggests
\begin{equation}
\Rmax^{(n)} = C_1-C_2\left(\tmax^{(n)}-\tmin^{(n)}\right) ,
\label{eq:linWald}
\end{equation}
while in the chaotic case  
\begin{equation}
\Rmax^{(n)} \propto {\left[1/\left(\tmax^{(n)}-\tmin^{(n)}\right)\right]} .
\end{equation}
{\newtext 
The linear relationship (\ref{eq:linWald}) was also reproduced in some 
stochastically forced nonlinear dynamo models
(\citealt{Pipin+Sokoloff:Waldmeierdyn}; \citealt{Pipin+Kosovichev:Waldmeierdyn};
\citealt{Pipin:GOdynamo}).
}

Note that based on the actual sunspot number series Waldmeier originally
proposed
\begin{equation}
\log\Rmax^{(n)} =C_1-C_2\left(\tmax^{(n)}-\tmin^{(n)}\right) , 
\end{equation}
%
while according to \cite{Dmitrieva:Waldmeierdynamo} the relation takes the
form
\begin{equation}
\log\Rmax^{(n)} \propto {\left[1/\left(\tmax^{(n)}-\tmin^{(n)}\right)\right]} . 
\end{equation}

At first glance, these logarithmic empirical relationships seem to be more
compatible with the relation~(\ref{eq:stochRf}) predicted by the stochastic
models. These, on the other hand, do not actually reproduce the Waldmeier
effect, just a general asymmetric profile and an amplitude--frequency
correlation. At the same time, inspection of the the left hand panel in
Fig.~\ref{fig:waldmeier} shows that the data is actually not incompatible with
a linear or inverse rise time--amplitude relation, especially if the anomalous
cycle~19 is ignored as an outlier. (Indeed, a logarithmic representation is
found not to improve the correlation coefficient -- its only advantage is that
cycle~19 ceases to be an outlier.) All this indicates that nonlinear dynamo
models may have the potential to provide a satisfactory quantitative explanation
of the Waldmeier effect, but more extensive comparisons will need to be done,
using various models and various representations of the relation. 
{\newtext 
In one such exploratory study for instance \cite{Nagy_Py:Waldmoscill} found that
solar-like parameter correlations can be obtained in a stochastically forced van
der Pol oscillator but only if the perturbations are applied to the nonlinearity
parameter rather than to the damping. In another study \cite{Karak+:Waldmeier}
found that in a stochastically forced flux transport dynamo perturbing the
poloidal field amplitude is not sufficient to induce solar-like parameter
correlations, and perturbations to the meridional flow speed are also needed.
}

\newsubsection{Approaches to solar cycle prediction}
 \label{sect:classif}
 
As the SSN series is a time series it is only natural that time series analysis
methods have been widely applied in order to predict its future variations,
including the amplitude of an upcoming cycle. As a group, however, time series
methods have not been particularly successful in attaining this goal. In
addition, time series analysis is a purely mathematical tool offering little
physical insight into the processes driving cycle-to-cycle variations. In view
of this, time series methods (or extrapolation methods) have been relegated to a
later section of this review, after dealing with the currently much more lively
field of the physically more insightful and more successful alternative
approaches: precursor schemes and model-based forecasts.

In the 1st edition of this review the model-based approach, then still very new, was
discussed well separated from precursor methods, in a section following the
discussion of the time series approach. In the time elapsed since the 1st edition,
however, a major surge of activity in surface flux transport (SFT) modelling,
new developments in dynamo models and in empirical precursors have made it
harder to draw a clear line between the precursor and model based approaches.
Indeed, there seem to be at least ``five shades of grey'' arching between
archetypical examples of these two categories:
\begin{descr}
\item{(a)} Internal empirical precursors 
{\revone (relying only on the SSN series)}
\item{(b)} External empirical precursors 
{\revone (relying on other activity indicators)}
\item{(c)} Physical[ly motivated] precursors
\item{(d)} Forecasts based on SFT models
\item{(e)} Forecasts based on dynamo models
\end{descr}
Accordingly, the present edition has been reorganized so the section discussing
model-based predictions immediately follows the section on precursors: the two
topics have been separated, somewhat arbitrarily, between the classes (c) and
(d) above, i.e., the term ``model-based'' is reserved for methods employing
detailed quantitative models rather than empirical or semiempirical correlations
based on qualitative physical ideas.



\section{Precursor methods}
\label{sect:precursor}

\begin{nquote}
{\sl ``Jeder Fleckenzyklus mu{\ss} als ein abgeschlossenes Ganzes, als ein
Ph{\"a}nomen f{\"u}r sich, aufgefa{\ss}t werden, und es reiht sich einfach
Zyklus an Zyklus.''}\\ 
\strut\hfill{\citep{Gleissberg:book}}
\end{nquote}

In the most general sense, precursor methods rely on the value of some measure
of solar activity or magnetism at a specified time to predict the amplitude of
the following solar maximum. The precursor may be any proxy of solar activity or
other indicator of solar and interplanetary magnetism. Specifically, the
precursor may also be the value of the sunspot number at a given time.

In principle, precursors might also herald the activity level at other phases of
the sunspot cycle, in particular the minimum. Yet the fact that practically all
the good precursors found need to be evaluated at around the time of the minimum
and refer to the next maximum is not simply due to the obvious greater interest
in predicting maxima than predicting minima. Correlations between minimum
parameters and previous values of solar indices have been looked for, but the
results were overwhelmingly negative (e.g.,~\citealt{Tlatov:polar.precursors}).
This indicates that the sunspot number series is \textit{not} homogeneous and
Rudolf Wolf's instinctive choice to start new cycles with the minimum rather
than the maximum in his numbering system is not arbitrary -- for which even more
obvious evidence is provided by the butterfly diagram. Each numbered solar cycle
is a consistent unit in itself, while solar activity seems to consist of a
series of much less tightly intercorrelated individual cycles, as suggested by
Wolfgang Gleissberg in the motto of this section.

In Sect.~\ref{sect:memory} we have seen that there 
{\revone may be some} evidence for
a \textit{long-term memory} underlying solar activity. In addition to the evidence
reviewed there, systematic long-term statistical trends and periods of solar
activity, such as the secular and supersecular cycles (to be discussed in
Sect.~\ref{sect:spectral}), also attest to a secular mechanism underlying solar
activity variations and ensuring some degree of long-term coherence in activity
indicators. However, as we noted, this long-term memory is of limited importance
for cycle prediction due to the large, apparently haphazard decadal variations
superimposed on it. What the precursor methods promise is just to find a system
in those haphazard decadal variations -- which clearly implies a different type
of memory. As we already mentioned in Sect.~\ref{sect:memory}, there is
obvious evidence for an \textit{intracycle memory} operating \textit{within} a
single cycle, so that forecasting of activity in an ongoing cycle is currently a
much more successful enterprise than cycle-to-cycle forecasting. As we will see,
this intracycle memory is one candidate mechanism upon which precursor
techniques may be founded, via the Waldmeier effect.


The controversial issue is whether, in addition to the intracycle memory, there
is also an \textit{intercycle memory} at work, i.e., whether behind the apparent
stochasticity of the cycle-to-cycle variations there is some predictable
pattern, whether some imprint of these variations is somehow inherited from one
cycle to the next, or individual cycles are essentially independent. The latter
is known as the ``outburst hypothesis'': consecutive cycles would then represent
a series of ``outbursts'' of activity with stochastically fluctuating amplitudes
(\citealt{Halm, Waldmeier:effect, Vitinsky:book2}; see also
\citealt{deMeyer:impulse} who calls this ``impulse model''). Note that
cycle-to-cycle predictions in the strict temporal sense may be possible even in
the outburst case, as solar cycles are known to overlap. Active regions
belonging to the old and new cycles may coexist for up to three years or so
around sunspot minima; and high latitude ephemeral active regions oriented
according to the next cycle appear as early as 2\,--\,3 years after the maximum
(\citealt{Tlatov_:ER} -- the so-called extended solar cycle).

In any case, it is undeniable that for cycle-to-cycle predictions, which are our
main concern here, the precursor approach seems to have been the relatively most
successful, so its inherent basic assumption must contain an element of truth --
whether its predictive skill is due to a ``real'' cycle-to-cycle memory
(intercycle memory) or just to the overlap effect (intracycle memory).

The two precursor types that have received most attention are polar field
precursors and geomagnetic precursors. A link between these two categories is
forged by a third group, characterizing the interplanetary magnetic field
strength or ``open flux''.
{\newtext 
In terms of the classification outlined in Sect.~\ref{sect:classif}
above, all these belong to the category (c) of physically motivated precursors.
But before considering these approaches, we start by discussing categories (a)
and (b): the \textit{empirical precursors} based on the chance discovery of
correlations between certain solar parameters and cycle amplitudes. These
parameters involved may also be \textit{external} to the SSN series (b); but first 
of all we will focus on the most obvious precursor type:
\textit{internal empirical precursors} (a)}
--- the level of solar activity at some epoch before the next maximum.



\subsection{Cycle parameters as precursors and the Waldmeier effect}
\label{sect:minimax}

The simplest weather forecast method is saying that ``tomorrow the weather will
be just like today'' (works in about 2/3 of the cases). Similarly, a simple
approach of sunspot cycle prediction is correlating the amplitudes of
consecutive cycles. There is indeed a marginal correlation, but the correlation
coefficient is quite low (0.35). The existence of the correlation is related to
secular variations in solar activity, while its weakness is due to the
significant cycle-to-cycle variations.

\begin{figure}[htbp]
    \centerline{\includegraphics[width=0.5\textwidth]{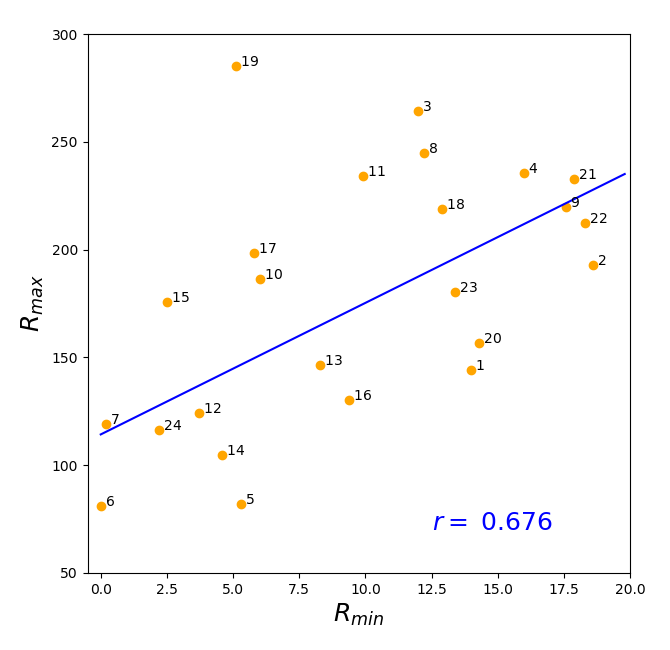}
    \includegraphics[width=0.5\textwidth]{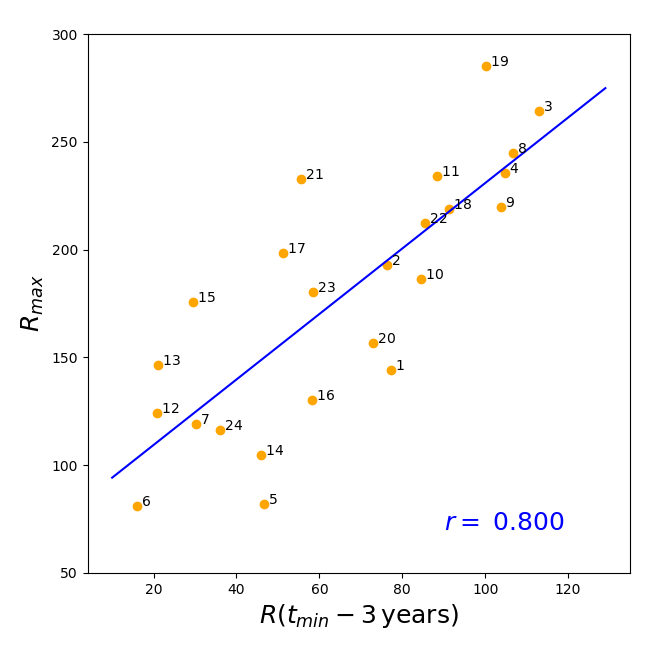}}
    \caption{Monthly smoothed sunspot number \R at cycle maximum
      plotted against the values of \R at the previous minimum (left) and 3
      years before the minimum (right). Cycles are labeled with their numbers.
      Blue solid lines are linear regressions to the data. 
      In the left-hand plot 
      Cycle~19 was treated as an outlier and omitted from the regression.
      The corresponding correlation coefficients are also displayed.}
    \label{fig:minimax}
\end{figure}

A significantly better correlation exists between the \textit{minimum} activity
level and the amplitude of the next maximum (Fig.~\ref{fig:minimax}). The
relation is linear (\citealt{Brown:minimax}), with a correlation coefficient of
0.68 (if the anomalous cycle 19 is ignored -- \citealt{Brajsa_:gleissbg}; see also
\citealt{Pishkalo}). The best fit is
\begin{equation}
\Rmax=114.3+6.1 \Rmin\,.
\label{eq:minimax}
\end{equation}
%

\cite{Cameron_:prediction} pointed out that the activity level three years before
the minimum is an even better predictor of the next maximum. {\revone The exact
value of the time shift producing optimal results may somewhat vary depending on
the base period considered: in the Version 2 SSN series for the whole series of
cycles 1--24 the highest correlation results for a time shift of 39 months while
considering only cycles 8--24, for which data are sometimes considered more
reliable, the best correlation is found at 32 months. Here in the right hand
panel of Fig.~\ref{fig:minimax}  we simply use the round value of of 3 years (36
months) as originally proposed.  The linear regression is
\begin{equation}
\Rmax=79+1.52 R(\tmin-3\,{\rm yr}) .
\label{eq:minimax3}
\end{equation}
This method, to be referred to as as ``minimax3'' for brevity, can only provide
an upper estimate for the expected amplitude of Cycle 25 as the minimum has not
been reached at the time of writing. As the minimum will take place no earlier
than May 2019, the value of the predictor is not higher than 44.8, which would
result in a cycle amplitude of 147. This already gives an indication that the
upcoming cycle will be weaker than the climatological average.}

As the epoch of the minimum of \R cannot be known with certainty until about a
year \textit{after} the minimum, the practical use of these methods is rather
limited: a prediction will only become available 2\,--\,3 years before the maximum,
and even then with the rather low reliability reflected in the correlation
coefficients quoted above. In addition, as convincingly demonstrated by
\cite{Cameron_:prediction} in a Monte Carlo simulation, these methods do not
constitute real cycle-to-cycle prediction in the physical sense: instead, they
are due to a combination of the overlap of solar cycles with the Waldmeier
effect. As stronger cycles are characterized by a steeper rise phase, the
minimum before such cycles will take place earlier, when the activity from the
previous cycle has not yet reached very low levels.

The same overlap readily explains the $\Rmax\mbox{\,--\,}t_{\mathrm{cycle},n}$
correlations discussed in Sect.~\ref{sect:Waldmeier}. These relationships may
also be used for solar cycle prediction purposes
(e.g.,~\citealt{Kane:cyclength}) but they lack robustness.  
The forecast is not only sensitive to the value of
$n$ used but also to the data set (relative or group sunspot numbers)
(\citealt{Vaquero_:cyclength}).
{\newtext 
Similar correlations between the properties of subsequent cycles were used by 
\cite{Li+:pred25} to give a prediction for Cycle~25.
}

  \begin{figure}[htbp]
    \centerline{\includegraphics[width=0.9\textwidth]{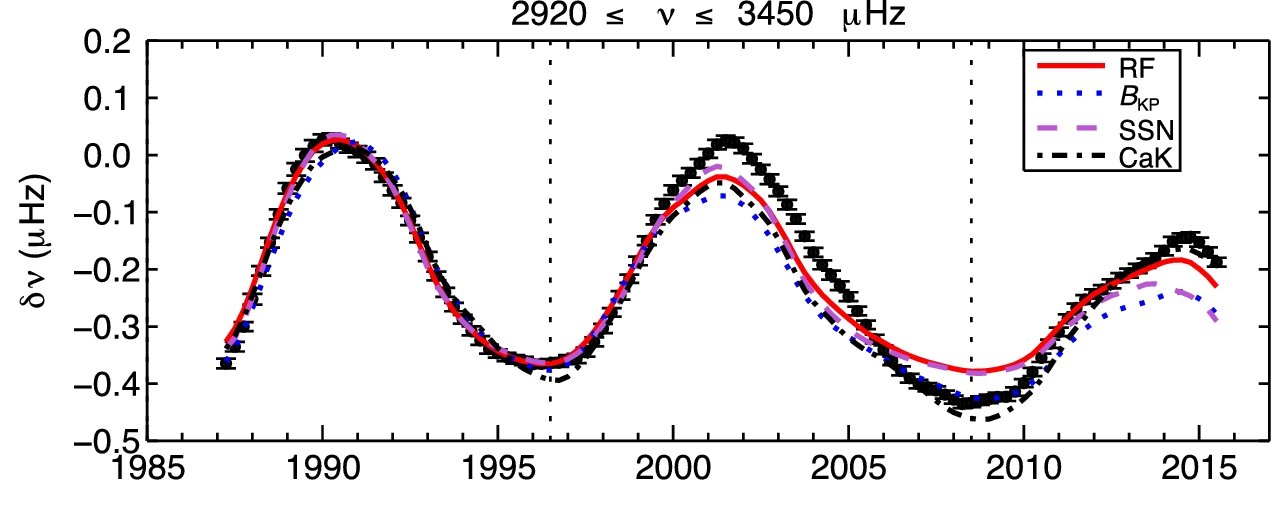}}
    \centerline{\includegraphics[width=0.9\textwidth]{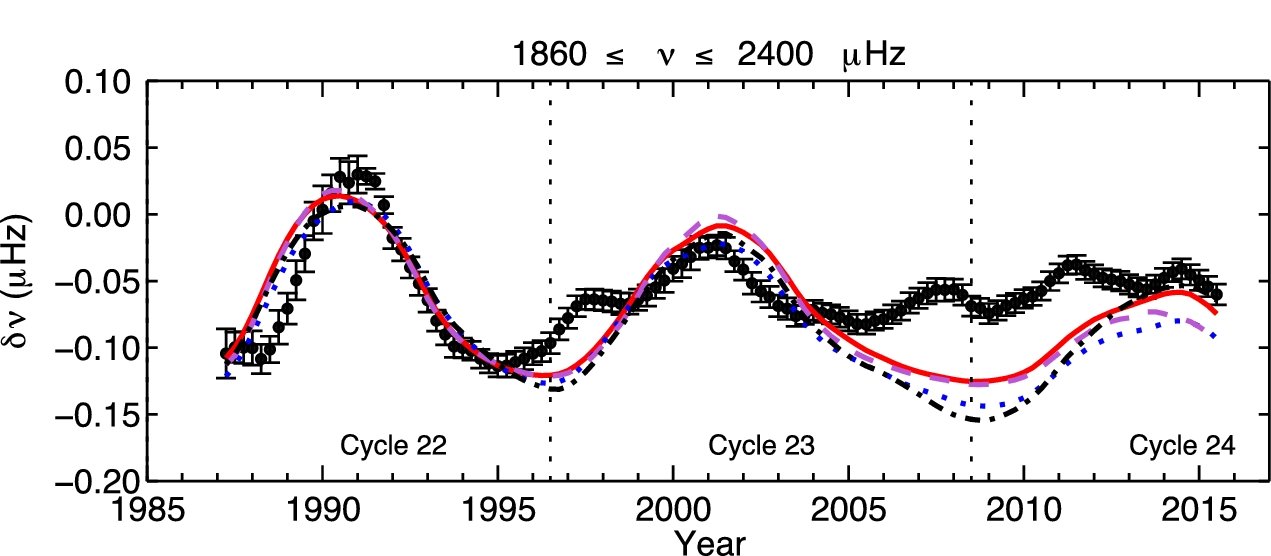}}
    \caption{Portents of changes to come 1: solar oscillations. Averaged
    frequency shifts (symbols with error bars) in the indicated frequency bands
    as a function of time. Renormalized data on the 10.7-cm radio flux (RF),  Ca
    K index, Kitt Peak global magnetic field strength index ($B_{\mathrm KP}$)
    and SSN v1 are plotted for comparison as shown in the legend. Vertical
    dotted lines mark cycle minima.
(From \citealt{Howe+:seismicportent})}
    \label{fig:seismicportent}
\end{figure}

\newsubsection{External empirical precursors}
\label{sect:portents}

{\revone  Cycle 24 peaked at an amplitude 35\,\% lower than Cycle 23. This was
the 4th largest intercycle drop in solar activity in the monthly SSN record. The
last such occurrence was rather different, following the single anomalously
strong cycle 19, while in the case of the drop after Cycle 23 a series of 7
strong cycles was ended. This is illustrated in Fig.~\ref{fig:Gleissbgfilt}
where the shaded area marks a $\pm 2\sigma$ band around the mean amplitude of
cycles 17--23 comprising the Modern Maximum. Cycle 24 clearly deviates downwards
from the Modern Maximum cycles by over $2\sigma$ on the low side.

The first two occurrences of similar intercycle drops in activity (following
cycles 4 and 11) are closer analogues of the recent events. These heralded the
two previous Gleissberg minima centuries ago. The circumstance that a new
Gleissberg minimum was already overdue suggests that this is indeed what we are
witnessing. This is in line with current indications that solar activity will
remain at Cycle 24 levels also in Cycle 25 (cf.~Sect.~\ref{sect:summary}). Yet
no work predating 2005 predicted this drop which mostly caught solar physicists
by surprise.

This} unexpected drop in the level of solar activity from Cycle~23 to 24 has
spurred efforts to find previously overlooked earlier signs of the coming change
in solar data. A number of such ``portents'' were indeed identified, as first
reviewed and correlated by \cite{Balogh+:portents}. 

A ``seismic portent'' was identified by \cite{Basu+:seismicportent}.
High-frequency solar oscillations, sampling the top of the solar convective
zone, have long been known to display frequency variations correlated with the
solar cycle. The analysis of \cite{Basu+:seismicportent} showed that for
[relatively] lower frequencies the amplitude of the frequency variation was
strongly suppressed in Cycle~23, compared with Cycle~22 or with the variation in
higher frequency modes (Fig.~\ref{fig:seismicportent}). This suggests that the
(presumably magnetically modulated) variations in the sound speed were limited
to the upper 3 Mm of the convective zone in Cycle~23, whereas in the previous
cycle they extended to deeper layers. Revisiting the issue,
\cite{Howe+:seismicportent} confirmed a change in the frequency response to
activity during Solar Cycle~23, with a lower correlation of the low-frequency 
shifts with activity in the last two cycles compared to Cycle~22.

  \begin{figure}[htbp]
   
\centerline{\includegraphics[width=\textwidth]{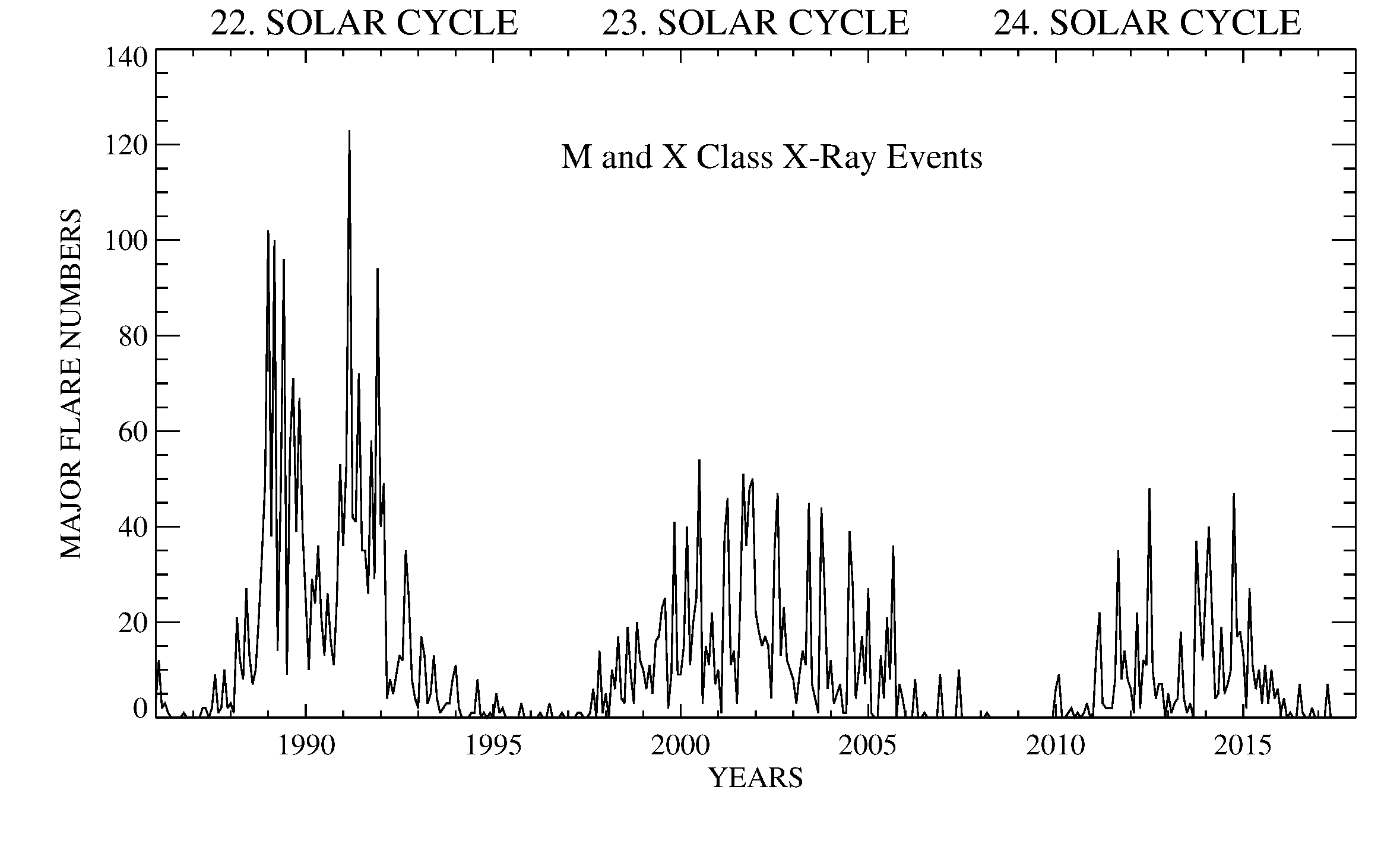}}
    \caption{Portents of changes to come 2: flare statistics.
    Occurrence rate of X and M class flares during the last three solar
    cycles. In terms of flare statistics, the Modern Maximum seems to have 
    ended \textit{before} Cycle~23 already.
    (Figure courtesy of A. \"Ozg\"u{\c c})}
    \label{fig:flareportent}
\end{figure}

A similar disproportionately strong suppression of Cycle~23 relative to Cycle~22
is seen in the occurrence rate of flares, especially of class X and M
(Fig.~\ref{fig:flareportent}), and also in the variation of the H$_\alpha$ flare
index. The suppression is rarely commented on yet clearly seen in the plots of
\cite{Atac+:2001}, \cite{Atac+:2006},  
\cite{Hudson+:flare.productivity}, or
\cite{Gao+Zhong}.\footnote{\cite{Hudson+:flare.productivity} also called attention
to the unusal temporal distribution of active region flare productivity during
Cycle~23: in the first half of the cycle the number of flares produced by a
flaring AR remained at an all-time low, then from the 2003 Halloween events it
suddenly rose.}  
{\revone It may be worth noting that the relationship between the SSN and the
soft X-ray background was also found to differ for different solar cycles by
\cite{Winter+:flarestat2}.}

This ``eruptivity portent'' is also manifested in the variation of the coronal
index (green coronal line emissivity), as seen in the plots of
\cite{Atac+:2006}. A curious disagreement is seen regarding the suppression of
the number of C class flares in Cycle~23: while the data of
\cite{Hudson+:flare.productivity} suggest a significant suppression of the
number of these flares, only slightly less than for M and X class flares,
\cite{Gao+Zhong} found a much less strong suppression for C flares compared to M
and X type flares. This is puzzling as both works are based on the same NOAA
data, the only apparent difference being the exclusion of flares close to the
limb by \cite{Hudson+:flare.productivity}.

  \begin{figure}[htbp]
    \centerline{\includegraphics[width=0.5\textwidth]{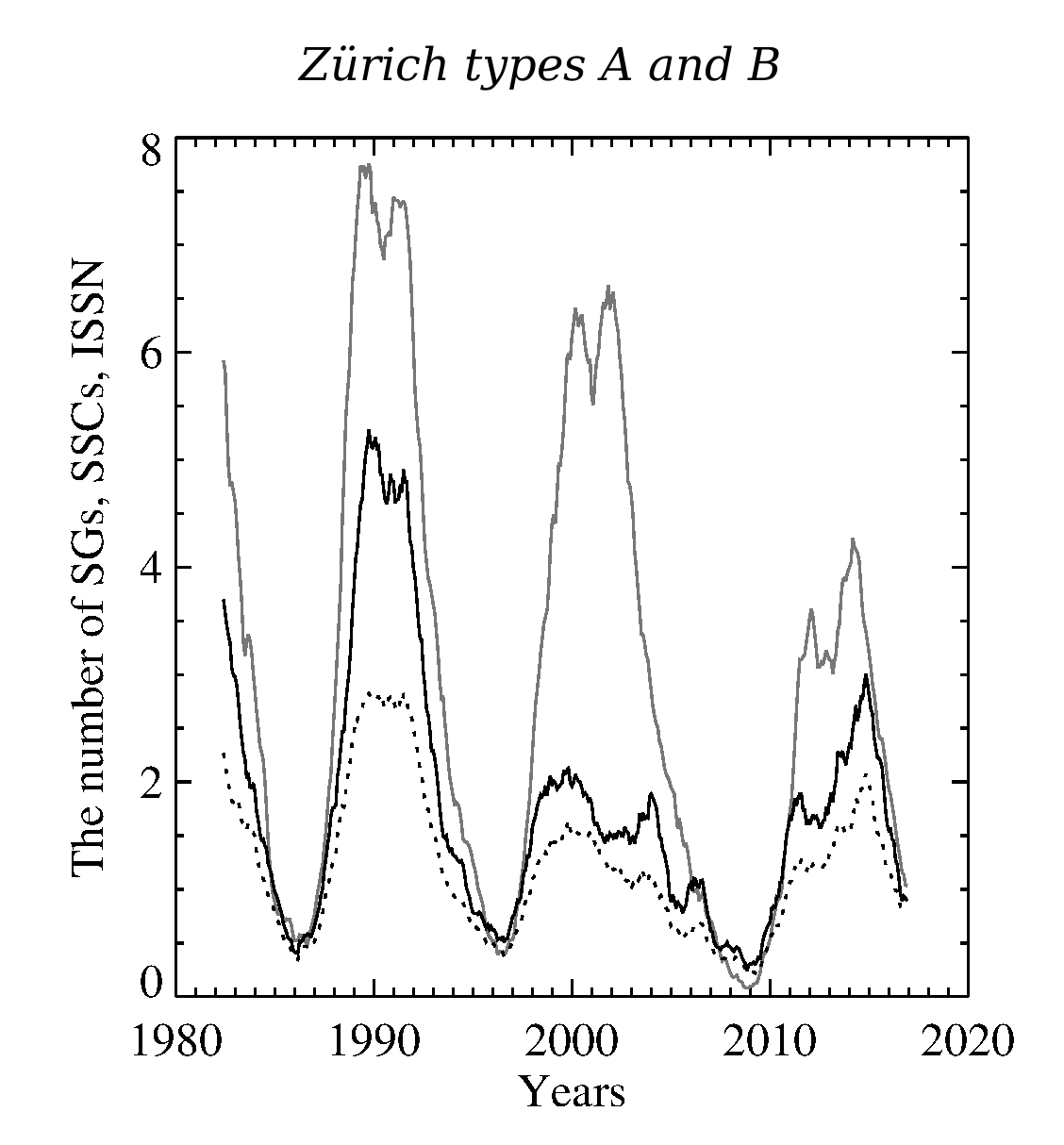}
    \includegraphics[width=0.5\textwidth]{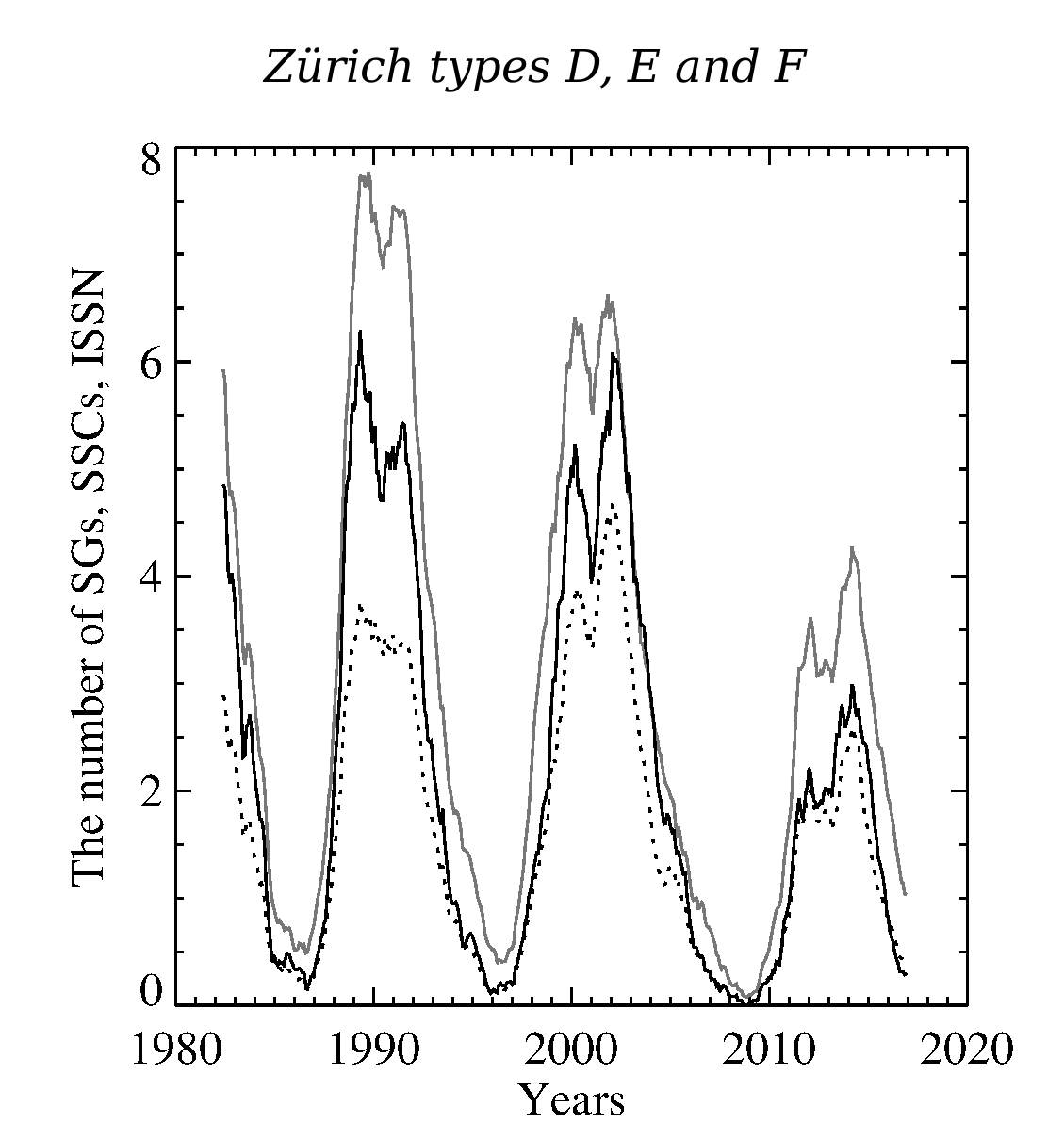}}
    \caption{Portents of changes to come 3: sunspot size statistics.
    Variation of sunspot counts (SSC, thick solid) and group counts 
    (SG, dotted) for small (Z\"urich types A and B, left) and large (types D to
    F, right) sunspot groups. SSN v1 is shown in grey for comparison.
    (Figure courtesy of A. Kilcik)}
    \label{fig:spotportent}
\end{figure}

The stronger suppression of larger flares might be interpereted as a relative
lack of large active regions harbouring sufficient magnetic energy to produce
such flares. Based on the expectation that the magnetic ``roots'' of larger ARs
reach deeper, this would also agree with the seismic portent. Indeed,
\cite{Howe+:seismicportent} explicitly speculate that the observed suppression of
the low frequency modulation in Cycle~23 is ``perhaps because a greater
proportion of activity is composed of weaker or more ephemeral regions''.

Apparently in line with the above reasoning, \cite{deToma+:bigspotslost}
reported
a strong suppression of the number of very large ($>700$ msh) sunspots and
sunspot groups in Cycle~23. But the situation is not so simple as
\cite{Kilcik+:smallspotslost1}, \cite{Lefevre+:smallspotslost} and
\cite{Kilcik+:smallspotslost2} found an apparently opposite trend: a strong
suppression in the number of very small ($< 17$ msh) spots or of sunspot groups
of Z\"urich type A and B (pores/pore pairs) while the number of larger, more
complex spots/groups is largely unaffected, 
{\revone or even slightly enhanced} (Fig.~\ref{fig:spotportent}). As the
contribution to plage areas, radio flux, TSI or disk-integrated magnetic flux
density is dominated by these large ARs, no significant suppression of Cycle~23
is detected in these proxies either (\citealt{Goker+:Serbian}).

These perplexing findings may also be linked to the apparent decrease of the
sunspot magnetic field strengths throughout Cycle~23
(\citealt{Livingston+:decreasingB}; \citealt{Nagovitsyn:2spotpops}). There is,
however, as yet no consensus regarding the reality of this trend
(\citealt{deToma+:spotfields}; \citealt{Watson+:spotfields}) or regarding to
what extent they are cycle related or due to secular trends
(\citealt{Norton+:spotfields}; \citealt{Rezaei+:spotfields},
\citeyear{Rezaei+:spotfields2}; \citealt{Nagovitsyn+:spotfields}).

Studies pointing to possible interrelationships between the various portents
discussed above include \cite{Kilcik+:flaringnonflaring} where a stronger
decrease in sunspot count in flaring AR was reported compared to non-flaring
regions. While local subsurface flow properties in AR, in particular vorticity,
have also been found to correlate with flare productivity
(\citealt{Mason+:subsurface.flows}; \citealt{Komm+:subsurface.flows.1},
\citeyear{Komm+:subsurface.flows.2}), the apparently only study of the
relationship between local disturbances of seismic properties (such as sound
speed) in AR and flare index led to inconclusive results
(\citealt{Lin:subsurface_eruptivity}).

\newsubsection{Polar precursor}
\label{sect:polar}

The polar precursor method, as first suggested by \cite{Schatten_:polar.prec},
is based on the correlation between the amplitude of a sunspot maximum with a
measure of the amplitude of the magnetic field near the Sun's poles at the
preceding cycle minimum. Its physical background is the plausible causal
relationship between the toroidal flux and the poloidal flux that serves as a
seed for the generation of toroidal fields by the winding up of field lines in a
differentially rotating convective zone. 

It is now widely agreed that, beside internal empirical precursor methods based
on the Waldmeier rule, the polar precursor method is currently the most reliable
way to forecast an upcoming solar cycle. As the first revision of this review
concluded, the polar precursor method ``has consistently proven its skill in all
cycles.'' It is now also widely agreed that the polar precursor stands behind
the apparent predicting skill of several other forecasting methods, including
geomagnetic precursors.

\begin{figure}[htbp]
   
\centerline{\includegraphics[width=\textwidth]{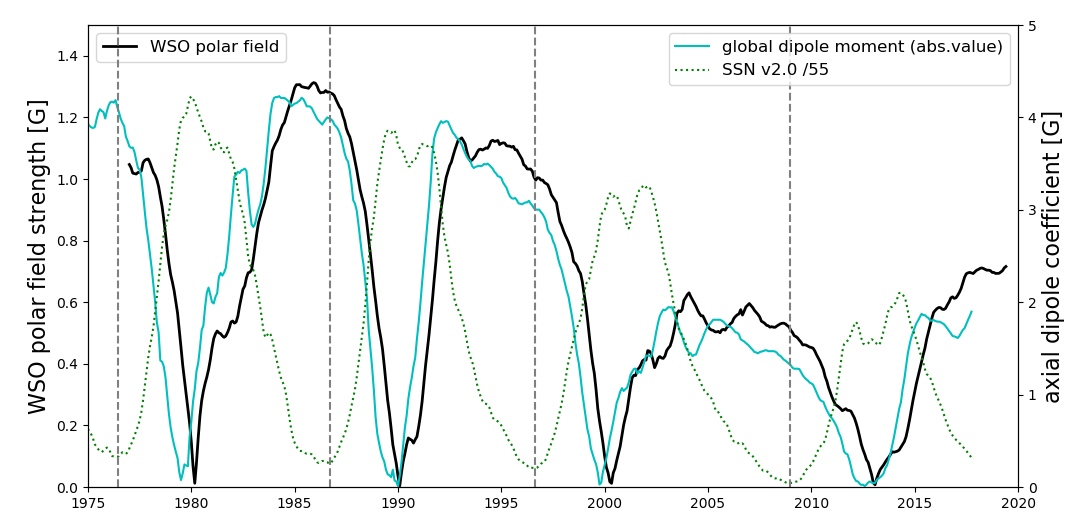}}
    \caption{The hemispherically averaged polar field amplitude from the WSO
    data set (black) and the overall dipole amplitude (cyan) as a function of
    time. The sunspot number series (green dotted) is shown for comparison, with
    an arbitrary rescaling. All curves were smoothed with a 13-month sliding
    window. Times of sunspot minima are marked by the dashed vertical lines.
    Global dipole amplitudes were obtained by courtesy of Jie Jiang and
    represent the average of values computed for all available data sets 
    {\revone out of a maximum of five (WSO, NSO, MWO, MDI, HMI)} at the
    given time.}
    \label{fig:polarfield}
\end{figure}

\subsubsection{Polar magnetic field data}

Observational data on magnetic fields near the Sun's poles were reviewed by
\cite{Petrie:LRSP}. Solar magnetograms have been available on a regular basis
from Mt. Wilson Observatory since 1974, from Wilcox Solar Observatory (WSO)
since 1976, and from Kitt Peak since 1976 (with a major change in the instrument
from KPVT to SOLIS in 2003). The most widely used set of direct measurements of
the magnetic field in the polar areas of the Sun is from the WSO series
\citep{Svalgaard_:polarfield, Hoeksema:SSR}. While these magnetograms have the
lowest resolution of the three sets, from the point of view of the
characterization of the polar fields this is not necessarily a disadvantage, as
integrating over a larger aperture suppresses random fluctuations and improves
the S/N ratio. As a result of the low resolution, the WSO polar field value is a
weighted average of the line-of-sight field in a polar cap extending down to
$\sim 55^\circ$ latitude on average (with significant annual variations due to
the $7^\circ$ tilt of the solar axis). 

The classic reference on the processing and analysis of WSO polar field data is
\cite{Svalgaard_:polarfield}. The inference from their analysis was that
assuming a form $B=B_0 f(\theta)$ with $f(\theta)=\cos^n(\theta)$ for the actual
mean magnetic field profile ($\theta$ is colatitude) inside the polar cap around
minimum, $n=8\pm 1$ while $B_0$ was around 10~G for Cycle~21 and the next two
cycles, being reduced to about half that value in Cycle~24. While one later
study (\citealt{Petrie+Patrikeeva}) points to a possibility that the value of $n$
may be even higher, up to 10, the ``canonical'' value  $n=8$ seems quite
satisfactory in most cases (e.g., Fig.~2 in
\citealt{Whitbread+:SFToptimization}).

Figure~\ref{fig:polarfield} shows the variation of the smoothed amplitude of the
WSO polar field, averaged over the two poles. (The presence of undamped residual
fluctuations on short time scales illustrates the unsatisfactory nature of the
13-month smoothing, applied here for consistency with the rest of this review. A
regularly updated plot of the WSO polar field with a more optimal smoothing 
(low-pass filter) is available from the WSO web 
site\footnote{\url{http://wso.stanford.edu/gifs/Polar.gif}}.)

Also shown is an alternative measure of the amplitude of the poloidal field
component, the \textit{axial dipole coefficient}, i.e., the amplitude of the
coefficient of the $Y_1^0$ term in a spherical harmonic expansion of the
distribution of the radial magnetic field strength over the solar disk:
\begin{equation}                       \label{eq:dipole}
    D(t) = \frac32 \int_0^{\pi} 
    \overline{B}(\theta,t)\cos\theta\sin\theta\, \mathrm{d}\theta.
 \label{eq:dipmom}
\end{equation}
where $\overline B$ denotes the azimuthally averaged radial magnetic field.

This formula assumes the use of the Schmidt quasi-normalization in the
definition of the spherical harmonics, widely used in solar physics and
geomagnetism (see, e.g., \citealt{Winch+:Schmidtnorm}). For direct comparison of the
amplitudes of harmonics of different degree, a full normalization is sometimes
preferred (e.g., in \citealt{DeRosa+:harmonics}): this results in a normalized
dipole coefficient $\hat D=(4\pi/3)^{1/2}D$. While (\ref{eq:dipole}) or even
(\ref{eq:polardipole}) are often loosely referred to as the ``solar dipole
moment'', it should also be kept in mind that the magnetic [dipole] moment, as
normally defined in physics, is related to $D$ as $(2\pi \Rsun^3/\mu_0) D$
where $\Rsun$ is the solar radius and $\mu_0$ is the vacuum permeability. 

The two curves in Fig.~\ref{fig:polarfield} are quite similar even in many of
their details: the polar field amplitude follows the variations of the dipole
coefficient with a phase lag of about a year. This is hardly surprising as the
hemispherically averaged polar field amplitude $|B_N-B_S|/2$ is clearly
proportional to the contribution to $D$ coming from the polar caps:
\begin{eqnarray}                       \label{eq:polardipole}
    D_{NS}(t) && = \frac32 \int_0^{\theta_c} 
    \overline B(\theta)\cos\theta\sin\theta\, \mathrm{d}\theta +
    \frac32 \int_{\pi-\theta_c}^\pi 
    \overline B(\theta)\cos\theta\sin\theta\, \mathrm{d}\theta \nonumber \\
    && = \frac32 (B_{0,N}-B_{0,S})  \int_0^{\theta_c} 
    \overline f(\theta)\cos\theta\sin\theta\, \mathrm{d}\theta .
\end{eqnarray}
As the polar field is formed by the poleward transport of magnetic fields at
lower latitudes, it is only to be expected that the variation of the polar cap
contribution will follow that of the overall dipole with some time delay.

Indeed, based on the good agreement of the two curves in
Fig.~\ref{fig:polarfield}, $(B_N-B_S)/2$ may simply be used as a simple
measure of the amplitude of the dipole (on an arbitrary scale); on similar
grounds, $(B_N+B_S)/2$ may be considered a measure of the quadrupole
component (e.g., \citealt{Svalgaard_:prediction24}; 
\citealt{Munozjara+:polar.precursor}).

We note that the dipole moment in our figure is an average of all
available values from different magnetogram data sets for a given
date; however, there is a quite good overall agreement among the
values from different data sets (see Fig.~9 in
\citealt{Jiang+:1cycle}).\footnote{With the exception of NSO data (not
shown in that plot), which in Cycle~21 significantly deviate from the
others.}

The behaviour of the curves in Fig.~\ref{fig:polarfield} further shows that
the times of dipole reversal are usually rather sharply defined. Based on the 4
reversals seen in the plot, the overall dipole is found to reverse $3.44 \pm
0.18$ years after the minimum, while the polar contribution to the dipole
reverses after $4.33 \pm 0.36$ years. (The formal errors given are $1\sigma$.)
The low scatter in these values suggests that the cycle phase of dipole reversal
may be a sensitive test of SFT and dynamo models.

In contrast to reversal times, maxima of the dipole amplitude are much less well
defined (occurring $7.27 \pm 1.38$ and $8.33 \pm 1.08$ years after minimum for
the two curves). The curves display broad, slightly slanting plateaus covering 3
to 5 years (\citealt{Iijima+:plateau}); the dipole amplitude at the time of
solar minimum is still typically $84\pm 12\%$ (global dipole) and $90\pm
6\%$ (polar fields) of its maximal value, reached years earlier. This kind of
slanting profile is actually good news for cycle prediction as it opens the way
to guess the dipole amplitude at the time of minimum, used as a predictor, years
ahead. For example, using the rather flat and low preceding maximum in polar field
strength, \cite{Svalgaard_:prediction24} were able to predict a relatively weak
Cycle~24 (predicted Version~1 peak SSN value $75\pm 8$ vs.\ $67$ observed) as
early as 4~years before the sunspot minimum took place in December 2008! 

The potential use of the dipole amplitude as a precursor is borne out by the
comparison with the sunspot number curve in Fig.~\ref{fig:polarfield}. After
our arbitrary rescaling of the SSN, its maxima in each cycle are roughly at
level with the preceding plateaus of the solid curves. This certainly seems to
indicate that the suggested physical link between the precursor and the cycle
amplitude is real.

\begin{table}[htbp]
\caption{Predictors based on magnetic field measurements and their forecasts.
{\revone Forecast limits have been calculated adding/subtracting the scatter (as
calculated nominally from the 4 data points) to/from the value given by the fit.}}
    \label{table:polarprecursors}
\begin{tabular}{lrrrrr}
\hline\\
Cycle & Amplitude & Previous WSO & WSO field  &  Previous &  Dipole coef. $D$ \\  
     &  (SSN v2) & field maximum & at minimum  &  $D$ maximum &  at minimum  \\  
\hline\\
21 &  232.9  &   $[1.07]$  & $1.05$   &	$4.19$   &     $4.10$  \\
22 &   212.5 &     $1.31$  &    $1.28$  &       $4.23$  &       $3.98$ \\
23 &   180.3 &     $1.13$  &    $1.00$  &       $3.96$  &       $3.01$ \\
24 &   116.4 &     $0.63$  &    $0.52$  &       $1.95$  &       $1.33$ \\
25 & 	     &	   $>0.72$  & $<0.72$ &	$>1.93$ &  $<1.93$ \\
\hline\\
Fit coef. & &  $177.4$  & $188.8$  &  $51.3$ &	$57.5$\\
Scatter     & &  $ 25.9$  & $ 24.8$  &  $16.8$ &	$21.9$\\
Cycle~25 &&&&&\\    
\ forecast  &  &  $>102$   & $<161$   &  $>82$  &	$<133$ \\
\hline\\
						        
\end{tabular}
\end{table}

The flatness of the maxima of the polar field imply that the precursor
normally cannot be strongly affected by the exact time when it is
evaluated. The cycle overlap effect combined with the Waldmeier
relation, affecting the timing of the minimum, is therefore unlikely
to explain the predictive skill of the polar precursor: we are here
dealing with a real physical precursor (as also argued by
\citealt{Charbonneau+:realpolarprecursor}).

The ``polar precursor'' may thus be interpreted in four different
ways. The precursor may be the value of the global dipole moment or of
the contribution of polar fields to this dipole moment only (i.e., the
WSO field); and either of these may be evaluated at cycle minimum, or
a few years earlier when they reach their respective maxima.
Table~\ref{table:polarprecursors} lists these precursor values for
individual solar cycles, compared with the actual cycle amplitude. A
homogeneous linear fit with one free parameter to the precursor--cycle
amplitude relation yields the fit coefficient values given in the
lower part of the table. The nominal random scatter is also
indicated. Precursor values for Cycle~25 have been evaluated in late
2018; as it is not yet clear whether the maximum of the dipole moment
has passed or when the minimum will take place, the
forecasts based on these values are to be interpreted as lower/upper
estimates, respectively. Taking into account the given values of the
scatter, a combination of these results implies that Cycle~25 should
peak in the range {\revone 102--133 with a nominal probability of $\sim 70\,$\%,
i.e.\ the next cycle should have an amplitude similar to Cycle 24.
}

\begin{figure}[htbp]
   
\centerline{\includegraphics[width=\textwidth]{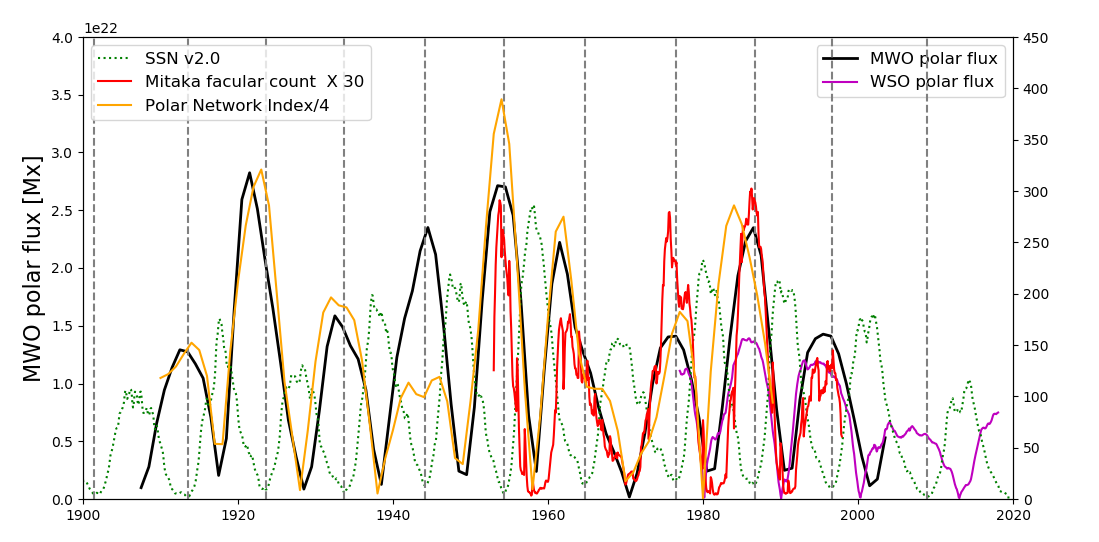}}
    \caption{Temporal variation of hemispherically averaged polar magnetic
    fluxes resulting from the Mt.Wilson polar facular counts (black), WSO polar
    flux (magenta), rescaled Mitaka polar facular counts (red), and rescaled
    Kodaikanal polar network index {\revone (yellow)}. 
    The sunspot number series (green dotted) is
    shown for comparison, with an arbitrary rescaling. Flux values are based on
    the calibration in \cite{Munozjara+:polarfacproxy}. All data refer to the
    area poleward of $70^\circ$ latitude. All curves were smoothed with a
    13-month sliding window except the annually sampled Mt.Wilson and Kodaikanal
    data where a 121 filter was applied. Times of sunspot minima are marked by
    the dashed vertical lines.}
    \label{fig:polarfac}
\end{figure}

\subsubsection{Proxy reconstructions of the polar magnetic field}

Despite the plausibility of a physical link between polar magnetic fields and
cycle amplitude, the shortness of the available direct measurement series
represents a difficulty when it comes to finding a convincing statistical link
between these quantities. A way to circumvent this difficulty is offered by the
availability of proxy data for polar magnetism spanning much longer time scales.
Indeed, Schatten's original suggestion of a polar field precursor
(\citealt{Schatten_:polar.prec}) on a generic physical basis was supported by
such proxy studies. It is remarkable that despite the very limited available
experience, proxy-based forecasts using the polar field method  proved to be
consistently in the right range for Cycles~21, 22, and 23
\citep{Schatten_:pred22, Schatten_:pred23}.

The types of proxies of solar polar magnetism were reviewed by
\cite{Petrie+:polarcycle}. In the present subsection we focus on polar faculae,
which are currently considered the best photospheric proxy data for the 
reconstruction of polar magnetic field/flux. The interplanetary and geomagnetic
precursors discussed in the following sections, however, may also be
interpreted as indirect proxies of the solar polar magnetic field.
 
High resolution observations by the Hinode space observatory confirm that, like
all other magnetic fields in the solar photosphere, polar fields are highly
intermittent, nearly all of the flux being concentrated in isolated strong
magnetic field concentrations \citep{Tsuneta_:polar.landscape}. These magnetic
elements are observed as bright facular points in white light and in some
spectral lines. The number density of these polar faculae is then related to the
intensity of the polar magnetic field, while their total number above a certain
latitude is related to the total magnetic flux (\citealt{Sheeley1964}). This
conclusion was indeed confirmed by \cite{Li_:polarfac} and, more recently, by
\cite{Tlatov:polar.precursors}.

Perhaps the most carefully analyzed polar facular data set is the series of
observations in the Mt. Wilson Observatory. These data were validated against
MDI observations and then calibrated to WSO and MDI magnetic measurements by
\cite{Munozjara+:polarfacproxy}. This resulted in a time series of the solar
polar magnetic flux (poleward of $70^\circ$ latitude) for each hemisphere,
covering the period 1906--2014. Owing to the varying tilt of the solar axis, the
data are available with an annual cadence and with a time shift of 0.5 year
between the hemispheres. This reconstructed polar flux was subsequently
correlated with cycle amplitudes, confirming the usefulness of the polar
precursor method (\citealt{Munozjara+:polar.precursor};
\citeyear{Munozjara+:polarfac.precursor}). Hemispherically averaged data show a
highly significant correlation, albeit with a large scatter ($r=0.69$ at 96\%).
This imperfect correlation is reflected in Fig.~\ref{fig:polarfac} in the
wildly varying ratio between the maxima of the black curve and the subsequent
maxima of the green dotted curve.

The poor correlation can be improved by separating the hemispheres, but some
outlier points appear, apparently obeying an alternative relationship. Arguing
that outliers correspond to cycles where the hemispheric asymmetry of polar
fields exceeds a threshold, \cite{Munozjara+:polar.precursor} finally arrived at
four (two for each hemisphere) linear empirical relationships between the
reconstructed polar magnetic flux at minimum and the amplitude of the next
cycle. The suggested relation correctly reproduced the observed amplitude of
Cycle~24 (predicted Version~1 SSN $77\pm 16$ vs.\ $67$ observed). 

Another relevant data set is the polar facular counts recorded in the National
Astronomical Observatory of Japan (NAOJ) at Mitaka Observatory during the period
1954--1996 (\citealt{Li_:polarfac}). A third series of polar facular data,
originating from Kodaikanal Observatory Ca K spectroheliograms, was compiled by
\cite{Priyal+:polarnetwork}. (As these are chromospheric features, the authors
prefer to call their index a polar network index.) Major disagreements between
these data sets are seen in  Fig.~\ref{fig:polarfac} indicating that the use
of polar facular proxies is still not on very firm grounds. This is further
shown by a comparison of the deducted MWO polar flux with polar magnetic fluxes
computed from WSO polar field data using the calibration formula derived in
\cite{Munozjara+:polarfacproxy} (magenta curve in the plot).  It is apparent
that prior to the calibration period 1996--2006 the reconstructed MWO polar
fluxes are systematically higher than the measured WSO fluxes, suggesting
problems with these calibrations.  In particular, while WSO polar field
strengths peak at roughly the same amplitude in cycles 21, 22 and 23, in
agreement with the comparable amplitudes of the subsequent sunspot cycles, the
different polar facular counts indicate greatly different amplitudes for these
minima, and they are also mutually incompatible.

Polar faculae are not the only long-term data base relevant for the study of
high-latitude magnetic fields. $\mathrm{H}_\alpha$ synoptic charts are 
available from various observatories from as early as 1870. As
$\mathrm{H}_\alpha$ filaments and filament channels lie on the magnetic neutral
lines, these maps can be used to reconstruct the overall topology, if not the
detailed map, of the large-scale solar magnetic field. While in between the
neutral lines only the polarity of the field can be considered known, the
variation of the global dipole moment may be tolerably well estimated even by
fixing the field amplitude at a constant value. As higher order multipoles decay
rapidly with distance, a potential field model fitted to the given synoptic map
will then yield an acceptable representation of the polar field at the source
surface. With this approach \cite{Makarov+:dipoctu} computed the multipole
coefficients of the solar magnetic field, reconstructing the polar field
strength at the source surface back to 1915. They also introduced the so-called
dipole-octupole index (aka $A$-index): the sum of the axial dipole and octupole
magnetic moments as a simple measure of the polar field amplitude.  The method
was later applied by \cite{Obridko_Shelting} to predict the amplitude of Cycle
24: their forecast proved to be within 10\% of the actual value. In turn,
\cite{Tlatov:polar.precursors} has shown that several indices of the polar
magnetic field during the activity minimum, determined from these charts,
correlate well with the amplitude of the incipient cycle.

\subsubsection{Extending the range of the polar precursor: early
forecasts for Cycle~25}

The polar  precursor is customarily evaluated at the cycle minimum, offering a
prediction over a time span of 3\,--\,4 years, comparable to the rise time of
the next cycle. Table~\ref{table:polarprecursors}, however, shows that using the
maximal value of the global dipole moment results in a somewhat lower scatter
around the mean relationship. This opens the possibility of making a prediction
several years before the minimum. 

The forecasting potential of the global dipole may also provide the ground for
the findings of \cite{Hawkes+Berger} who proposed the ``helicity flux'' as a
cycle precursor. Perhaps more aptly called \textit{helicity input rate} by the
differential rotation, their helicity flux is defined by a weighted hemispheric
surface integral of (a functional of) the radial magnetic field, where the
weight function is fixed by the differential rotation profile. In such an
integral, which is not unlike equation (\ref{eq:dipmom}), the dipole component
will naturally give a dominant contribution as the higher order terms change
sign over each hemisphere, which largely cancels their effect. The authors found
that the helicity input rate anticipates the sunspot numbers with a time shift
of 4.5--6.9 years (depending on the cycle considered). This seems to be in line
with the variable time delays between maxima of the global dipole moment and the
next cycle maximum (Fig.~\ref{fig:polarfield}). Their prediction for the
amplitude of Cycle~25 is 117 --- similar or slightly stronger than Cycle~24.

Similarly, using the brightness temperature of the 17 GHz microwave emission as
a proxy for the field strength, \cite{Gopal+:microwpred} found that the
correlation between this proxy and the sunspot number is maximal for time shifts
of $\sim 4$--$6$ years (depending on cycle and hemisphere). Their forecast for
Cycle~25 is a smoothed SSN of 89 for the S hemisphere and 59 for the N
hemisphere (the latter being a lower bound as the proxy had not reached its
maximum at the time of publication). These values are again comparable to or
slightly higher than the amplitudes for Cycle~24.

In this context it is interesting to note that \cite{Makarov:polfacpred89} and
\cite{Makarov_:polfacpred96} found that the number of polar faculae observed at
Kislovodsk anticipates the next sunspot cycle with a time lag of 5\,--\,6 years
on average in cycles 20--22; even short term annual variations or ``surges'' of
sunspot activity were claimed to be discernible in the polar facular record. An
apparently conflicting result was obtained by \cite{Li_:polarfac}, who found
using the Mitaka data base that the best autocorrelation results with a time
shift of about 4~years only. The discrepancy may perhaps be related to the cycle
dependence of these time shifts, as partly different periods were considered.

In theory it is conceivable that successful forecasts might be attempted even
earlier. After all, the polar field starts to build up at the time of polar
reversal, about 5--6 years before the next minimum. \cite{Petrovay+:greenlpred}
explored this possibility in a study of the coronal green lime emission at high
latitudes.  They tried to correlate various features of the characteristic
``rush to the poles'' (RTTP) feature of this emission around the polar reversal
with properties of the following sunspot cycle and found a significant
correlation between the speed of the RTTP and the time from the reversal to the
next maximum. From this they predict that Cycle~25 will most likely peak in late
2024. Combining this date with the minimax3 method discussed in Section
\ref{sect:minimax} above, the cycle is amplitude is estimated as 130, and the
minimum is expected for 2019.

It is worth noting here that, in addition to the start of the increase of
the polar field 5--6 years before the minimum, early precursors at high
latitudes may be expected also on completely different grounds. The concept of
the \textit{extended solar cycle} implies that small ephemeral bipoles belonging to
an upcoming solar cycle appear at high latitudes and start to migrate
equatorward years before the first spots of the new cycle are observed. Thus,
early signs of the equatorward propagating toroidal flux ring at high latitudes
may give hints on the amplitude of an upcoming cycle \citep[cf.\ also][]{Badalyan_}. (From a formal point of
view this would be then an internal cycle precursor, related to the one based on
the Waldmeier rule.) 

It is not impossible that some of the very early precursors suggested above, if
real, may be partly explained by this effect. For example,
\cite{Makarov_:polfacpred96} considered all faculae poleward of $50^\circ$ 
latitude. \textit{Bona fide} polar faculae, seen on Hinode images to be knots of
the unipolar field around the poles, are limited to higher latitudes, so the
wider sample may consist of a mix of such ``real'' polar faculae and small
bipolar ephemeral active regions. These latter are known to obey an extended
butterfly diagram, as confirmed by \cite{Tlatov_:ER}: the first bipoles of the
new cycle appear at higher latitudes about 4~years after the activity maximum. 

In interpreting  high-latitude migration patterns it should, however, be taken
into account that it is not yet clear how far the wings of the butterfly
diagram can actually extended backwards, i.e., to what extent a high latitude
equatorward propagating branch is contiguous with the low latitude branch or is
an unrelated phomenon (cf. the discussion in \citealt{Cliver:extendedcycle} and 
\citealt{Petrie+:polarcycle}).

The high-latitude torsional oscillation pattern is usually considered the
most pregnant manifestation of the extended solar cycle. This pattern has been
unusually week during Solar Cycle~24, apparently suggesting significant further
weakening of solar activity (\citealt{Howe+:torsoprecursor1}). However, later
observations (\citealt{Howe+:torsoprecursor2}; \citealt{Komm+:torsoprecursor})
indicate that the low-latitude equatorward branch of the torsional oscillation
is actually stronger in Cycle~24 than it was in Cycle~23, if measured against
the same background flow.

{\oldtext 
In order to obtain a precursor that varies smoothly enough to be useful also
between successive minima, \cite{Schatten_:SoDA} introduced a new activity
index, the ``Solar Dynamo Amplitude'' (SoDA) index, combining the polar field
strength with a traditional activity indicator (the 10.7~cm radio flux
\textit{F}10.7). Around minimum, SoDA is basically proportional to the polar
precursor and its value yields the prediction for \textit{F}10.7 at the next
maximum; however, it was constructed so that its 11-year modulation is
minimized, so theoretically it should be rather stable, making predictions
possible well before the minimum.
}
 It remains to be seen whether SoDA actually
improves the predictive skill of the polar precursor, to which it is more or
less equivalent in those late phases of the solar cycle when forecasts start to
become reliable. Using the SoDA index \cite{Pesnell+:SoDA2018} predicted Cycle~25
to peak at $R=134\pm 25$ in 2025.

\subsection{Geomagnetic and interplanetary precursors}
\label{sect:geomg}

Relations between the cycle related variations of geomagnetic indices and solar
activity were noted long ago. It is, however, important to realize that the
overall correlation between geomagnetic indices and solar activity, even after
13-month smoothing, is generally far from perfect. This is due to the fact that 
the Sun can generate geomagnetic disturbances in two ways:

\begin{figure}[htbp]
   
\centerline{\includegraphics[width=\textwidth]{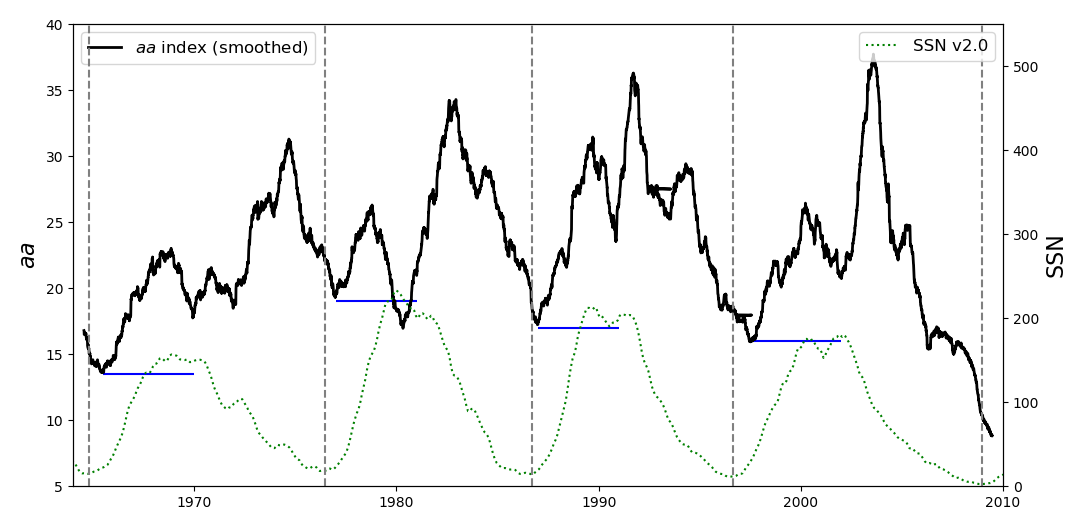}}
    \caption{Illustration of Ohl's method in cycles 20--23. Shown is the
    variation of the daily geomagnetic $aa$ index (solid black) and the monthly
    SSN (dotted green), both smoothed with a 13-month sliding window. Horizontal
    blue lines mark the value of the predictor in each cycle. Times of sunspot
    minima are marked by the dashed vertical lines.}
    \label{fig:aaindex}
\end{figure}

\begin{descr}
\item{(a)} By material ejections (such as CMEs or flare particles) hitting the
terrestrial magnetosphere. This effect is obviously well correlated with solar
activity, with no time delay, so this contribution to geomagnetic disturbances
peaks near, or a few years after, sunspot maximum. (Note that the occurrence of
the largest flares and CMEs is known to peak some years after the sunspot
maximum -- see Figure~16 in \citealt{Hathaway:LRSP2}.)
\item{(b)} By a variation of the strength of the general interplanetary magnetic
field and of solar wind speed. Geomagnetic disturbances may be triggered by the
alternation of the Earth's crossing of interplanetary sector boundaries (slow
solar wind regime) and its crossing of high speed solar wind streams while well
within a sector. The amplitude of such disturbances will clearly be higher for
stronger magnetic fields. The overall strength of the interplanetary magnetic
field, in turn, depends mainly on the total flux present in coronal holes, as
calculated from potential field source surface models of the coronal magnetic
field. At times of low solar activity the dominant contribution to this flux
comes from the two extended polar coronal holes, hence, in a simplistic
formulation this interplanetary contribution may be considered linked to the
polar magnetic fields of the Sun, which in turn is a plausible precursor
candidate as we have seen in the previous subsection. As the polar field
reverses shortly after sunspot maximum, this second contribution often
introduces a characteristic secondary minimum in the cycle variation of
geomagnetic indices (cf.~Fig.~\ref{fig:aaindex}, somewhere around the 
maximum of the curve.
\end{descr}

The component (a) of the geomagnetic variations actually \textit{follows}
sunspot activity with a variable time delay. Thus a geomagnetic precursor based
on features of the cycle dominated by this component has relatively little 
practical utility. This would seem to be the case, e.g., with the forecast
method first proposed by \cite{Ohl66}, who noticed that the minimum amplitudes
of the smoothed geomagnetic \textit{aa} index are correlated to the amplitude of
the next sunspot cycle (see also \citealt{Du_:Ohlmethod}{\newtext,
\citealt{Du:Ohlmethod}}).
{\revone This is clearly just a secondary consequence of the previously 
discussed correlations between solar indices at cycle minimum (occurring
somewhat {\it earlier} than the minimum in $aa$, see Fig.~\ref{fig:aaindex}) and
the amplitude of the next maximum.
While the polar precursor is more directly relevant and is available sooner,
the amplitude of the $aa$ index at minimum is still useful as a proxy as these
data have been available since 1868.}

An indication that the \textit{total} geomagnetic activity, resulting from both
mechanisms does contain useful information on the expected amplitude of the next
solar cycle was given by \cite{Thompson:geomg.prec}, who found that the total
number of disturbed days in the geomagnetic field in cycle $n$ is related to the
sum of the amplitudes of cycles $n$ and $n+1$ \citep[see also][]{Dabas_}.

A method for separating component (b) was proposed by \cite{Feynman} who
correlated the annual \textit{aa} index with the annual mean sunspot number and found a
linear relationship between \R and the \textit{minimal} value of \textit{aa} for years
with such \R values. She interpreted this linear relationship as representing
the component (a) discussed above, while the amount by which \textit{aa} in a given
year actually exceeds the value predicted by the linear relation would be the
contribution of type (b) (the ``interplanetary component''). The interplanetary
component usually peaks well ahead of the sunspot minimum and the amplitude of
the peak seemed to be a good predictor of the next sunspot maximum. However, it
is to be noted that the assumption that the ``surplus'' contribution to \textit{aa}
originates from the interplanetary component only is likely to be erroneous,
especially for stronger cycles. It is known that the number of large solar
eruptions shows no unique relation to \R: in particular, for  \R~\textgreater~100 their
frequency may vary by a factor of 3 \citep[see Figure~18 in][]{Hathaway:LRSP2}, so
in some years they may well yield a contribution to \textit{aa} that greatly exceeds
the minimum contribution. A case in point was the ``Halloween events'' of 2003,
that very likely resulted in a large false contribution to the derived
``interplanetary'' \textit{aa} index \citep{Hathaway:Halloween}. As a result, the
geomagnetic precursor method based on the separation of the interplanetary
component predicted an unusually strong cycle~24 ($R_m\sim 150$), in contrast to
most other methods, including Ohl's method and the polar field precursor, which
suggested a weaker than average cycle ($R_m\sim 80\mbox{\,--\,}90$). 

{\newtext An alternative method for the separation of the interplanetary component, based
on the use of the use of the $F_{10.7}$ index to model the variation of the
activity-related component, was proposed by \cite{Pesnell:newgeomg}.}

In addition to the problem of neatly separating the interplanetary contribution
to geomagnetic disturbances, it is also wrong to assume that this interplanetary
contribution is dominated by the effect of polar magnetic fields at all times
during the cycle.  Indeed, \cite{Wang_Sheeley:geomg.precursor} pointed out that
the interplanetary magnetic field amplitude at the Ecliptic is related to the
equatorial dipole moment of the Sun that does not survive into the next cycle,
so despite its more limited practical use, Ohl's original method, based on the
minima of the \textit{aa} index is physically better founded, as the polar dipole
dominates around the minimum.
{\newtext (Cf. also \citealt{Ng}.)}
 The total amount of open interplanetary flux,
more closely linked to polar fields, could still be determined from geomagnetic
activity if the interplanetary contribution to it is further split
into:
\begin{descr} 
\item{(b1)} A contribution due to the varying solar wind speed (or to the
interplanetary magnetic field strength anticorrelated with it), which in turn
reflects the strength of the equatorial dipole. 
\item{(b2)} Another contribution due to the overall interplanetary field
strength or open magnetic flux, which ultimately reflects the axial dipole.
\end{descr} 
Clearly, if the solar wind speed contribution (b1) could also be subtracted, a
physically better founded prediction method should result.  While in situ
spacecraft measurements for the solar wind speed and the interplanetary magnetic
field strength do not have the necessary time coverage, 
\cite{Svalgaard_:LDV1,Svalgaard_:LDV2}  and \cite{Rouillard_:openflux} devised a
method to reconstruct the variations of both variables from geomagnetic
measurements alone. {\newtext Building on their results,
\cite{Wang_Sheeley:geomg.precursor} arrived at a prediction 
for the maximum amplitude of solar cycle~24 which agreed well with that of
\cite{Bhatt_}, who applied a modified form of Ohl's method. Both forecasts proved to
be somewhat too high (by $\sim 20$--$25\,$\%, a little over $1\sigma$).}


The open magnetic flux can also be derived from the extrapolation of solar
magnetograms using a potential field source surface model. The magnetograms
applied for this purpose may be actual observations or the output from surface
flux transport models, using the sunspot distribution (butterfly diagram) and
the meridional flow as input. Such models indicate that the observed latitude
independence of the interplanetary field strength (``split monopole'' structure)
is only reproduced if the source surface is far enough ($>10\,R_\odot$) and the
potential field model is modified to take into account the heliospheric current
sheet (current sheet source surface model, \citealt{Schussler_:openflux,
Jiang_:openflux}). The extrapolations are generally found to agree well with in
situ measurements where these are available. {\newtext A comprehensive review of
this topic is given in \cite{Lockwood:LRSP}.}

{\newtext

\subsection{The quest for a precursor of the polar precursor}
\label{sec:quest}

Just as the suggestion of a polar precursor was based on a qualitative physical
understanding of the process generating the strong toroidal magnetic field that
gives rise the observed active regions, an extension of the temporal range of
our forecasting capability would clearly benefit from a similar qualitative
physical understanding of how the strong polar fields prevalent
around sunspot minimum are formed.

Magnetograph observations rather clearly indicate that the polar field is built
up as a result of the the poleward transport of trailing polarity flux from
active regions, while much of the leading polarity flux cancels with its
counterpart on the other hemisphere by cross-equatorial diffusion. In the
currently widely popular Babcock--Leighton scenario the poleward transport is
interpreted as a combination of turbulent diffusion and advection by a poleward
meridional flow.\footnote{Note that the real situation may well be more complex than
this simple scenario suggests: in numerical simulations of spherical turbulent
dynamos latitudinal transport by pumping effects is quite often prevalent (see,
e.g., \citealt{Racine+:alphatensor}, \citealt{Simard+:alphatensor},
\citealt{Warnecke+:alphatensor}).} This suggests that the buildup of the polar
field may be controlled by  either of two effects:\\
(a) variations of the poleward flow speed \\
(b) variations in the tilt angles {\revone and/or latitudinal distribution} of 
bipolar active regions  which ultimately
determine the net flux imbalance in the meridional direction.

In the following subsections these two influencing factors are considered in
turn.

}

\subsubsection{Photospheric flow variations}
\label{sect:flows}

{\newtext
Considering the effect of meridional flow variations on intercycle variations is a
delicate task as such
}
changes are also associated with the normal course of
the solar activity cycle, the overall flow at mid-latitudes being slower before
and during maxima and faster during the decay phase. Therefore, it is just the
cycle-to-cycle variation in this normal pattern that may be associated with the
activity variations between cycles. In this respect it is of interest to note
that the poleward flow in the late phases of Cycle~23 seems to have had an
excess speed relative to the previous cycle \citep{Hathaway_Rightmire}. If this
were a latitude-independent amplitude modulation of the flow, then most flux
transport dynamo models (e.g., \citealt{Belucz+:merid.circ}) would predict a
stronger than average polar field at the minimum, contrary to observations. On
the other hand, in the surface flux transport model of \cite{Wang_:varying.circ}
an increased poleward flow actually results in weaker polar fields, as it lets
less leading polarity flux to diffuse across the equator and cancel there. As
the analysis by \cite{Munozjara} has shown, the discrepancy resulted from
the form of the Babcock--Leighton source term in flux transport dynamo models,
and it can be remedied by substituting a pair of opposite polarity flux rings
representing each individual AR as source term instead of the $\alpha$-term.
With this correction, 2D flux transport dynamos and surface flux transport
models agree in predicting a weaker polar field for faster meridional flow.

It is known from helioseismology {\newtext and magnetic correlation
tracking} that meridional flow speed fluctuations follow
a characteristic latitudinal pattern associated with torsional oscillations and
the butterfly diagram, consisting of a pair of axisymmetric bands of latitudinal
flows converging towards the activity belts, migrating towards the equator, and
accompanied by similar high-latitude poleward branches 
{\newtext  
(\citealt{Snodgrass+Dailey}, \citealt{Chou+Dai}, \citealt{Beck+:circ.belts}, 
\citealt{Liang+:seismic.flows}, \citealt{Lin+Chou:cycledep.flows}).
}
 This suggests 
interpreting the unusual meridional flow speeds observed during cycle~23 as an
increased amplitude of this migrating modulation, rather than a change in the
large-scale flow speed \citep{Cameron_:circ.belts}. In this case, the flows
converging on the activity belts tend to inhibit the transport of following
polarities to the poles, resulting in a lower polar field 
(\citealt{Jiang_:merid.flow}; note, however, that \citealt{Svanda} find
no change in the flux transport in areas with increased flows). It is
interesting to note that the torsional oscillation pattern, and thus presumably
the associated meridional flow modulation pattern, was shown to be fairly well
reproduced by a microquenching mechanism due to magnetic flux emerging in the
active belts \citep{Petrovay_Dajka:torso}. 
{\newtext  
Alternatively, the modulation pattern may also be thermally induced
(\citealt{Spruit:torso}) or it may result from  large-scale magnetic
field torques (\citealt{Passos+:magn.circulation},
\citealt{Hazra+:merid.flow}, \citealt{Ruzdjak:difrot.activity}). }
This suggests that stronger cycles may be associated with a stronger
modulation pattern, introducing a nonlinearity into the flux transport
dynamo model (\citealt{Jiang_:merid.flow},
{\newtext 
\citealt{Cameron+:inflowtilt}, \citealt{Karak+:mcquenching}).
The relationship between activity level and flow modulation, however,
seems more complex than a simple proportionality 
(\citealt{Komm+:seismic.inflows}). In particular, the modulation signal in the
Cycle~24 activity belt seems to be too strong in comparison with the low
amplitude of the sunspot cycle.
}

In addition to a variation in the amplitude of migrating flow modulations, their
migration speed may also influence the cycle. \cite{Howe_:torso.precursor}
pointed out that in the minimum of Cycle~24 the equatorward drift of the
torsional oscillation shear belt corresponding to the active latitude of the
cycle was slower than in the previous minimum. They suggested that this slowing
may explain the belated start of cycle~24.

{\newtext 
Under the assumption that meridional flow modulations are the main factor
controlling the buildup of the poloidal field from AR sources, 
\cite{Hung2015}, \cite{Hung2017} suggest an inverse approach to derive flow
variations from magnetic data. As, however, we will see in the next subsection,
the validity of the underlying assumption is open to question.

In summary: while magnetically induced modulations of the meridional flow and
their effect on flux transport may be a potentially important nonlinear
feedback mechanism controlling intercycle activity variations, the limited
observational record and the apparent complexities of the interplay have as yet
not permitted their use as a precursor.
}

\newsubsubsection{Active region tilts}

As luck would have it, soon after the first version of this review was finished,
a paper was published that prompted a flurry of activity in a completely new 
field. In the paper \cite{Dasi-Espuig+} analysed sunspot group catalogues
extracted from the Mt.Wilson and Kodaikanal photoheliograms for solar cycles
15--21, focusing on the distribution of tilt angles of the longitudinal axis of
bipolar active regions to the azimuthal direction. Area-weighted averages of the
tilt angles of sunspot groups were calculated in latitudinal bins, then
normalized by the latitude to yield a tilt parameter. Two effects emerged from
the analysis:\\
(1) Tilt quenching (TQ):  an anticorrelation between the amplitude of a solar
cycle and its mean tilt parameter. \\
(2) Tilt precursor (TP): the product of the amplitude of a sunspot cycle with
its mean tilt parameter turned out to be a good predictor of the amplitude of
the \textit{subsequent} cycle.  \cite{Munozjara+:polarfac.precursor} further
demonstrated that this product is also a good predictor of the polar magnetic
flux at minimum (as reconstructed from polar facular counts), suggesting that
the predictive potential of this method is based on the role of tilt angles in
controlling the amount of net flux transported towards the poles.

The combination of TP with a TQ relationship upon which random fluctuations
in the tilt are superposed implies that intercycle variations in solar activity
will be controlled by a nonlinear feedback mechanism, into which a stochastic
element is incorporated. This realization prompted intense activity in the
development of model-based cycle prediction, to be discussed in the following
section.

Ironically, in parallel with the major impact of the
\cite{Dasi-Espuig+} paper on theoretical work, it was soon subjected
to criticism on observational grounds. \cite{Ivanov:noTQ} repeated the
analysis, now including the Pulkovo sunspot group catalogue (covering
a shorter period, cycles 18--21). The TQ effect in the Mt.Wilson data
set was found to depend crucially on the low value of the mean tilt for
the anomalous strong cycle 19. In the Kodaikanal data the effect
appeared more robustly but it still seems to depend on high tilt
values in cycles 15 and 16, for which the Mt.Wilson set yields lower
values. The Pulkovo data are consistent with the Kodaikanal series but
they only start with Cycle 18, so no definitive conclusion was drawn
from them. 

In their hemispherically separated analysis of the Mt.Wilson data
\cite{McClintock+Norton:MtWtilts} found that in Cycle 19 a strong suppression of
the tilt was only present in the Southern hemisphere. Accordingly, the TQ effect
is only seen in the South.

\cite{Kitchat+:tilts} analysed the Pulkovo data set, focusing on the TP effect.
Instead of considering average tilts, they evaluated the area-weighted latitude
difference between leading and trailing subgroups and averaged this quantity for
each cycle in their data set. While data were available for 3 cycles only, they
confirmed the good correlation between this predictor and a measure (the
dipole--octupole index) of the amplitude of the polar magnetic field in the next
minimum.

A fourth tilt database was compiled by \cite{Baranyi:tilts} based on the
Debrecen sunspot catalogues, while a fifth set of tilts was measured by 
\cite{Isik:tilts} from solar drawings made at Kandilli Observatory for cycles
19--24. Overall, the results for these cycles seem to compare well for the
Kodaikanal, Pulkovo, Debrecen and Kandilli data, but as cycles 15 and 16 are
only covered by the Kodaikanal data, it is no surprise that the TQ effect is not
clearly seen in the shorter Debrecen and Kandilli data sets.

All the previously considered data sets were based on sunspot positions alone,
without magnetic polarity information. Tilts of active regions taking into
account the magnetic polarities of spots were determined by \cite{Li+Ulrich}
from Mt.Wilson and MDI magnetograms. \cite{McClintock+Norton:DPD} compared these
measurements with the Debrecen tilt data, focusing on anti-Hale regions which
are the major reason for the discrepancies. The occurrence rate of anti-Hale
regions was found to be 8.5\%. Tilts of active regions taking into account the
magnetic polarities of spots were recently also determined by
\cite{Tlatova+:drawingtilts} from the archive of solar \textit{drawings} at
Mt.Wilson Observatory: these drawings include polarity information since 1917.
Their results concerning cycle dependence of tilts are inconclusive.

In summary, the suspected tilt quenching and tilt precursor effects have opened
new directions in sunspot cycle forecasting, but the evidence is still
controversial, especially for TQ. These inconclusive results underline the
importance of obtaining more data and, especially, longer data sets.
Contributions towards this goal like that of \cite{SenthamizhPavai}, who
determined tilts for cycles 7--10 from sunspot drawings by Schwabe, are
therefore highly valuable and may be key for a later clarification of these
issues. 

{\revone In view of the doubts regarding the reality of TQ, attention has
recently started to turn towards alternatives. One such alternative, proposed by J.
Jiang, might be {\it latitude quenching:} the mean latitude of active regions in
a given cycle was found to be positively correlated with cycle amplitudes
(\citealt{Jiang+:1700a}). From higher latitudes a lower fraction of leading flux
will manage to diffuse across the equator, leaving less trailing flux to
contribute to the polar fields. Therefore, the correlation found represents a
negative feedback effect ---hence the name latitude quenching.
}

{\revone

\newsubsubsection{Alternatives to tilt quenching: inflows and latitude quenching}

In view of the doubts regarding the reality of TQ, attention has
recently started to turn towards alternatives.}

The tilts of active regions are usually attributed to the effect of Coriolis
force on the rising magnetic flux loops ({\revone \citealt{Schmidt:ARdynamics}; 
 \citealt{DSilva+:ARtilts}; \citealt{Fisher+:ARtilts}; } 
\citealt{Pevtsov:twist.review};). The
tilt angle varies inversely with the initial field strength, making the
suggested tilt quenching effect quite plausible: stronger toroidal fields should
simply come with weaker tilt. (Note that the tilt also depends on the entropy of
the rising loop: a cycle-dependent variation in thermal properties of the layer
where toroidal flux is stored may therefore also explain TQ, as shown by
\citealt{Isik:Tpert}.)

An alternative explanation for tilt quenching was put forward by
\cite{Jiang_:merid.flow} and \cite{Cameron+:inflowtilt} who pointed out that the
meridional inflows directed towards the activity belts, discussed in
Sect.~\ref{sect:flows} will tend to reduce the AR tilts.

On the other hand, the activity-related meridional flow pattern may partly
originate from the superposition of more localized circular inflows towards
individual active regions (\citealt{Haber+:inflows}, \citealt{Svanda:inflows},
\citealt{Loptien+:inflows}). It is unclear whether such concentric flows can
exert a torque on the flux loops that may reduce their tilt; however this may
not even be necessary: simulations by \cite{MartinBelda+:inflows} suggested
that the hindering effect on the inflows on the separation of leading and
trailing polarity fluxes is sufficient to significantly reduce the amplitude of
the dipolar seed field being built up for the next cycle. This inflow effect
might then even completely substitute TQ as the main nonlinearity mechanism
controlling intercycle variations

{\revone Another alternative to TQ, proposed by J. Jiang, might be {\it latitude
quenching:} the mean latitude of active regions in a given cycle was found to be
positively correlated with cycle amplitudes (\citealt{Jiang+:1700a}). From
higher latitudes a lower fraction of leading flux will manage to diffuse across
the equator, leaving less trailing flux to contribute to the polar fields.
Therefore, the correlation found represents a negative feedback effect ---hence
the name latitude quenching. }



\section{Model-based predictions}
\label{sec:Model-Based-Predictions}

\newsubsection{Surface flux transport models}
\label{sec:SFT}

Surface flux transport (SFT) models describe the transport of
magnetic flux across the solar surface, modelling it as an
advective-diffusive transport:

\begin{eqnarray}
\pdv Bt &=& -\Omega(\theta)\pdv B\phi -\frac 1{\Rsun\sin\theta}
  \pdv{}\theta[\varv(\theta) B\sin\theta ] \nonumber\\
   &&+ \frac\eta{\Rsun^2\sin\theta}
  \left[ \pdv{}\theta \left(\sin\theta \pdv B\theta\right) +\frac 1{\sin\theta}
  \ppd B\phi \right] - B/\tau +S(\theta,\phi,t) 
  \label{eq:SFT}
\end{eqnarray}
where $\theta$ and $\phi$ are colatitude and longitude, $B$ is the radial
component of the magnetic flux density, $\eta$ is the turbulent magnetic
diffusivity, $\varv$ is the speed of the meridional flow, and $\Omega(\theta)$ 
is the differential rotation profile. 

Equation (\ref{eq:SFT}) can be interpreted as the radial component of the
induction equation where the neglected radial terms have been replaced by the
source term $S$ describing flux emergence and the heuristic decay term $B/\tau$,
supposed to represent vertical diffusion. The latter term is only occasionally
used, to improve agreement with observations.

SFT models were first developed in the 1980s for the interpretation of the then
newly available synoptic magnetogram record. This ``age of enlightenment'', to
use the expression of Sheeley's (\citeyear{Sheeley:LRSP}) historical review,
was followed by less intense activity in the field until, from about 2010, a
renaissance of SFT modelling ensued. The revival was prompted by the increasing
acceptance of the polar precursor as the most reliable physical precursor
technique: SFT models offered a way to ``predict the predictor'', promising to
increase the temporal scope of forecasts.

For detailed discussions of the advance made in SFT modelling we refer the
reader to the reviews by \cite{Jiang+:SFTreview} and \cite{Wang:SFTreview}.
Here we only present brief highlights of the main results from the past decade.

\subsubsection{Parameter optimization}

For an application of the method, the parameters in equation
(\ref{eq:SFT}) such as $\eta$, $\varv(\theta)$ or $\tau$ 
{\revone need to be specified}. 
This problem has reopened the issue of parameter optimization
(\citealt{Lemerle1}, \citealt{Virtanen+:SFT}, 
\citealt{Whitbread+:SFToptimization}). Choosing the right parameters
is a nontrivial task as the outcome depends on the data used for
calibration and on the choice of the merit function for the
optimization.

Recently \cite{Petrovay+Talafha} presented the results of a
large-scale systematic study of the parameter space in an SFT model
where the source term representing the net effect of tilted flux
emergence was chosen to represent a typical, average solar cycle as
described by observations, comparing the results with observational
constraints on the spatiotemporal variation of the polar magnetic
field. It was found that without a significant decay term in the SFT
equation (i.e., for $\tau >10$ yr) the global dipole moment reverses
too late in the cycle for all flow profiles and  parameters, providing
independent supporting evidence for the need of a decay term, even in
the case of identical cycles.  An allowed domain is found to exist for
$\tau$ values in the 5--10 yr range for all flow profiles considered.
Generally higher values of $\eta$ (500--800) are preferred though some
solutions with lower $\eta$ are still allowed.

\subsubsection{Nonlinear feedback effects}

Nonlinear feedback mechanisms such as TQ (\citealt{Cameron+:tiltprecursor}), a
variable modulation of the meridional flow in the form of inflow belts flanking
the active latitudes (\citealt{Jiang_:merid.flow},
\citealt{Cameron+:inflowbelts}), concentric inflows around individual AR
(\citealt{MartinBelda1}, \citeyear{MartinBelda2}) or latitude quenching
(\citealt{Jiang+:1700b}) have been considered in many
studies. With the right parameterization, feedback was found to allow magnetic
field reversals even when a stronger cycle is followed by a considerably weaker
cycle. [A difficulty in obtaining reversal in such situations was a main
motivation for the introduction of the decay term in equation
(\ref{eq:SFT}).]

It is a curious circumstance that TQ currently has stronger support from models
than from observations. Indeed, \cite{Cameron+:tiltprecursor} found that a
combination of TQ with an SFT model results in polar field strengths that
approximately correctly predict the amplitude of the next cycle for Cycles
15/16--21/22. The same holds also if the TQ results in the model as a
consequence of meridional inflow belts (\citealt{Cameron+:inflowbelts}). 

\subsubsection{Extending the polar precursor}

Sunspot observations are available for several centuries, while proxy indicators
of the polar field strength start from the late 19th century only. Using sunspot
records to reconstruct the source term in SFT models, the polar field may also
be reconstructed for cycles where no polar field proxies are known. Such a
programme was carried out by \cite{Jiang+:1700a} who reconstructed the butterfly
diagram from the sunspot number record for the period since 1700;  properties of
the butterfly ``wings'' were correlated with the amplitudes and lengths of
solar cycles based on the GPR sunspot catalogue, then these correlations were
used for reconstruction in earlier times. In a subsequent work
\citep{Jiang+:1700b}, the same authors used this butterfly diagram as a source in
an SFT model to determine polar fields. The derived polar field values were
found to correlate rather well with the amplitude of the subsequent cycle,
thereby extending the period of time for which we have evidence for the polar
precursor.

A deficiency of the source reconstruction based on sunspot catalogues
is the lack of information on magnetic polarities and on the
distribution of weaker plage fields. A promising new method based on
the use of {Ca\,{\sc ii} K} synoptic maps combined with available sunspot
magnetic field measurements was recently successfully tested by 
\cite{Virtanen+:reconstruction.method}.

\subsubsection{The importance of fluctuations: rogue active regions}

The nonlinear feedback effects discussed above define an essentially
deterministic mechanism of intercycle variations. While on the level of the
spatiotemporal distribution of individual active regions there may be numerous
different realizations of a sunspot cycle with a given sunspot number profile,
one might expect that the variation of at least the statistical average taken
over all realizations can be reliably predicted.

Starting from 2013, however, it was gradually realized that the behaviour of an
individual realization (like the real Sun) can strongly deviate from the
statistical expectations. The magnetic flux of the largest ARs in a solar cycle
is comparable to the flux in the polar caps where the polar magnetic field is
concentrated around the minimum. The amplitude of the polar field built up
during a cycle is therefore highly sensitive to the exact balance of leading vs.
trailing polarity flux transported to the poles and cancelled by cross-equator
diffusion. Major plumes on the observational magnetic butterfly diagram were
identified by \cite{Cameron+:crosseq} as originating from large 
cross-equatorial AR where cancellation between the two polarities is minimal
due to their advection in opposite directions by the meridional flow diverging
on the equator. Such cross-equatorial plumes were incorporated into an SFT model
by  \cite{Cameron+:crosseqSFT}.

The role of random scatter in active region properties was
further investigated by \cite{Jiang+:scatter} who showed that the dipole
contribution of a single active region drops quite fast with heliographic
latitudes. For those low-latitude active regions with an unusually large
contribution to the global dipole \cite{Nagy+:rogue} introduced the name
``rogue'' AR. Representing an active region as a simple bipole with tilt angle
$\alpha$ relative to the azimuthal direction, it is straightforward to show from
equation ({\ref{eq:dipmom}) that its contribution to the axial dipole is 
\begin{equation}\label{eq:thenumber}
  \delta D_{\mathrm{BMR}} = \frac 3{4\pi\Rsun^2} F d  \sin\alpha \sin\theta ,
  \label{eq:dipcontr}
\end{equation}
where $F$ is the magnetic flux and $d$ is the angular separation of the two
polarities. Large values of $F$, $d$ and/or $\alpha$ are therefore conditions
for a significant ``dynamo effectivity'' of an AR, i.e., a significant
contribution to the polar field built up in the cycle. However, equation
(\ref{eq:dipcontr}) only gives the \textit{initial} contribution to the dipole
moment. For AR further from the equator cross-equatorial cancellation of the
leading polarity will be less efficient, fluxes of both polarities will be
largely transferred to the pole and their net effect will mostly cancel: this is
the reason why the \textit{final} dipole contributions and therefore the dynamo
effectivity of AR also drops quite fast with heliographic latitude.

\subsubsection{Explaining the end of the Modern Maximum}

Beside the general investigations discussed, a special objective of SFT
modelling efforts was to correctly ``hindcast'' the unusually weak polar fields
in the minimum of Cycle~24 that brought the Modern Maximum to an end. Initial
efforts (\citealt{Yeates:cyc23sft}, \citealt{Upton+:MCvar}) encountered 
difficulties in reproducing the polar field, until \cite{Jiang+:cyc24hindcast}
were finally able to correctly reproduce the evolution of the polar field by
incorporating in their source term individual observed active regions (modelled
as idealized bipoles, but with tilt values, fluxes and separations derived from
observations).  After carefully excluding recurrent ARs from the source term
they found that the chief responsibility for the deviation of the polar flux
from its expected value lies with a low number of large low-latitude rogue AR
with non-Hale or non-Joy orientations.

In a similar research \cite{Upton+:cyc24hindcast} focused on the predictability
of the evolution of the axial dipole moment. From some selected instant onwards,
they substitute actual ARs with the ARs of another solar cycle of similar
amplitude, and they find that the dipole evolution can be well predicted over
$\sim 3$ years. It should be noted, however, that this study does not cover the
period 2003--2006 which seems to have been crucial in the development of an
anomalously low dipole moment at the end of Cycle~23. Predictability was also
considered by \cite{Whitbread+:dipcontr} who addressed the issue how many AR
need to be taken into account to reproduce the dipole moment at the end of a
cycle. 
{\revone Ordering active regions by decreasing dipole moment contribution they
found that, despite the fact that the polar magnetic flux is comparable to the
flux a single large AR, the dipole moment at the end of a cycle cannot be
reproduced without accounting for the net contribution from hundreds of active
regions during the cycle.}

The tilt of active regions is a manifestation of the writhe of the
underlying flux loop, and writhe is one form of helicity {\revone 
(cf.~\citealt{Petrovay+:helicity}), while nonzero helicity} is a
condition for free magnetic energy available for eruptions. On this
ground \cite{Petrovay+:IAUS340} tentatively suggested that there may
be a large overlap between rogue AR and flare/CME-productive AR. A
relevant study was recently undertaken by \cite{Jiang+Baranyi} who
found that flare productivity and dynamo effectivity of ARs are
governed by different parameters. Flare productivity primarily depends
on the structural complexity of ARs, large flares being much more
common $\delta$-spots, while the dipole moment contribution of an AR,
which ultimately determines its effect on the dynamo, is determined by
the latitudinal separation of polarities. So while there is
indeed a large overlap between the flare-productive ARs and ``rogue''
or exceptional ARs, the two characteristics do not necessarily go
hand in hand.

\subsubsection{Forecasting Cycle~25}

The buildup of the polar dipole moment during the ongoing Cycle~24 has been
followed with keen attention. Researchers analysed a number of important
episodes (e.g., \citealt{Petrie+Ettinger}). \cite{Yeates+vDG} examined the origin
of a prominent poleward surge in the magnetic butterfly diagram in 2010--11 by a
combination of analysis of observational data and SFT simulations, concluding
that the episode is not expected to have a major imact of the dipole buildup.
\cite{Sun+:reversal24} presented an observational analysis of the polar reversal
process in Cycle~24. This is of particular interest owing to the ill-defined
nature of polarity reversal in the N hemisphere: the field strength here
lingered around zero for well over two years until it finally started to
increase towards the end of 2014. 
{\revone (It may be worth noting that this was correctly predicted in the SFT
simulation of \citealt{Upton+:cyc24hindcast}.)} 
As a result, the phase shift between the hemispheres has changed sign: while
activity peaked first on the N hemisphere in the last few cycles, indications
are that in Cycle~25 activity will first peak in the South {\revone (cf.~also 
\citealt{Labonville+})}. The phase shift was
the consequence of a few surges of opposite polarity that hindered the growth of
flux in the N polar region.

Going further than focusing on selected events, a number of researchers
attempted to predict Cycle~25 by incorporationg all ARs of Cycle~24 in the
source term of an SFT simulation and modelling further evolution under some more
or less plausible assumptions. 

Allowing a random scatter in AR tilts and also in the time-dependent meridional
flow \cite{Hathaway+:pred25} 
{\revone and \cite{Upton+:pred25update} arrived at a prediction
similar to or marginally lower than Cycle~24.}
A similar conclusion 
{\revone (Cycle 25 somewhat weaker than Cycle 24)} 
was reached by \cite{Iijima+:plateau} who, based on the
plateau-like nature of the dipole moment maximum discussed in
Sect.~\ref{sect:polar} above, assume no further contributions to the dipole
moment. 

Considering 50 different random realizations drawn from a statistical ensemble
of ARs \cite{Cameron+:pred25} predict that Cycle~25 will be similar or slightly
stronger than Cycle~24. In a similar later study with improved technical details
\cite{Jiang+:1cycle} arrived at the prediction that Cycle~25 will peak in the
range 93 to 159 (see also \citealt{Jiang+Cao:pred25}).

\subsection{The solar dynamo: a brief summary of current models}

While attempts to predict future solar cycles on the basis of the empirical
sunspot number record have a century-old history, predictions based on physical
models of solar activity only started {\newtext one solar cycle} ago. The
background of this new trend is, however, not some significant improvement in
our understanding of the solar dynamo. Rather, it is the availability of
increasingly fast new computers that made it possible to fine-tune the
parameters of certain dynamo models to reproduce the available sunspot record to
a good degree of accuracy and to apply data assimilation methods (such as those
used in terrestrial weather prediction) to these models. This is not without
perils. On the one hand, the capability of multiparametric models to fit a
multitude of observational data does not prove the conceptual correctness of the
underlying model. On the other hand, in chaotic or stochastic systems such as
the solar dynamo, fitting a model to existing data will not lead to a good
prediction beyond a certain time span, the extent of which can only be
objectively assessed by ``postdiction'' tests, i.e., checking the models
predictive skill by trying to ``predict'' previous solar cycles and comparing
those predictions to available data. Apparently successful postdiction tests
have led some groups to claim a breakthrough in solar cycle prediction owing to
the model-based approach \citep{Dikpati_:prediction, Kitiashvili_:prediction}.
Yet, as we will see in the following discussion, a closer inspection of these
claims raises many questions regarding the role that the reliance on a
particular physical dynamo model plays in the success of their predictions.

Extensive summaries of the current standing of solar dynamo theory are given in
the reviews by \cite{Petrovay:SOLSPA}, \cite{Ossendrijver:dynamo.review},
\cite{Solanki_:RPP}, \cite{Charbonneau:LRSP2} and \cite{Cameron+:dynrev2017}. 
As explained in detail in those reviews, all current models {\newtext claiming
to acceptably represent the solar dynamo} are based on the mean-field theory
approach wherein a coupled system of partial differential equations governs the
evolution of the toroidal and poloidal components of the large-scale magnetic
field. {\newtext Until recently, the large-scale field was} assumed to be
axially symmetric in practically all models. In some nonlinear models
the averaged equation of motion, governing large-scale flows is also coupled
into the system. 

In the simplest case of homogeneous and isotropic turbulence, where  the scale
$l$ of turbulence is small compared to the scale $L$ of the mean variables
(scale separation hypothesis), the dynamo equations have the form
\begin{equation}
  \pdv{\vc B}t=\nabla\times(\vc U\times\vc B+\alpha\vc B) 
  -\nabla\times(\eta_T\times\nabla\vc B)\,.
  \label{eq:dynamo}
\end{equation}
Here $\vc B$ and $\vc U$ are the large-scale mean magnetic field and flow
speed, respectively; $\eta_T$ is the magnetic diffusivity (dominated by the
turbulent contribution for the highly conductive solar plasma), while $\alpha$
is a parameter related to the non-mirror symmetric character of the magnetized
plasma flow. 

In the case of axial symmetry the mean flow $\vc U$ may be split into a
meridional circulation $\vc U_c$ and a differential rotation characterized by
the angular velocity profile $\Omega_0(r,\theta)$:
\[ \vc U=\vc U_c + r\sin\theta\,\Omega_0\,\vc e_\phi\,,   \]
where $r$, $\theta$, $\phi$ are spherical coordinates and $\vc e_\phi$ is the
azimuthal unit vector. Now introducing the shear 
\[ {\vc\Omega}=r\sin\theta\,\nabla\Omega_0  , \qquad
\Omega=-\sgn{\dv{\Omega_0}r}\cdot|\vc\Omega| , \]
assuming $\alpha\ll\Omega L$ and ignoring spatial derivatives of $\alpha$ and
$\eta_T$, Eq.~(\ref{eq:dynamo}) simplifies to the pair
\begin{equation}
  \pdv At=\alpha B-(\vc U_c\cdot\nabla)A-(\nabla\cdot\vc U_c)A+\eta_T\,\nabla^2 A \,, 
  \label{eq:pol} 
\end{equation}
\begin{equation} \pdv Bt=\Omega\,\pdv Ax 
   -(\vc U_c\cdot\nabla)B -(\nabla\cdot\vc U_c)B
   +\eta_T\,\nabla^2 B     \label{eq:tor}  ,  
\end{equation}
where $B$ and $A$ are the toroidal (azimuthal) components of the magnetic field
and of the vector potential, respectively, and $\pdv Ax$ is to be evaluated in
the direction $90^{\circ}$ clockwards of $\vec\Omega$ (along the isorotation
surface) in the meridional plane. These are the classic $\alpha\Omega$ dynamo
equations, including a meridional flow.

In the more mainstream solar dynamo models the strong toroidal field is now
generally thought to reside near the bottom of the solar convective zone.
Indeed, it is known that a variety of flux transport mechanisms such as pumping
\citep{Petrovay:NATO} remove magnetic flux from the solar convective zone on a
timescale short compared to the solar cycle.  Following earlier simpler
numerical experiments, MHD numerical simulations have indeed demonstrated this
pumping of large scale magnetic flux from the convective zone into the
tachocline below, where it forms strong coherent toroidal fields
(\citealt{Browning_:dynsimu.pumping}, \citealt{Fan+Fang},
\citealt{Warnecke+:alphatensor}). As this layer is also where rotational shear is
maximal, it is plausible that the strong toroidal fields are not just stored but
also generated here, by the winding up of poloidal field.\footnote{\newtext Some
doubts concerning this deep-seated storage of toroidal flux recently arose as in
at least one MHD simulation (\citealt{Nelson+:wreath}) the toroidal flux was
found to be floating in a wreath-like structure in the middle of the convective
envelope.} The two main groups of dynamo models, interface dynamos and flux
transport dynamos, differ mainly in their assumptions about the site and
mechanism of the  $\alpha$-effect responsible for the generation of a new
poloidal field from the toroidal field. 

In interface dynamos $\alpha$ is assumed to be concentrated near the bottom of
the convective zone, in a region adjacent to the tachocline, so that the dynamo
operates as a wave propagating along the interface between these two layers.
While these models may be roughly consistent and convincing from the
physical point of view, they have only had limited success in reproducing the
observed characteristics of the solar cycle, such as the butterfly diagram. 

Flux transport dynamos, in contrast, rely on the Babcock--Leighton mechanism for
$\alpha$, arising due to the action of the Coriolis force on emerging flux
loops, and they assume that the corresponding $\alpha$-effect is concentrated
near the surface. They keep this surface layer \textit{incommunicado} with the
tachocline by introducing some arbitrary unphysical assumptions (such as very
low diffusivities in the bulk of the convective zone). The poloidal fields
generated by this surface $\alpha$-effect are then advected to the poles and
there down to the tachocline by the meridional circulation -- which,
accordingly, has key importance in these models. The equatorward deep return
flow of the meridional circulation is assumed to have a significant overlap with
the tachocline (another controversial point), and it keeps transporting the
toroidal field generated by the rotational shear towards the equator. By the
time it reaches lower latitudes, it is amplified sufficiently for the flux
emergence process to start, resulting in the formation of active regions and, as
a result of the Babcock--Leighton mechanism, in the reconstruction of a poloidal
field near the surface with a polarity opposed to that in the previous 11-year
cycle. While flux transport models may be questionable from the point of view of
their physical consistency, they can be readily fine-tuned to reproduce the
observed butterfly diagram quite well. {\newtext This ready adaptability made
flux transport dynamos hugely popular in the research community
(\citealt{Charb:ftdynamos}, \citealt{Karak:ftdynrev}). In flux transport dynamos
the $\alpha$ term is usually nonlocal (generating poloidal field at the surface
out of toroidal field at the bottom of the convective zone).}

It should be noted that while the terms ``interface dynamo'' and ``flux
transport dynamo'' are now very widely used to describe the two main approaches,
the more generic terms ``advection-dominated'' and ``diffusion-dominated'' would
be preferable in several respects. This classification allows for a
continuous spectrum of models depending on the numerical ratio of advective and
diffusive timescales (for communication between surface and tachocline). In
addition, even at the two extremes, classic interface dynamos and
circulation-driven dynamos are just particular examples of advection or
diffusion dominated systems with different geometrical structures.

\subsection{Is model-based cycle prediction feasible?}

As it can be seen even from the very brief and sketchy presentation given above,
all current solar dynamo models are based on a number of quite arbitrary
assumptions and depend on a number of free parameters, the functional form and
amplitude of which is far from being well constrained. For this reason, 
\cite{Bushby_Tobias} rightfully say that all current solar dynamo models are
only of ``an illustrative nature''. This would suggest that as far as solar
cycle prediction is concerned, the best we should expect from dynamo models is
also an ``illustrative'' reproduction of a series of solar cycles with the same
kind of long-term variations (qualitatively and, in the statistical sense,
quantitatively) as seen in solar data. Indeed, \cite{Bushby_Tobias} demonstrated
that even a minuscule stochastic variation in the parameters of a particular
flux transport model can lead to large, unpredictable variations in the cycle
amplitudes. And even in the absence of stochastic effects, the chaotic nature of
nonlinear dynamo solutions seriously limits the possibilities of prediction, as
the authors find in a particular interface dynamo model: even if the very same
model is used to reproduce the results of one particular run, the impossibility
of setting initial conditions exactly representing the system implies that
predictions are impossible even for the next cycle. Somewhat better results are
achieved by an alternative method, based on the phase space reconstruction of
the attractor of the nonlinear system -- this is, however, a purely empirical
time series analysis technique for which no knowledge of the detailed underlying
physics is needed. (Cf.\ Sect.~\ref{sect:nonlin} above.)

Despite these very legitimate doubts regarding the feasibility of model-based
prediction of solar cycles, in recent years several groups have claimed to be
able to predict {\newtext upcoming solar cycles} on the basis of dynamo models
with a high confidence. So let us consider these claims.

\subsection{Explicit models}
\label{sect:explicit}

\subsubsection{Axisymmetric models}
\label{sec:model_axi}

The {\newtext first attempt at} model-based solar cycle prediction was the work
of the solar dynamo group in Boulder \citep{Dikpati:forecast1st, 
Dikpati_:prediction}. Their model is a flux transport dynamo,
advection-dominated to the extreme. The strong suppression of diffusive effects
is assured by the very low value (less than 20~\sqkms) assumed for the turbulent
magnetic diffusivity in the bulk of the convective zone. As a result, the
poloidal fields generated near the surface by the Babcock--Leighton mechanism
are only transported to the tachocline on the very long, decadal time scale of
meridional circulation. The strong toroidal flux residing in the low-latitude
tachocline, producing solar activity in a given cycle is thus the product of the
shear amplification of poloidal fields formed near the surface about 2\,--\,3
solar cycles earlier, i.e., the model has a ``memory'' extending to several
cycles. The mechanism responsible for cycle-to-cycle variation is assumed to be
the stochastic nature of the flux emergence process. In order to represent this
variability realistically, the model drops the surface $\alpha$-term completely
(a separate, smaller $\alpha$ term is retained in the tachocline); instead, the
generation of poloidal field near the surface is represented by a source term,
the amplitude of which is based on the sunspot record, while its detailed
functional form remains fixed. 

\cite{Dikpati_:prediction} found that, starting off their calculation by fixing
the source term amplitudes of sunspot cycles~12 to 15, they could predict the
amplitudes of each subsequent cycle with a reasonable accuracy, provided that
the relation between the relative sunspot numbers and the toroidal flux in the
tachocline is linear, and that the observed amplitudes of all previous cycles
are incorporated in the source term for the prediction of any given cycle. For
cycle~24 the model predicted peak smoothed annual relative sunspot
numbers of 150 (v1) or more. Elaborating on their model, they proceeded to apply
it separately to the northern and southern hemispheres, to find that the model
can also be used to correctly forecast the hemispheric asymmetry of solar
activity \citep{Dikpati_:hemisph.prediction}.

Even though, ultimately, the prediction proved to be off by about a factor of 2,
the extraordinary claims of this pioneering research prompted a hot debate in
the dynamo community. Besides the more general, fundamental doubt regarding the
feasibility of model-based predictions (see Sect.~\ref{sect:spectral} above),
more technical concerns arose. In order to understand the origin of the
predictive skill of the Boulder model, \cite{Cameron_:prediction} studied a
version of the model simplified to an axially symmetric SFT model, wherein only
the equation for the radial field component is solved as a function of time and
latitude. The equation includes a source term similar to the one used in the
Boulder model. As the toroidal flux does not figure in this simple model, the
authors use the 
{\revone cross-equatorial flux $\Phi$ 
(the amount of magnetic flux crossing the equator in a given time)}
as a proxy, arguing that this may be
more closely linked to the amplitude of the toroidal field in the upcoming
cycle than the polar field. They find that $\Phi$ indeed correlates quite well
(correlation coefficients $r\sim0.8\mbox{\,--\,}0.9$, depending on model
details) with the amplitude of the next cycle, as long as the form of the
latitude dependence of the source term is prescribed and only its amplitude is
modulated with the observed sunspot number series (``idealized model''). But
surprisingly, the predictive skill of the model is completely lost if the
prescribed form of the source function is dropped and the actually observed
latitude distribution of sunspots is used instead (``realistic model'').
\cite{Cameron_:prediction} interpreted this by pointing out that $\Phi$ is mainly
determined by the amount of very low latitude flux emergence, which in turn
occurs mainly in the last few years of the cycle in the idealized model, while
it has a wider temporal distribution in the realistic model. The conclusion is
that the root of the apparently good predictive skill of the truncated model
(and, by inference, of the Boulder model it is purported to represent) is
actually just the good empirical correlation between late-phase activity and the
amplitude of the next cycle, discussed in Sect.~\ref{sect:minimax} above. This
correlation is implicitly ``imported'' into the idealized flux transport model
by assuming that the late-phase activity is concentrated at low latitudes, and
therefore gives rise to cross-equatorial flux which then serves as a seed for
the toroidal field in the next cycle. So if \cite{Cameron_:prediction} are
correct, the predictive skill of the Boulder model is due to an empirical
precursor and is thus ultimately explained by the good old Waldmeier effect (cf.
Sect.~\ref{sect:Waldmeier}) {\newtext In view of the fact that no version of
the Boulder model  with a modified source function incorporating the realistic
latitudinal distribution of sunspots in each cycle was ever presented, this
explanation seems to be correct, despite the fact that the effective diffusivity
represented by the sink term in the reduced model is 
significantly higher than in the Boulder model, and consequently, the reduced
model will have a more limited memory, cf.\ \cite{Yeates_:prediction}.}

Another flux transport dynamo code, the Surya code, originally developed by A.~
Choudhuri and coworkers in Bangalore, has also been utilized for prediction
purposes. The crucial difference between the two models is in the value of the
turbulent diffusivity assumed in the convective zone: in the Bangalore model
this value is 240~\sqkms, 1\,--\,2 orders of magnitude higher than in the
Boulder model, and within the physically plausible range
\citep{Chatterjee_:model}. As a result of the shorter diffusive timescale, the
model has a shorter memory, not exceeding one solar cycle. As a consequence of
this relatively rapid diffusive communication between surface and tachocline,
the poloidal fields forming near the surface at low latitudes due to the
Babcock--Leighton mechanism diffuse down to the tachocline in about the same
time as they reach the poles due to the advection by the meridional circulation.
In these models, then, polar magnetic fields are not a true physical precursor
of the low-latitude toroidal flux, and their correlation is just due to their
common source. In the version of the code adapted for cycle prediction
\citep{Choudhuri_:prediction, Jiang_:prediction}, the ``surface'' poloidal field
(i.e., the poloidal field throughout the outer half of the convection zone) is
rescaled at each minimum by a factor reflecting the observed amplitude of the
Sun's dipole field. The model showed reasonable predictive skill for the last
three cycles for which data are available, and could even tackle hemispheric
asymmetry \citep{Goel_:hemispheric}. For Cycle~24, the predicted amplitude was
30\,--\,35\% lower than Cycle~23.

{\newtext

In retrospect, these first attempts at model-based solar cycle prediction are
now generally seen as precursor methods in disguise. \cite{Cameron_:prediction}
convincingly argued that the apparent predictive (postdictive) skill of the
Boulder model was related to the ``minimax'' family of internal precursors,
based on the Waldmeier law combined with the overlap of consecutive cycles (see
Sect.~\ref{sect:minimax}). The predictive skill of the Surya model, on the other
hand, is based on the polar precursor: essentially, the polar field amplitude at
minimum is set to its observed value, and the model simply mechanically  winds
up the poloidal field into a toroidal field, ensuring a proportionality.

A more sophisticated approach to dynamo based predictions was spurred by the
realization of the importance of nonlinearities like TQ and stochastic effects
in SFT models . In particular, the significance of the effect of individual
active regions in the buildup of poloidal fields gave an impetus to the
development of non-axisymmetric models where individual AR can be treated.
Nevertheless, \cite{Kitchat+:stochBL} recently constructed an axisymmetric flux
transport dynamo where stochastic effects were still retained as fluctuations of
the $\alpha$ parameter. It was found that with a correlation time on the order
of a month the model is able to mimick the effects of rogue AR such as
intercycle variations and even the triggering of grand minima, as in the model
of \cite{Nagy+:rogue}.

}

\newsubsubsection{Nonaxisymmetric models}
\label{sec:model_nonaxi}

The first dynamo model to explicitly deal with individual AR was constructed by
\cite{Yeates+:ARdyn}. Emerging flux loops in this model were created by imposed
upflows with properties calibrated to reproduce observed AR characteristics.
Another 3D code capable of dealing with individual AR, the STABLE code was
developed in Boulder (\citealt{Miesch+Dikpati:STABLE}, \citealt{Miesch+Tewelde}). 

The third such code is the {2\,$\times$\,2D} code of the Montreal group
(\citealt{Lemerle1}, \citealt{Lemerle2}). This latter code ingenuously combines
an axially symmetric flux transport dynamo code with a 2D SFT code. The dynamo
code couples to the SFT code by an emergence function specifying the locations
and properties of the randomly created  bipolar source regions for the SFT,
based on the distribution of the toroidal field. The azimuthally averaged
magnetic field resulting from the SFT component, in turn, provides the upper
boundary condition for the flux transport model. The model is computationally
highly efficient, optimizable, and it was carefully calibrated to reproduce the 
observed statistics of active regions in Cycle~21. It can be run for hundreds or
thousands of cycles, thereby providing a data base for analysis that far exceeds
the observational record of solar cycles. While the model's behaviour also
displays some differences relative to the real Sun (esp. in the phase
relationship of the poloidal and toroidal fields), intercycle variations
comparable to those seen in the real Sun are displayed by the solutions,
including Dalton-like minima and grand minima (\citealt{Lemerle2},
\citealt{Nagy+:rogue}).

Investigations in close analogy with this were also undertaken using the STABLE
model. \cite{Karak+Miesch:3DBL} considered the effect of tilt quenching and
tilt scatter, and found that the scatter can be a main factor behind long-term
activity variations, while even a subtle tilt quenching effect is sufficient to
limit the growth of the dynamo. \cite{Karak+Miesch:grandminima} found that
fluctuations in AR tilt can be responsible for both the onset of and recovery
from grand minima. While these results are in general agreement with those
obtained with other codes, in another work based on the STABLE model
\cite{Hazra+:weirdpaper} reported that the dynamo effectivity of individual ARs
increases with heliographic latitude, in contradiction to other results and to
expectations based on the importance of cross-equatorial diffusion for the
removal of leading polarity flux. It may be that diffusion of weak magnetic flux
through the surface might be responsible for these results --- an effect that, by
construction, is avoided in the {2\,$\times$\,2D} model and which can also be
suppressed by introducing downwards pumping of the magnetic field
(\citealt{Karak+Cameron:pumping}).

\begin{figure}[htbp]
   
\centerline{\includegraphics[width=\textwidth]{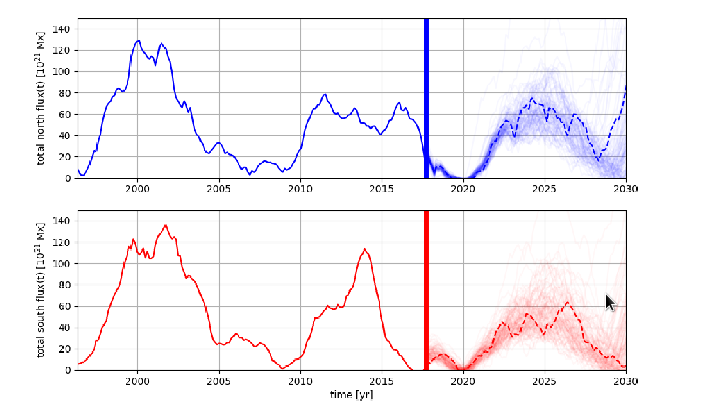}}
    \caption{Model ensemble predictions from the {2\,$\times$\,2D} dynamo model,
    driven by observational data up to November 2017. 
    {\revone The time variation of the total magnetic flux in the Northern (top)
    and Southern (bottom) hemisphere are shown in units of $10^{21}\,$Mx.}
    (Courtesy of P.~Charbonneau).}
    \label{fig:pred_charb2019}
\end{figure}

These nonaxisymmetric dynamo models have now reached a level where actual
physics-based prediction is within reach. A new generation of  dynamo-based
cycle predictions is heralded by the work of {\revone \cite{Labonville+}} who
used data assimilation in the {2\,$\times$\,2D} to arrive to a state closely
mimicking the current state of the solar dynamo, then run the code further to
see its future development. Repeating these runs with many different random
realizations of the AR emergences the authors generated  an ensemble of future
solar dynamo models (Fig.~\ref{fig:pred_charb2019}).
{\revone This enabled them to give
both a mean forecast and to attribute meaningful errors to their forecast.
They predicted a peak $R$ value of $89^{+29}_{-14}$ for cycle 25, the maximum
occurring in $2025.3^{+0.89}_{-1.05}$. The forecast is also broken down to
hemispheres: activity in the North is predicted to peak 6 months later but at a
20\,\% higher level than in the South.}

\subsection{Truncated models}
\label{sect:truncated}

The ``illustrative'' nature of current solar dynamo models is nowhere more
clearly on display than in truncated or reduced models where some or all of the
detailed spatial structure of the system is completely disregarded, and only
temporal variations are explicitly considered. This is sometimes rationalised as
a truncation or spatial integration of the equations of a more realistic
inhomogenous system; in other cases, no such rationalisation is provided,
representing the solar dynamo by an infinite, homogeneous or periodic turbulent
medium where the amplitude of the periodic large-scale magnetic field varies
with time only.

In the present subsection we deal with models that do keep one spatial variable
(typically, the latitude), so growing wave solutions are still possible
--- these models, then, are still dynamos even though their spatial
structure is not in a good correspondence with that of the solar dynamo.

This approach in fact goes back to the classic migratory dynamo model of
\cite{Parker:migr.dynamo} who radially truncated (i.e., integrated) his
equations to simplify the problem. Parker seems to have been the first to
employ a heuristic relaxation term of the form $-B_r/\tau_d$ in the poloidal
field equation to represent the effect of radial diffusion; here,
$\tau_d=d^2/\eta_T$ is the diffusive timescale across the thickness $d$ of the
convective zone. His model  was generalized by
\cite{Moss_:stochastic.dynamo1} and \cite{Moss_:stochastic.dynamo2} to the case
when the $\alpha$-effect includes an additive stochastic noise, and nonlinear
saturation of the dynamo is achieved by $\alpha$-quenching. These authors do
not make an attempt to predict solar activity with their model but they can
reasonably well reproduce some features of the very long term solar activity
record, as seen from cosmogenic isotope studies.

{\newtext 
The other classic reduced dynamo model is that of \cite{Leighton69}, the first
mathematical formulation of the flux transport dynamo concept. An updated
version of this dynamo, recently developed by \cite{Cameron+:updatedLeighton} is
a very promising, simple and versatile tool to reproduce many of the observed
features of the solar dynamo with simple parameterizations. While it has not
yet been utilized for cycle prediction, it has already proved to be useful in
understanding long-term activity variations such as the dominant periods in
hemispheric asymmetry (\citealt{Schussler+:asymmetry}).
}

Another radially truncated model, this time formulated in a Cartesian system, is
that of \cite{Kitiashvili_:nonlin.dynamo}. In this model stochastic effects are
not considered and, in addition to using an $\alpha$-quenching recipe, further
nonlinearity is introduced by coupling in the Kleeorin--Ruzmaikin equation
\citep{Zeldovich_:mg.book} governing the evolution of magnetic helicity, which
in the hydromagnetic case contributes to $\alpha$. Converting the toroidal field
strength to relative sunspot number using the Bracewell transform,
Eq.~(\ref{eq:Bracewell}), the solutions reproduce the asymmetric profile of the
sunspot number cycle. For sufficiently high dynamo numbers the solutions become
chaotic, cycle amplitudes show an irregular variation. Cycle amplitudes and
minimum--maximum time delays are found to be related in a way reminiscent of the
Waldmeier relation. 

Building on these results, \cite{Kitiashvili_:prediction} attempted to predict
solar cycles using a data assimilation method. The approach used was the
so-called Ensemble Kalman Filter method. Applying the model for a
``postdiction'' of the last 8 solar cycles yielded astonishingly good results,
considering the truncated and arbitrary nature of the model and the fundamental
obstacles in the way of reliable prediction discussed above.  The question may
arise whether the actual physics of the model considered has any significant
role in this prediction, or we are dealing with something like the phase space
reconstruction approach discussed in Sect.~\ref{sect:nonlin} where
basically any model with an attractor that looks reasonably similar to that of
the actual solar dynamo would do. Either way, the method is remarkable.
{\newtext Its prediction for cycle~24 essentially proved to be correct. An early
prediction for Cycle~25 (\citealt{Kitiashvili:cyc25pred}) yielded  a maximum
occurring in 2023/24 at a level $R=90\pm15$ (v2 values), with the cycle starting
in 2019/20. It is noteworthy that a forecast considering observational data up
to the previous minimum (2008) already yielded quite similar results for the
amplitude, although the timing of the maximum would have been expected earlier.
Understanding the origin of this impressive apparent predictive skill certainly
deserves more in-depth research.}

\subsection{The Sun as an oscillator}
\label{sect:oscillator}

An even more radical simplification of the solar dynamo problem ignores any
spatial dependence in the solutions completely, concentrating on the time
dependence only. Spatial derivatives appearing in Eqs.~(\ref{eq:pol}) and
(\ref{eq:tor}) are estimated as $\nabla\sim 1/L$ and the resulting terms $U_c/L$
and $\eta_T/L^2$ as $1/\tau$ where $\tau$ is a characteristic time scale. This
results in the pair 
\begin{equation} 
  \dot A=\alpha B -A/\tau \,,
  \label{eq:A2} 
\end{equation}
\begin{equation} 
  \dot B=({\Omega}/L) A -B/\tau \,,
  \label{eq:B2} 
\end{equation} 
which can be combined to yield 
\begin{equation}
  \ddot B=\frac{D-1}{\tau^2} B -\frac 2\tau\dot B \,,
  \label{eq:sunosc}
\end{equation} 
where $D=\alpha\Omega\tau^2/L$ is the dynamo number. For $D<1$,
Eq.~(\ref{eq:sunosc}) clearly describes a damped linear
oscillator. For $D>1$, solutions have a non-oscillatory character. The
system described by Eq.~(\ref{eq:sunosc}), then, is not only not
a true dynamo (missing the spatial dependence) but it does not even
display growing oscillatory solutions that would be the closest
counterpart of dynamo-like behaviour in such a system. Nevertheless,
there are a number of ways to extend the oscillator model to allow for
persistent oscillatory solutions, i.e., to turn it into a
\textit{relaxation oscillator}:

(1) The most straightforward approach is to add a \textit{forcing term}
$+\sin(\omega_0t)$ to the r.h.s.\ of Equations~(\ref{eq:sunosc}). Damping would cause
the system to relax to the driving period $2\pi/\omega_0$ if there were no
stochastic disturbances to this equilibrium. \cite{Hiremath:oscillator}
fitted the parameters of the forced and damped oscillator model to each observed
solar cycle individually; then in a later work \citep{Hiremath} he applied
linear regression to the resulting series to provide a forecast (see
Sect.~\ref{sect:regression} above).

(2) Another trick is to account for the $\pi/2$ phase difference between
poloidal and toroidal field components in a dynamo wave by introducing a \textit{phase factor} $i$ into the first term on the r.h.s.\ of Eq.~(\ref{eq:B2}).
This can also be given a more formal derivation as equations of this form result
from the substitution of solutions of the form $A\propto e^{ikx}$, $B\propto
e^{i(kx+\pi/2)}$ into the 1D dynamo equations. This route, combined with a
nonlinearity due to magnetic modulation of differential rotation described by a
coupled third equation, was taken by \cite{Weiss_:chaotic.dynamo}. Their model
displayed chaotic behaviour with intermittent episodes of low activity similar
to grand minima. 
{\newtext
The generic normal form model recently introduced  by 
\cite{Cameron+:normalform} is also loosely related to this approach.
}

(3) \cite{WilmotSmith_} showed that another case where dynamo-like behaviour can
be found in an equation like~(\ref{eq:sunosc}) is if the missing effects of
finite communication time between parts of a spatially extended system are
reintroduced by using a \textit{time delay} $\Delta t$, evaluating the first
term on the r.h.s.\ at time $t-\Delta t$ to get the value for the l.h.s.\ at
time $t$.
{\newtext
This time-delay approach has been further developed by \cite{Hazra+:timedelay}
and \cite{Turner+Ladde}.
}

(4) Yet another possibility is to introduce a \textit{nonlinearity} into the model
by assuming $D=D_0[1-f(B)]$ where $f(B=0)=0$ and $f\geq 0$ everywhere. (Note
that any arbitrary form of $\alpha$- or $\Omega$-quenching can be cast in the
above form by series expansion.) The governing equation then becomes one of a
nonlinear oscillator: 
\begin{equation} 
\ddot B=\frac{D_0-1}{\tau^2} B -\frac 2\tau\dot B-\frac{D_0-1}{\tau^2} B f(B) \,. 
  \label{eq:sunosc2}
\end{equation} 

In the most commonly assumed quenching mechanisms the leading term in $f(B)$ is
quadratic; in this case Eq.~(\ref{eq:sunosc2}) describes a \textit{Duffing
oscillator} \citep{Kanamaru:Duffing}. For large positive dynamo numbers,
$D_0\gg 1$, then, the large nonlinear term dominates for high values of $B$, its
negative sign imposing oscillatory behaviour; yet the origin is a repeller so
the oscillation will never be damped out.  The Duffing oscillator was first
considered in the solar context by \cite{Palus_Novotna}. 

{\newtext 
More generally, a complete spatial truncation of the nonlinear $\alpha\Omega$
dynamo equations with a dimensional analysis of some of the terms (see review by
\citealt{Lopes+:oscillrev}) can be shown  to give rise to a nonlinear oscillator
equation of the form
\begin{equation}
\label{nemlinoegy}
\ddot B = -\omega^2 B - \mu(\xi B^2-1)\dot{B} -\gamma B^3
\end{equation}
where $B$ is the amplitude of the toroidal magnetic field, and the
parameters $\mu$, $\xi$ and $\gamma$ may be expressed by the dynamo
parameters (dynamo number, meridional flow amplitude, nonlinearity
parameters). This is a combination of the
van der Pol and Duffing oscillators, the two most widely studied
nonlinear oscillator problems.
}
Under certain
conditions on the parameters, it is reduced to a \textit{van der Pol
oscillator} \citep{Adomian, Mininni+:vanderpol1, Kanamaru:vdPol}:
\begin{equation} 
  \ddot \xi= -\xi +\mu(1-\xi^2)\dot\xi \,,
  \label{eq:vanderPol}
\end{equation} 
with $\mu>0$. From this form it is evident that the problem is equivalent to
that of an oscillator with a damping that increases with amplitude; in fact, for
small amplitudes the damping is negative, i.e., the oscillation is self-excited.

These simple nonlinear oscillators were among the first physical systems where
chaotic behaviour was detected (when a periodic forcing was added). Yet,
curiously, they first emerged in the solar context precisely as an alternative
to chaotic behaviour. Considering the mapping of the solar cycle in the
differential phase space $\{B, dB/dt\}$, \cite{Mininni_:vanderpol} got the
impression that, rather than showing signs of a strange attractor, the SSN
series is adequately modelled by a van der Pol oscillator with stochastic
fluctuations.  This concept was further developed by \cite{Lopes_Passos:dalton}
who fitted the parameters of the oscillator to each individual sunspot cycle.
{\newtext
Subsequently, this parameter fitting was also exploited for cycle prediction
purposes (\citealt{Passos:oscillpred}).
}
The parameter $\mu$ is related to the meridional flow speed and the fit
indicates that a slower meridional flow may have been responsible for the Dalton
minimum. This was also corroborated in an explicit dynamo model (the Surya code)
--- however, as we discussed in Sect.~\ref{sect:flows}, this result of flux
transport dynamo models is spurious and the actual effect of a slower meridional
flow is likely to be opposite to that suggested by the van der Pol oscillator
model.

In an alternative approach to the problem, \cite{Nagovitsyn:reconstr} attempted
to constrain the properties of the solar oscillator from its
amplitude--frequency diagram, suggesting a Duffing oscillator driven at two
secular periods. While his empirical reconstruction of the amplitude--frequency
plot may be subject to many uncertainties, the basic idea is certainly
noteworthy.

{\newtext
A further twist to the oscillator representation of the solar cycle is to
consider a system of two coupled nonlinear oscillators (Kuramoto model). This
approach has been taken in a series of papers by Blanter et al.
(\citeyear{Blanter1}, \citeyear{Blanter2}). The two variables were taken to
represent the sunspot numbers and the geomagnetic $aa$ index, considered to be
proxies for the toriodal and poloidal field amplitudes, respectively. In a later
work (\citealt{Blanter3}) a coupled oscillator model was employed to account for the
time variation of hemispheric asymmetries of solar activity, as known from
observations (\citealt{Norton+Gallagher}; \citealt{Norton+:asym.rev}).
}

In summary: despite its simplicity, the oscillator representation of the solar
cycle is a relatively new development in dynamo theory, and its obvious
potential for forecasting purposes {\newtext is yet to be fully} exploited.



\section{Extrapolation methods}
\label{sec:Extrapolation-Methods}

In contrast to precursor methods, extrapolation methods only use the time series
of sunspot numbers (or whichever solar activity indicator is considered) but
they generally rely on more than one previous point to identify trends that can
be used to extrapolate the data into the future. They are therefore also known
as \textit{time series analysis} or, for historic reasons, \textit{regression
methods}.

A cornerstone of time series analysis is the assumption that the time series is
\textit{homogeneous}, i.e., the mathematical regularities underlying its
variations are the same at any point of time. This implies that a forecast for,
say, three years ahead has equal chance of success in the rising or decaying
phase of the sunspot cycle, across the maximum or, in particular, across the
minimum. In this case, distinguishing intracycle and intercycle memory effects,
as we did in Sects.~\ref{sect:memory} and \ref{sect:precursor}, would be
meaningless. This concept of solar activity variations as a continuous process
stands in contrast to that underlying precursor methods, where solar cycles are
thought of as individual units lasting essentially from minimum to minimum,
correlations within a cycle being considerably stronger than from one cycle to
the next. While, as we have seen, there is significant empirical evidence for
the latter view, the possibility of time homogeneity cannot be discarded out of
hand. Firstly, if we consider the time series of global parameters (e.g.,
amplitudes) of cycles, homogeneity may indeed be assumed fairly safely. This
approach has rarely been used for the directly observed solar cycles as their
number is probably too low for meaningful inferences --- but the long data sets
from cosmogenic radionuclides are excellent candidates for time series analysis.

In addition, there may be good reasons to consider the option of homogeneity of
solar activity data even on the scale of the solar cycle.  Indeed, in dynamo
models the solar magnetic field simply oscillates between (weak) poloidal and
(strong) toroidal configuration: there is nothing inherently special about
either of the two, i.e., there is no \textit{a priori} reason to attribute a
special significance to solar minimum. While at first glance the butterfly
diagram suggests that starting a new cycle at the minimum is the only meaningful
way to do it, there may be equally good arguments for starting a new cycle at
the time of polar reversal. 
{\newtext
And even though SFT and dynamo models strongly suggest that spatial information
regarding, e.g., the latitudinal distributions of sunspots may well be essential
for cycle prediction, some studies point to a possibility to reconstruct this
spatial information from time series alone (\citealt{Jiang+:1700a};
\citealt{Mandal+:bfly.reconstr}).
}
There is, therefore, plenty of motivation to try and
apply standard methods of time series analysis to sunspot data. 

Indeed, as the sunspot number series is a uniquely homogeneous and long data
set, collected over centuries and generated in {\newtext what has long been
perceived to be} a fairly carefully controlled
manner, it has become a favorite testbed of time series analysis methods and is
routinely used in textbooks and monographs for illustration purposes
\citep{Yule, Box_Jenkins, Wei, Tong}. This section will summarize
the various approaches, proceeding, by and large, from the simplest towards the
most complex.

\subsection{Linear regression}
\label{sect:regression}

Linear (auto)regression means representing the value of a time series
at time $t$ by a linear combination of values at times $t-\Delta t$, $t-2\Delta
t$, $\dots$, $t-p\Delta t$.  Admitting some random error $\epsilon_n$, the value
of \R in point $n$ is
\[
R_n={R_0}+\sum_{i=1}^p c_{n-i} R_{n-i} +\epsilon_n\,,
\]
where $p$ is the order of the autoregression and the $c_i$'s are weight
parameters. A further twist on the model admits a propagation of errors from the
previous $q$ points:
\[
R_n={R_0}+\sum_{i=1}^p c_{n-i} R_{n-i} +\epsilon_n +\sum_{i=1}^q
d_{n-i}\epsilon_{n-i}\,.
\]
This is known as the ARMA (AutoRegressive Moving Average) model.

Linear regression techniques have been widely used for solar activity prediction
during the course of an ongoing cycle. Their application for cycle-to-cycle
prediction has been less common and successful \citep{Lomb_Andersen,
  Box_Jenkins, Wei}.

\cite{Brajsa_:gleissbg} applied an ARMA model to the series of annual values
of \R. A successful fit was found for $p=6$, $q=6$. Using this fit, the next
solar maximum was predicted to take place around 2012.0 with an amplitude 
$90 \pm 27$, and the following minimum occurring in 2017.

Instead of applying an autoregression model directly to SSN data,
\cite{Hiremath} applied it to a forced and damped harmonic oscillator model
claimed to well represent the SSN series. This resulted in a predicted amplitude
of $110 \pm 10$ for Solar Cycle~24, with the cycle starting in mid-2008 and
lasting 9.34 years.


\subsection{Spectral methods}
\label{sect:spectral}

\begin{nquote}
{\sl ``\dots the use of any mathematical algorithm to derive hidden periodicities
from the data always entails the question as to whether the resulting cycles are
not introduced either by the particular numerical method used or by the time
interval analyzed.''}\\
\strut\hfill\citep{deMeyer:impulse}
\end{nquote}

Spectral analysis of the sunspot number record is used for prediction under the
assumption that the main reason of variability in the solar cycle is a long-term
modulation due to one or more periods.  

The usual approach to the problem is the purely formal one of representing the
sunspot record with the superposition of eigenfunctions forming an orthogonal
basis. From a technical point of view, spectral methods are a complicated form
of linear regression. The analysis can be performed by any of the widely used
means of harmonic analysis:

(1) {Least squares (LS) frequency analysis} (sometimes called ``Lomb--Scargle
periodogram'') consists in finding by trial and error the best fitting sine
curve to the data using the least squares method, subtracting it
(``prewhitening''), then repeating the procedure until the residuals become
indistinguishable from white noise. The first serious attempt at sunspot cycle
prediction, due to \cite{Kimura}, belonged to this group. The analysis resulted
in a large number of peaks with dubious physical significance. The prediction
given for the upcoming cycle~15 failed, the forecasted amplitude being $\sim$~60
while the cycle actually peaked at 105 {\revone (SSN v1 values)}. However, it is
interesting to note that Kimura correctly predicted the long term strengthening
of solar activity during the first half of the 20th century! LS frequency
analysis on sunspot data was also performed by \cite{Lomb_Andersen}, with
similar results for the spectrum.

(2) \textit{Fourier analysis} is probably the most commonly used method of spectral
decomposition in science. It has been applied to sunspot data from the beginning
of the 20th century \citep{Turner1, Turner2, Michelson}.
\cite{Vitinsky:book2} judges Fourier-based forecasts even less reliable than LS
periodogram methods. Indeed, for instance \cite{Cole} predicted cycle~21 to
have a peak amplitude of 60, while the real value proved to be nearly twice
that. 

(3) The \textit{maximum entropy method (MEM)} relies on the Wiener--Khinchin
theorem that the power spectrum is the Fourier transform of the autocorrelation
function. Calculating the autocorrelation of a time series for $M\ll N$ points
and extrapolating it further in time in a particular way to ensure maximal
entropy can yield a spectrum that extends to arbitrarily low frequencies despite
the shortness of the data segment considered, and also has the property of being
able to reproduce sharp spectral features (if such are present in the data in
the first place). A good description of the method is given by \cite{Ables},
accompanied with some propaganda for it --- see \cite{numrec} for a more
balanced account of its pros and cons. The use of MEM for sunspot number
prediction was pioneered by \cite{Currie:max.entropy}.  Using maximum entropy
method combined with multiple regression analysis (MRA) to estimate the
amplitudes and phases, \cite{Kane:MEM24} arrived at a prediction of 80 to 101
for the maximum amplitude of cycle~24 {\revone (v1 values)}. It should be noted
that the same method yielded a prediction \citep{Kane:MEM23} for cycle~23 that
was far off the mark.

(4) \textit{Singular spectrum analysis (SSA)} is a relatively novel method for the
orthogonal decomposition of a time series. While in the methods discussed above
the base was fixed (the trigonometric functions), SSA allows for the
identification of a set of othogonal eigenfunctions that are most suitable for
the problem. This is done by a principal component analysis of the covariance
matrix $r_{ik}=\langle R_i R_{i+k}\rangle$. SSA was first applied to the sunspot
record by \cite{Rangarajan:SSA} who only used this method for pre-filtering
before the application of MEM. \cite{Loskutov:SSA} who also give a good
description of the method, already made a prediction for cycle~24: a peak
amplitude of 117 {\revone (v1 value)}.The forecast was later corrected slightly
downwards to 106 \citep{Kuzanyan:predict.poster}.

The dismal performance of spectral predictions with the methods (1)\,--\,(3)
indicates that the sunpot number series cannot be  well represented by the
superposition of a limited number of fixed periodic components. Instead,

\begin{itemize}
\item the periods may be time dependent,
\item the system may be quasiperiodic, with a significant finite width of the
periodic peaks (esp.\ the 11-year peak),
\item there may be non-periodic (i.e., chaotic or stochastic) components in the
behaviour of the system, manifested as a continuous background in the spectrum.
\end{itemize}

In practice, all three effects suggested above may play some part. The first
mentioned effect, time dependence, may in fact be studied within the framework
of spectral analysis. MEM and SSA are intrinsically capable of detecting or
representing time dependence in the spectrum, while LS and Fourier analysis can
study time dependence by sliding an appropriate data window across the period
covered by observations. If the window is Gaussian with a width proportional to
the frequency we arrive at the popular \textit{wavelet analysis}.  This method
was applied to the sunspot number series by \cite{Ochadlick_},
\cite{Vigouroux_}, \cite{Frick_:wavelet}, \cite{Fligge_}, and \cite{Li_:wavelet}
who could confirm the existence and slight variation of the 11-year cycle
and the Gleissberg-cycle. Recently, \cite{Kollath_Olah:1} called attention to a
variety of other generalized time dependent spectral analysis methods, of which
the pseudo-Wigner transform yields especially clear details
(see Fig.~\ref{fig:Kollath}). The time varying character of the basic periods
makes it difficult to use these results for prediction purposes but they are
able to shed some light on the variation as well as the presistent or
intermittent nature of the periods determining solar activity. 

In summary, it is fair to say that forecasts based on harmonic analysis are
notoriously unreliable.  The secular variation of the basic periods, obeying as
yet unknown rules, would render harmonic analysis practically useless for the
prediction of solar cycles even if solar activity could indeed be described by a
superposition of periodic functions. Although they may be potentially useful for
very long term prediction (on centennial scales)\footnote{But even here care is
needed not to read more into the data than they contain, as discussed by 
\cite{Usoskin:debate}}, when it comes to
cycle-to-cycle forecasts the best we can hope from spectral studies is
apparently an indirect contribution, by constraining dynamo models with the
inambiguously detected periodicities. 

In what remains from this subsection, we briefly review what these apparently
physically real periods are and what impact they may have on solar cycle
prediction.

\subsubsection{The 11-year cycle and its harmonics}

As an example of the period spectrum obtained by these methods, in
Fig.~\ref{fig:ssn_powsp0} we present the FFT based power spectrum
estimate of the smoothed sunspot number record. Three main features
are immediately noticed:

\begin{itemize}
\item The dominant 11-year peak, with its sidelobes and its
5.5-year harmonic. 
\item The 22-year subharmonic, representing the even--odd rule. 
\item The significant power present at periods longer than 50~years,
associated with the Gleissberg cycle.
\end{itemize}

The dominant peak in the power spectrum is at $\sim$~11~years. Significant power
is also present at the first harmonic of this period, at 5.5~years. This is
hardly surprising as the sunspot number cycles, as presented in
Fig.~\ref{fig:SSNrecord}, have a markedly asymmetrical profile. It is a
characteristic of Fourier decomposition that in any periodic series of cycles
where the profiles of individual cycles are non-sinusoidal, all harmonics of the
base period will appear in the spectrum. 

  \begin{figure}[htbp]
\centerline{\includegraphics[width=0.9\textwidth]{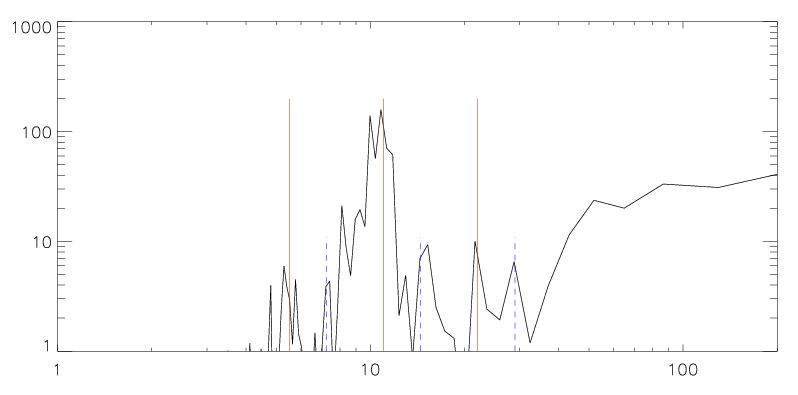}}
\caption{Power spectrum of the smoothed monthly sunspot number series for the
period 1749\,--\,2008. Solid vertical bars mark the 11-year period, its first
harmonic and subharmonic; dashed vertical bars are drawn at a fiducial period of
14.5 years, its harmonic and subharmonic.}
\label{fig:ssn_powsp0}
\end{figure}

Indeed, were it not for the 13-month smoothing, higher harmonics could also be
expected to appear in the power spectrum. It has been proposed
\citep{Krivova_Solanki:1.3yr} that these harmonics are detected in the sunspot
record and that they may be related to the periodicities of $\sim$~1.3~years
intermittently observed in solar wind speed \citep{Richardson_:1.3yr,
  Paularena_:1.3yr, Szabo_:1.3yr, Mursula_Zieger:1.3yr,
  Lockwood:1.3yr} and in the internal rotation velocity of the Sun
\citep[][Sect.~10.1]{Howe:LRSP}. An analoguous intermittent 2.5 year
variation in the solar neutrino flux \citep{Shirai} may also belong to
this group of phenomena. It may be worth noting that, from the other
end of the period spectrum, the 154-day Rieger period in solar flare
occurrence \citep{Rieger_, Bai_Cliver} has also been tentatively
linked to the 1.3-year periodicity. Unusually strong excitation of
such high harmonics of the Schwabe cycle may possibly be explained by
excitation due to unstable Rossby waves in the tachocline
\citep{Zaqarashvili_}.

The 11-year peak in the power spectrum has substantial width, related to the
rather wide variation in cycle lengths in the range 9\,--\,13 years. Yet
Fig.~\ref{fig:ssn_powsp0} seems to suggest the presence of a well detached
second peak in the spectrum at a period of $\sim$~14 years. The presence of a
distinct peak at the first harmonic and even at the subharmonic of this period
seems to support its reality. Indeed, peaks at around 14 and 7 years were
already found by other researchers \citep[e.g.,][]{Kimura,
  Currie:max.entropy} who suggested that these may be real secondary
periods of sunspot activity. 

The situation is, however, more prosaic. Constraining the time interval
considered to data more recent than 1850, from which time the sunspot number
series is considered to be more reliable, the 14.5-year secondary peak and its
harmonics completely disappear. On the other hand, the power spectrum for the
years 1783\,--\,1835 indicates that the appearance of the 14.5-year secondary peak
in the complete series is almost entirely due to the strong predominance of this
period (and its harmonic) in that interval. This interval covers the unusually
long cycle~4 and the Dalton minimum, consisting of three consecutive unusually
weak cycles, when the ``normal'' 11-year mode of operation was completely
suppressed.

As pointed out by \cite{Py:Rio}, this probably does not imply that the Sun was
operating in a different mode during the Dalton minimum, the cycle length being
14.5~years instead of the usual 11 years. Instead, the effect may be explained
by the well known inverse correlation between cycle length and amplitude, which
in turn is the consequence of the strong inverse correlation between rise rate
and cycle amplitude (Waldmeier effect), combined with a much weaker or
nonexistent correlation between decay rate and amplitude (see
Sect.~\ref{sect:Waldmeier}). The cycles around the
Dalton minimum, then, seem to lie at the low amplitude (or long period) end of a
continuum representing the well known cycle length--amplitude relation,
ultimately explained by the Waldmeier effect. 

  \begin{figure}[htbp]
\centerline{\includegraphics[width=0.6\textwidth]{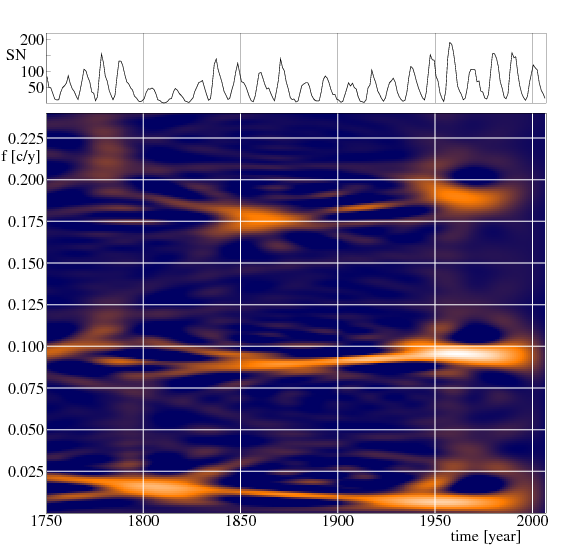}}
\caption{Pseudo-Wigner power distribution in the sunspot number record, with
time on the abscissa and frequency on the ordinate. The three horizontal bands
of high power correspond, from bottom to top, to the Gleissberg cycle, the
11-year cycle and its first harmonic. The sunspot number curve is shown on top
for guidance (figure courtesy of Z.~Koll\'ath).}
\label{fig:Kollath}
\end{figure}

A major consequence of this is that the detailed distribution of peaks
varies significantly depending on the interval of time considered. Indeed,
\cite{Kollath_Olah:1} recently applied time dependent harmonic analysis to the
sunspot number series and found that the dominant periods have shown 
systematic secular changes during the past 300~years
(Fig.~\ref{fig:Kollath}). For instance, the basic period seems to
have shortened from 11~years to 10~years between 1850 and 1950, with
some moderate increase in the last 50~years. (This is consistent with
the known anticorrelation between cycle length and amplitude,
cf.\ Sect.~\ref{sect:Waldmeier}.)

\subsubsection{The even--odd (a.k.a.\ Gnevyshev--Ohl) rule}
\label{sect:evenodd}

A cursory look at Fig.~\ref{fig:SSNrecord} shows that solar cycles often
follow an alternating pattern of higher and lower maxima. In this apparent
pattern, already noticed by the early observers \citep[e.g,][]{Turner:evenodd},
odd cycles have been typically stronger than even cycles in the last two
centuries.

This even--odd rule can be given two interpretations: a ``weak'' one of a general
tendency of alternation between even and odd cycles in amplitude, or a
``strong'' one of a specific numerical relation between the amplitudes of
consecutive cycles. 

Let us first consider the rule in its weak interpretation. At first sight the
rule admits many exceptions, but the amplitude of solar cycles depends on the
particular measuring method used. Exceptions from the even--odd alternation rule
become less common if a long term trend (calculated by applying a 12221 or 121
filter, see Sect.~\ref{sect:secular}) is subtracted from the data
\citep{Charbonneau:evenodd}, or if integrated cycle amplitudes (sums of annual
mean sunspot numbers during the cycle) are used \citep{Gnevyshev_Ohl}. 

In fact, as evident from, e.g., the work of \cite{Mursula_:evenodd} where cycle
amplitudes are based on group sunspot numbers and the amplitude of a cycle is
defined as the sum of the annual GSN value over the course of the cycle, the
odd--even alternation may be considered as \textit{strictly} valid
with only four exceptions:

\begin{itemize}
\item In the pairs 7\,--\,8 and 17\,--\,18, odd cycles are followed by stronger even
cycles at the end of Dalton minimum and at the beginning of the Modern Maximum.
These exceptions could be made to disappear by the subtraction of the long term
trend as suggested by \cite{Charbonneau:evenodd}. 
\item The pair 22\,--\,23 represents another apparent break of the weak even--odd
rule which is not easily explained away, even though the relative difference is
smaller if the Kislovodsk sunspot number series is used
\citep{Nagovitsyn_:evenodd}. The possibility is obviously there that the
subtraction of the long term trend may resolve the problem but we have no way
to tell in the near future.
\item Prior to cycle~5, the phase of the alternation was opposite, even cycles
being stronger than odd cycles. As cycle~4 is known to have been anomalously
long anyway (the so-called ``phase catastrophe'' in the solar cycle,
\citealt{Vitinsky_:book}) and its decaying phase is not well covered by
observations \citep{Vaquero}, this gave rise to the suggestion of a ``lost solar
cycle'' between cycles~4 and 5 (\citealt{Usoskin_:lostcycle}). This
cycle, however, would have been even more anomalous than cycle 4 and despite
intensive searches in historic data the evidence is still not quite conclusive
(\citealt{Krivova_:lostcycle}; see, however,
\citealt{Usoskin_:lostcycle.butterfly}).
\end{itemize}

The issue whether the even--odd rule can go through phase jumps or not is
important with respect to its possible origin. One plausible possibility is that
the alternation is due to the superposition of a steady primordial  magnetic
field component on the oscillatory magnetic field generated by the dynamo
\citep{Levy_Boyer}. In this case, any phase jump in the Gnevyshev--Ohl rule
should imply a phase jump in Hale's polarity rules, too.  Alternatively,
persistent even--odd alternation may also arise in nonlinear dynamos as a
period--2 limit cycle \citep{Durney:evenodd}; with a stochastic forcing
occasional phase jumps are also possible \citep{Charbonneau:evenodd,
  Charbonneau_:evenodd}.

While we have no information on this from the 18th century phase jump, we can be
certain that there was no such phase jump in polarities in the last two decades,
even though the even--odd rule seems to have been broken again. It will be
interesting to see when (and if) the even--odd rule settles in again, whether it
will have done so with a phase jump or not. For instance, if cycle~25 will again exceed
cycle~24 it would seem that no phase jump occurred and both theoretical  options
are still open. But if cycle~25 will represent a further weakening from
cycle~24, followed by a stronger cycle~26, a phase jump will have occurred,
which may exclude the primordial field origin of the rule if Hale's polarity
rules remain unchanged.

Let us now discuss the stronger interpretation of the even--odd rule. In the
first quantitative study of the relative amplitudes of consecutive cycles,
\cite{Gnevyshev_Ohl} found a rather tight correlation between the time
integrated amplitudes of even and subsequent odd cycles, while the correlation
between odd cycles and subsequent even cycles was found to be much less strong.
This gave rise to the notion that solar cycles come in ``two-packs'' as
even--odd pairs. \cite{Nagovitsyn_:evenodd} confirmed this puzzling finding on
the basis of data covering the whole period of telescopic observations (and
renumbering cycles before 1790 in accordance with the lost cycle hypothesis);
they also argue that cycle pair 22\,--\,23 does not deviate strongly from the
even--odd correlation curve so it should not be considered a ``real'' exception
to the even--odd rule.
{\newtext
\cite{Javaraiah:GOrule} analyzed the validity of the rule considering large
and small sunspot groups separately, and found that while for large groups the
rule holds with a few exceptions, for small groups a `reverse G-O rule' holds
where odd numbered cycles are consistently stronger than the preceding, rather
than the following even numbered cycle.
}

{\revone 
Shortly after its formulation by \cite{Gnevyshev_Ohl}, the (strong) even--odd
rule was used by \cite{Kopecky:pred19} to successfully predict the unusually
strong cycle~19. This made this rule quite popular for forecast purposes.} 
However, forecasts based on the
even--odd rule completely  failed for cycle~23, overpredicting the amplitude by
\textgreater~50\% \citep[see review by][]{Li_:pred.rev}. Taken together with the
implausibility of the suggested two-pack system, this shows that it is probably
wiser to take the position that ``extraordinary claims need extraordinary
evidence'' --- which is yet to be provided in the case of the ``strong''
even--odd rule.

Finally, in the context of the even--odd rule, it is also worth mentioning the
three-cycle regularity proposed by \cite{Ahluwalia:pred23}. Even though the
evidence presented for the alleged triadic pattern is not overwhelming, this
method resulted in one of the few successful predictions for the amplitude of
cycle~23.


\subsubsection{The Gleissberg cycle}
\label{sect:Gleissberg}

Besides the changes in the length of the 11-year cycle related to the
amplitude--cycle length correlation, even more significant are the variations in
the period of the so-called Gleissberg cycle \citep{Gleissberg:cycle}. This
``cycle'', corresponding to the 60\,--\,120 year ``plateau'' in
Fig.~\ref{fig:ssn_powsp0} was actually first noticed by Wolf, who placed it in
the range 55\,--\,80 years \citep[see][for a discussion of the history of the
studies of the Gleissberg cycle]{Richard:Gleissberg}. Researchers in the middle
of the 20th century characterized it as an 80\,--\,100 year variation.
Figure~\ref{fig:Kollath} explains why so widely differing periods were found in
different studies: the period has in fact shown a secular increase in the past
300 years, from about 50 years in the early 18th century, to a current value
exceeding 140 years. This increased length of the Gleissberg cycle also agrees
with the results of \cite{FDE_:midtermvar}.

The detection of $\sim$~100 year periods in a data set of 300 years is of course
always questionable, especially if the period is even claimed to be varying. 
However, the very clear and, most importantly, nearly linear secular trend seen
in Fig.~\ref{fig:Kollath} argues convincingly for the reality of the period in
question. This clear appearance of the period is due to the carefully optimized
choice of the kernel function in the time--frequency analysis, a method
resulting in a so-called pseudo-Wigner distribution (PWD). In addition, in their
study \cite{Kollath_Olah:1} presented a conscientious test of the
reliability of their methods, effectively proving that the most salient features
in their PWD are not artefacts. (The method was subsequently also applied to
stellar activity, \citealt{Kollath_Olah:2}.)  This is the most compelling evidence
for the reality of the Gleissberg cycle {\revone since 1750} yet presented.
{\newtext
Further evidence was more recently presented by \cite{LeMouel:SSA} using
singular spectrum analysis.
}

\subsubsection{Supersecular cycles}

For the 210-year Suess {\revone (known also as de Vries)} cycle,
\cite{McCracken_Beer:ICRC} presented further evidence for the temporally
intermittent nature of this marked peak in the spectrum of solar proxies. The
Suess cycle seems to have a role in regulating the recurrence rate of grand
minima. Grand minima, in turn, only seem to occur during \textless~1 kiloyear
intervals (``Sp\"orer events'') around the minima of the $\sim$~2400-year
Hallstatt cycle.

For further discussion of long term variations in solar activity we refer the
reader to the reviews by \cite{Beer_:SSR} and \cite{Usoskin:LRSP2}.

\subsection{Nonlinear methods}
\label{sect:nonlin}

\begin{nquote}
{\sl ``\dots every complicated question has a simple answer which is wrong. Analyzing a
time series with a nonlinear approach is definitely a complicated problem.
Simple answers have been repeatedly offered in the literature, quoting numerical
values for attractor dimensions for any conceivable system.''}\\
\strut\hfill\citep{Hegger_}
\end{nquote}

The nonlinearities in the dynamo equations readily give rise to chaotic
behaviour of the solutions. The long term behaviour of solar activity, with
phenomena like grand minima and grand maxima, is also suggestive of a chaotic
system. While chaotic systems are inherently unpredictable on long enough time
scales, their deterministic nature does admit forecast within a limited range.
It is therefore natural to explore this possibility from the point of view of
solar cycle prediction.

\subsubsection{Attractor analysis and phase space reconstruction: the
  pros \dots}
\label{sect:nldpro}

Assuming that the previous $(M-1)$ values of the sunspot number do in some way
determine the current expected value, our problem becomes restricted to an
$M$-dimensional \textit{phase space}, the dimensions being the current value and
the $(M-1)$ previous values. With a time series of length $N$, we have $N-M+1$
points fixed in the phase space, consecutive points being connected by a line.
This phase space trajectory is a sampling of the \textit{attractor} of the physical
system underlying the solar cycle (with some random noise added to it). The
attractor represents a \textit{mapping} in phase space which maps each point into
the one the system occupies in the next time step: if this mapping is known to
a good {\revone degree of precision}, it can be used to extend the trajectory
towards the future.

For the mapping to be known, $M$ needs to be high enough to avoid self-crossings
in the phase space trajectory (otherwise the mapping is not unique) but low
enough so that the trajectory still yields a good sampling of the attractor. The
lowest integer dimension satisfying these conditions is the \textit{embedding
dimension} $D$ of the attractor (which may have a fractal dimension itself).

Once the attractor has been identified, its mathematical description may be done
in three ways.

(1) \textit{Parametric fitting} of the attractor mapping in phase space. The simplest
method is the  piecewise linear fit suggested by \cite{Farmer_Sidorowich} and
applied in several solar prediction attempts, e.g., \cite{Kurths_Ruzmaikin}.
Using a method belonging to this group, \cite{Kilcik_} {\newtext gave a
correct prediction for Cycle~24.} Alternatively, a global
nonlinear fit can also be used: this is the method applied by \cite{Serre_Nesme}
as the first step in their global flow reconstruction (GFR) approach. 

(2) \textit{Nonparametric fitting.} The simplest nonparametric fit is to find the
closest known attractor point to ours (in the $(M-1)$-dimensional subspace
excluding the last value) and then using this for a prediction, as done  by
\cite{Jensen}. (This resulted in so large random forecast errors that it is
practically unsuitable for prediction.) 
{\newtext
A more refined approach is \textit{simplex projection analysis,} recently applied
by \cite{Singh+Bhargawa} for the problem of solar cycle prediction. (See also
\citealt{Sarp+:cyc25pred}.) A most remarkable extension of these methods was
presented by \cite{Covas:spatiotemp} who, instead of focusing on the time series
of SSN only, considered the problem of extending the whole spatiotemporal data
set of sunspot positions (butterfly diagram) into the future.
}
\textit{Neural networks,} discussed in more
detail in Sect.~\ref{sect:neural} below, are a much more sophisticated
nonparametric fitting device.

(3) Indirectly, one may try to find a set of differential equations describing a
system that gives rise to an attractor with properties similar to the observed.
In this case there is no guarantee that the derived equations will be unique, as
an alternative, completely different set may also give rise to a very similar 
attractor. This arbitrariness of the choice is not necessarily a problem from
the point of view of prediction as it is only the mapping (the attractor
structure) that matters. Such phase space reconstruction by a set of governing
equations was performed, e.g., by \cite{Serre_Nesme} or \cite{Aguirre_}. On the
other hand, instead of putting up with any arbitrary set of equations correctly
reproducing the phase space, one might make an effort to find a set with a
structure reasonably similar to the dynamo equations so they can be given a
meaningful physical interpretation. Methods following this latter approach were
discussed in Sects.~\ref{sect:truncated} and \ref{sect:oscillator}.

\subsubsection{\dots the cons \dots}

Finding the embedding dimension and the attractor structure is not a trivial
task, as shown by the widely diverging results different researchers arrived at.
One way to find the correct embedding dimension is the false nearest neighbours
method \citep{Kennel_}, essentially designed to identify self-crossings in the
phase space trajectory, in which case the dimension $M$ needs to be increased.
But self-crossings are to some extent inevitable, due to the stochastic
component superimposed on the deterministic skeleton of the system. 

As a result, the determination of the minimal necessary embedding dimension is
usually done indirectly. One indirect method fairly popular in the solar
community is the approach proposed by \cite{Sugihara_May} where the correct
dimension is basically figured out on the basis of how successfully the model,
fit to the first part of the data set, can ``predict'' the second part (using a
piecewise linear mapping).

Another widely used approach, due to \cite{Grassberger_Procaccia}, starts by
determining the correlation dimension of the attractor, by simply counting how
the number of neighbours in an embedding space of dimension $M\gg 1$ increases
with the distance from a point. If the attractor is a lower dimensional manifold
in the embedding space and it is sufficiently densely sampled by our data 
then the logarithmic steepness $d$ of this function should be constant over a
considerable stretch of the curve: this is the correlation dimension $d$. Now,
we can increase $M$ gradually and see at what value $d$ saturates: that value
determines the attractor dimension, while the value of $M$ where saturation is
reached yields the embedding dimension. 

The first nonlinear time series studies of solar activity indicators suggested a
time series spacing of 2\,--\,5 years, an attractor dimension $\sim$~2\,--\,3 and an
embedding dimension of 3\,--\,4 \citep{Kurths_Ruzmaikin, Gizzatullina_}. Other
researchers, however, were unable to confirm these results, either reporting
very different values or not finding any evidence for a low dimensional
attractor at all \citep{Calvo_, Price_:nochaos, Carbonell_:nochaos, Kilcik_,
Hanslmeier_Brajsa}. In particular, I would like to call attention to the paper
by \cite{Jensen}, which, according to ADS and WoS, has received a grand total of
zero citations up to 2010, yet it displays an exemplary no-nonsense approach
to the problem of sunspot number prediction by nonlinear time series methods.
Unlike so many other researchers, the author of that paper does not 
{\revone enforce a linear fit on} the logarithmic correlation integral
curve (his Figure~4); instead, he demonstrates on a simple example that the
actual curve can be perfectly well reproduced by a simple stochastic process.

These contradictory results obviously do not imply that the mechanism generating
solar activity is \textit{not} chaotic. For a reliable determination a long time
series is desirable to ensure a sufficiently large number of neighbours in a
phase space volume small enough compared to the global scale of the attractor.
Solar data sets (even the cosmogenic radionuclide proxies extending over
millennia but providing only a decadal sampling) are typically too short and
sparse for this. In addition, clearly distinguishing between the phase space
fingerprints of chaotic and stochastic processes is an unsolved problem of
nonlinear dynamics which is not unique to solar physics. A number of methods
have been suggested to identify chaos unambiguously in a time series but none of
them has been generally accepted and this topic is currently a subject of
ongoing research --- see, e.g., the work of \cite{Freitas_} which demonstrates that
the method of ``noise titration'', somewhat akin to the Sugihara--May algorithm,
is uncapable of distinguishing superimposed coloured noise from intrinsically
chaotic systems.

\subsubsection{\dots and the upshot}

Starting from the 1980s, many researchers jumped on the chaos bandwagon, applying
nonlinear time series methods designed for the study of chaotic systems to a
wide variety of empirical data, including solar activity parameters. From the
1990s, however, especially after the publication of the influential book by
\cite{Kantz_Schreiber}, it was increasingly realized that the applicability of
these nonlinear algorithms does not in itself prove the predominantly chaotic
nature of the system considered. In particular, stochastic noise superposed on a
simple, regular, deterministic skeleton can also give rise to phase space
characteristics that are hard to tell from low dimensional chaos, especially if
strong smoothing is applied to the data. As a result, the pendulum has swung in
the opposite direction and currently the prevailing view is that there is no
clear cut evidence for chaos in solar activity data \citep{Panchev_Tsekov}.

One might take the position that any forecast based on attractor analysis is
only as good as the underlying assumption of a chaotic system is: if that
assumption is unverifiable from the data, prediction attempts are pointless.
This, however, is probably a too hasty judgment. The potentially
most useful product of phase space reconstruction attempts is the inferences
they allow regarding the nature of the underlying physical system (chaotic or
not), even offering a chance to constrain the form of the dynamo equations
relevant for the Sun. As discussed in the {\revone previous} section, such
truncated models may be used for forecast directly, or alternatively, the
insight they yield into the mechanisms of the dynamo may be used to construct
more sophisticated dynamo models.

\subsubsection{Neural networks}
\label{sect:neural}

Neural networks are algorithms built up from a large number of small
interconnected units (``neurons'' or ``threshold logic units''), each of which
is only capable of performing a simple nonlinear operation on an input signal,
essentially described by a step function or its generalized (rounded) version, a
sigmoid function. To identify the optimal values of thresholds and weights
parameterizing the sigmoid functions of each neuron, an algorithm called ``back
propagation rule'' is employed which minimizes (with or without human guidance)
the error between the predicted and observed values in a process called
``training'' of the network. Once the network has been correctly trained, it is
capable of further predictions.

The point is that any arbitrary multidimensional nonlinear mapping may be
approximated by a combination of stepfunctions to a good degree --- so, as
mentioned in Sect.~\ref{sect:nldpro} above, the neural network can be used to
find the nonlinear mapping corresponding to the attractor of the given time
series. 

More detailed introductions to the method are given by \cite{Blais_Mertz},
\cite{Conway:neural}, and by \cite{Calvo_}; the latter authors were also the
first to apply a neural network for sunspot number prediction. Unfortunately,
despite their claim of being able to ``predict'' (i.e., postdict) some earlier
cycles correctly, their prediction for Cycle~23 was off by a wide margin
(predicted peak amplitude of 166 [v1] as opposed to 121 observed). One of the
neural network forecasts for Cycle~24 \citep{Maris_Oncica} was equally far off,
while another one \citep{Uwamahoro} yielded a more conservative value.
{\newtext
A prediction for Cycle~25 based on a version of the neural networks approach was
given by \cite{Attia+:neurofuzzy}, {\revone resulting in a cycle amplitude
slightly below that of cycle 24.}}
}



\section{Summary evaluation}
\label{sect:summary}

The performance of various forecast methods in Cycles 21\,--\,23 was
discussed by \cite{Li_:pred.rev} and \cite{Kane:pred23rev}. 
{\newtext
Predictions for Cycle~24 were presented in \cite{Edition1} (Table~1),
\cite{Pesnell} and \cite{Pesnell2012rev}; the experiences gained in this cycle
were discussed in \cite{Pesnell2016rev}. 
}

\textit{Precursor methods} {\newtext generally} stand out with their internally
consistent forecasts which for Cycles 21 and 22 proved to be correct. For
Cycle~23, these methods were still internally consistent in their prediction,
mostly scattering in a narrow range between\footnote{Version~1
sunspot numbers are used throughout this section, unless otherwise
indicated.} 150 and 170; however, the cycle amplitude proved to be considerably lower
($\Rmax=121$). It should be noted, however, that one precursor based prediction,
that of \cite{Schatten_:pred23} was significantly lower than the rest
($138 \pm 30$) and within $0.6\,\sigma$ of the actual value. 
{\newtext
For Cycle~24 most precursor methods again consistently indicated a
lower-than-average cycle amplitude in the range 70\,--\,100, except Feynman's
geomagnetic precursor method which mistakenly resulted in a very high value of
150. (The likely reasons were discussed in Sect.~\ref{sect:geomg} above.)   The
closest hit at the actual peak value of 67 [v1] or 116 [v2] was produced by the
Minimax3 and the polar field precursor methods {\revone (\citealt{Edition1};
\citealt{Svalgaard_:prediction24})}.
}
Indeed, the {\newtext polar precursor} method of \cite{Schatten_:pred22} and
\cite{Schatten_:pred23}, has consistently proven its skill in all cycles. As
discussed in Sect.~\ref{sect:polar}, this method is essentially based on the
polar magnetic field strength as precursor.

{\newtext 
\textit{Model based methods} are a new development that have only had limited
occasion to prove their skill.  For Cycle~24 only three conceptually different
such predictions were made, all of which were based on dynamo models. The
pioneering attempt by \cite{Dikpati_:prediction} proved to be way too high (see
Sect.~\ref{sec:model_axi} for a discussion of the possible reasons). The flux
transport dynamo based predictions of \cite{Choudhuri_:prediction} and
\cite{Jiang_:prediction} were close hits; however, as already mentioned, these
employed a technique (adjusting the dipole moment at the minimum to 
observations) which renders them essentially a polar field precursor method in
disguise. Another correct model-based forecast was given by
\cite{Kitiashvili_:prediction}; the good performance of this dynamo model,
seemingly rather far removed from physical reality, still needs to be understood
and it may possibly be equivalent to a phase space reconstruction method, as in
item (3) of Sect.~\ref{sect:nldpro}.
}

\textit{Extrapolation methods} as a whole have shown a much less impressive
performance. Overall, the statistical distribution of maximum amplitude values
predicted by ``real'' forecasts made using these methods (i.e., forecasts made
at or before the minimum epoch) for any given cycle does not seem to
significantly differ from the long term climatological average of the solar
cycle quoted in Sect.~\ref{sect:cycle} above. It would of course
be a hasty judgement to dismiss each of the widely differing individual
approaches comprised in this class simply due to the poor overall performance of
the group. In particular, some novel methods {\newtext introduced in the last 
decades,} such as SSA, {\newtext phase space reconstruction} or neural networks
have hardly had a chance to debut, so their further performance will be worth
monitoring in upcoming cycles.

{\revone The effect of the sunspot number revision on solar cycle prediction
methods is limited to numerical corrections of minor importance.}

\newsection{Forecasts for Cycle~25}
\label{sect:cyc25}

Table~\ref{table:pred} presents a collection of forecasts for the amplitude of
Cycle~25, without claiming completeness. The objective was to include one or two
representative forecasts from each category. As the time of the
minimum starting Cycle~25 is yet to be established, all these forecast
qualify as ``early'', in the sense that the most well-established
methods, relying on precursor values evaluated at the time of minimum,
cannot yet be applied.

It is clear also from this table that the issue of cycle prediction is
less contentious for Cycle~25 than it was for Cycle~24. The
overwhelming majority of forecasts agree that the amplitude of Cycle
25 is most likely to lie within $\pm 20\%$ of Cycle~24, i.e., no
major change in the level of solar activity is expected. The remaining
controversy mostly concerns where in this range the cycle will peak.  
Dynamo based predictions indicate that Cycle~25 will peak at somewhat
lower values than Cycle~24, while precursor techniques and SFT
modelling suggest a cycle amplitude comparable to or slightly higher
than the previous cycle. Two recent neural network based forecasts
yield a weak cycle peaking quite early. 

Following the development of the sunspot number during the next few
years will be most interesting in the light of these predictions, and
it may shed more light on the strong and weak points in our
understanding of the roots of solar activity variations.


\begin{acknowledgement}
This work was supported by the Hungarian National Research,
Development and Innovation Fund (grant no. NKFI K-128384) and by the
European Union's Horizon 2020 research and innovation programme under
grant agreement No.~739500.

Some figures (or their input data) have been provided by J.~Jiang, A.~Kilcik and
P.~Charbonneau, as noted in the captions. Source for International Sunspot
Numbers: WDC-SILSO, Royal Observatory of Belgium, Brussels. Wilcox Solar
Observatory data used in this study was obtained via the web site
{\url{wso.stanford.edu}}, courtesy of J.~T.~Hoeksema. The Wilcox Solar
Observatory is currently supported by NASA. 

\end{acknowledgement}

\begin{landscape}
\begin{table}
\caption{A selection of early forecasts for Cycle~25}
\label{table:pred}
{\small
\begin{tabular}{lcccl}
\toprule
Category & Minimum & Maximum & Peak amplitude & Reference \\
\midrule
Internal precursors & 2019.9 & 2023.8 & 175 [154--202] &  \cite{Li+:pred25} \\
External precursor  &  &  &  &  \\
 \ polar precursor & & & 117\,$\pm$\,15 & Table~\ref{table:polarprecursors} here\\
 \ polar precursor & & & 136\,$\pm$\,48 & \cite{Pesnell+:SoDA2018} \\
 \ helicity  & & &  117 & \cite{Hawkes+Berger} \\
 \ SoDA & & 2025.2\,$\pm$\,1.5  & 120\,$\pm$\,39 & based on \cite{Pesnell+:SoDA2018} \\
 \ rush-to-the-poles & 2019.4 &  2024.8 &  130  &  \cite{Petrovay+:greenlpred} \\
Model-based: SFT  &  &  &  &  \\
 \ SFT    & & & 124\,$\pm$\,31  & \cite{Jiang+:1cycle} \\
 \ AFT & 2020.9 & & 110 &  \cite{Upton+:pred25update} \\
Model-based: dynamo  &  &  &  &  \\
 \ $2{\times}2$D & 2020.5\,$\pm$\,0.12 &2027.2\,$\pm$\,1.0  & 89$^{+29}_{-14}$ & \cite{Labonville+} \\
 \ Truncated & 2019--20 & 2024\,$\pm$\,1 & 90\,$\pm$\,15 & \cite{Kitiashvili:cyc25pred} \\
Spectral  &  &  &  &  \\
 \ wavelet decomposition tree & & 2023.4 & 132 & \cite{Rigozo+} \\
Attractor analysis  &  &  &  &  \\
 \ simplex projection analysis & & 2024.0\,$\pm$\,0.6 &103\,$\pm$\,25 & \cite{Singh+Bhargawa} \\
 \ simplex proj./time-delay & & 2023.2\,$\pm$\,1.1 & 154\,$\pm$\,12 & \cite{Sarp+:cyc25pred} \\
Neural networks  &  &  &  &  \\
 \ neuro-fuzzy    & & 2022 & 90.7\,$\pm$\,8 & \cite{Attia+:neurofuzzy} \\
 \ spatiotemporal & & 2022--23 & 57\,$\pm$\,17 &	\cite{Covas2019} \\
\midrule
 Cycle~24 [comparison] & 2008.9 & 2014.3 & 116 & \\
\bottomrule
\end{tabular}
\\
{\footnotesize 
In the case of SFT models, forecasts were obtained by multiplying the amplitude
of Cycle~24 with the predicted \% increase in polar field strength between the
two minima. Errors resulting from a natural scatter in the polar field--cycle
ampliude relation are therefore not included in the error range given.}
}
\end{table}
\end{landscape}




\bibliography{refs_v1,refs_v2}

\end{document}